\documentclass[11pt]{cernrep}
\usepackage{graphicx}
\usepackage{here}

\def\lsim{\mathrel{\mathop
  {\hbox{\lower0.5ex\hbox{$\sim$}\kern-0.8em\lower-0.7ex\hbox{$<$}}}}}
\def\gsim{\mathrel{\mathop
  {\hbox{\lower0.5ex\hbox{$\sim$}\kern-0.8em\lower-0.7ex\hbox{$>$}}}}}

\begin{document}

\newcommand{\trh}{T_{\rm rh}}
\newcommand{\teff}{T_{\rm eff}}
\newcommand{\delcp}{\delta_{_{\rm CP}}}
\newcommand{\mh}{m_{_{\rm H}}}
\newcommand{\mw}{m_{_{\rm W}}}
\newcommand{\alphaw}{\alpha_{_{\rm W}}}
\newcommand{\ncs}{N_{_{\rm CS}}}

\title{ASTROPHYSICS AND COSMOLOGY}
\author{Juan Garc\'\i a-Bellido}
\institute{Theoretical Physics Group, Blackett Laboratory, 
Imperial College of Science, \\
Technology and Medicine, Prince Consort Road, London SW7 2BZ, U.K.}
\maketitle
\begin{abstract}
These notes are intended as an introductory course for experimental
particle physicists interested in the recent developments in
astrophysics and cosmo\-logy. I will describe the standard Big Bang
theory of the evolution of the universe, with its successes and
shortcomings, which will lead to inflationary cosmology as the paradigm
for the origin of the global structure of the universe as well as the
origin of the spectrum of density perturbations responsible for
structure in our local patch. I will present a review of the very rich
phenomenology that we have in cosmology today, as well as evidence for
the observational revolution that this field is going through, which
will provide us, in the next few years, with an accurate determination
of the parameters of our standard cosmological model.
\end{abstract}

\section{GENERAL INTRODUCTION}

Cosmology (from the Greek: {\em kosmos}, universe, world, order, and
{\it logos}, word, theory) is probably the most ancient body of
knowledge, dating from as far back as the predictions of seasons by
early civilizations. Yet, until recently, we could only answer to some
of its more basic questions with an order of magnitude estimate. This
poor state of affairs has dramatically changed in the last few years,
thanks to (what else?) raw data, coming from precise measurements of a
wide range of cosmological parameters. Furthermore, we are entering a
precision era in cosmology, and soon most of our observables will be
measured with a few percent accuracy. We are truly living in the Golden
Age of Cosmology. It is a very exciting time and I will try to
communicate this enthusiasm to you.

Important results are coming out almost every month from a large set of
experiments, which provide crucial information about the universe origin
and evolution; so rapidly that these notes will probably be outdated
before they are in print as a CERN report. In fact, some of the results
I mentioned during the Summer School have already been improved,
specially in the area of the microwave background anisotropies.
Nevertheless, most of the new data can be interpreted within a coherent
framework known as the standard cosmological model, based on the Big
Bang theory of the universe and the inflationary paradigm, which is with
us for two decades. I will try to make such a theoretical model
accesible to young experimental particle physicists with little or no
previous knowledge about general relativity and curved space-time, but
with some knowledge of quantum field theory and the standard model of
particle physics.

\section{INTRODUCTION TO BIG BANG COSMOLOGY}

Our present understanding of the universe is based upon the successful
hot Big Bang theory, which explains its evolution from the first
fraction of a second to our present age, around 13 billion years
later. This theory rests upon four strong pillars, a theoretical
framework based on general relativity, as put forward by Albert
Einstein~\cite{Einstein} and Alexander A. Friedmann~\cite{Friedmann} in
the 1920s, and three robust observational facts: First, the expansion of
the universe, discovered by Edwin P. Hubble~\cite{Hubble} in the 1930s,
as a recession of galaxies at a speed proportional to their distance
from us. Second, the relative abundance of light elements, explained by
George Gamow~\cite{Gamow} in the 1940s, mainly that of helium, deuterium
and lithium, which were cooked from the nuclear reactions that took
place at around a second to a few minutes after the Big Bang, when the
universe was a few times hotter than the core of the sun. Third, the
cosmic microwave background (CMB), the afterglow of the Big Bang,
discovered in 1965 by Arno A. Penzias and Robert W. Wilson~\cite{Wilson}
as a very isotropic blackbody radiation at a temperature of about 3
degrees Kelvin, emitted when the universe was cold enough to form
neutral atoms, and photons decoupled from matter, approximately 500,000
years after the Big Bang. Today, these observations are confirmed to
within a few percent accuracy, and have helped establish the hot Big
Bang as the preferred model of the universe.

\subsection{Friedmann--Robertson--Walker universes}

Where are we in the universe? During our lectures, of course, we were in
\v{C}asta Papierni\v{c}ka, in ``the heart of Europe'', on planet Earth,
rotating (8 light-minutes away) around the Sun, an ordinary star 8.5
kpc\footnote{One parallax second (1 pc), {\em parsec} for short,
corresponds to a distance of about 3.26 light-years or $3\times10^{18}$
cm.} from the center of our galaxy, the Milky Way, which is part of the
local group, within the Virgo cluster of galaxies (of size a few Mpc),
itself part of a supercluster (of size $\sim 100$ Mpc), within the
visible universe ($\sim {\rm few}\times1000$ Mpc), most probably a tiny
homogeneous patch of the infinite global structure of space-time, much
beyond our observable universe.

Cosmology studies the universe as we see it. Due to our inherent
inability to experiment with it, its origin and evolution has always
been prone to wild speculation. However, cosmology was born as a science
with the advent of general relativity and the realization that the
geometry of space-time, and thus the general attraction of matter, is
determined by the energy content of the universe~\cite{Weinberg},
\begin{equation}\label{EinsteinEquations}
G_{\mu\nu}\equiv R_{\mu\nu} - {1\over2}g_{\mu\nu}R = 8\pi G\,T_{\mu\nu}
+ \Lambda\,g_{\mu\nu} \,.
\end{equation}
These non-linear equations are simply too difficult to solve without
some insight coming from the symmetries of the problem at hand: the
universe itself. At the time (1917-1922) the known (observed) universe
extended a few hundreds of parsecs away, to the galaxies in the local
group, Andromeda and the Large and Small Magellanic Clouds: The universe
looked extremely anisotropic. Nevertheless, both Einstein and Friedmann
speculated that the most ``reasonable'' symmetry for the universe at
large should be {\em homogeneity} at all points, and thus {\em
isotropy}. It was not until the detection, a few decades later, of the
microwave background by Penzias and Wilson that this important
assumption was finally put onto firm experimental ground. So, what is
the most general metric satisfying homogeneity and isotropy at large
scales? The Friedmann-Robertson-Walker (FRW) metric, written here in
terms of the invariant geodesic distance $ds^2=g_{\mu\nu}dx^\mu dx^\nu$
in four dimensions, $\mu=0,1,2,3$, see Ref.~\cite{Weinberg},\footnote{I
am using $c=1$ everywhere, unless specified.}
\begin{equation}\label{FRWmetric}
ds^2 = dt^2 - a^2(t)\left[{dr^2\over1-K\,r^2} +
r^2(d\theta^2 + \sin^2\theta\,d\phi^2)\right]\,,
\end{equation}
characterized by just two quantities, a {\em scale factor} $a(t)$,
which determines the physical size of the universe, and a constant $K$,
which characterizes the {\em spatial} curvature of the universe,
\begin{equation}\label{SpatialCurvature}
{}^{(3)}\!R = {6K\over a^2(t)}\,. \hspace{3cm}
\left\{\begin{array}{lr}K=-1&\hspace{1cm}{\rm OPEN}\\
K=0&\hspace{1cm}{\rm FLAT}\\K=+1&\hspace{1cm}{\rm CLOSED}
\end{array}\right.
\end{equation}
Spatially open, flat and closed universes have different geometries.
Light geodesics on these universes behave differently, and thus could in
principle be distinguished observationally, as we shall discuss later.
Apart from the three-dimensional spatial curvature, we can also compute
a four-dimensional {\em space-time} curvature,
\begin{equation}\label{SpacetimeCurvature}
{}^{(4)}\!R = 6{\ddot a\over a} + 6\left({\dot a\over a}\right)^2 + 
6{K\over a^2}\,. 
\end{equation}
Depending on the dynamics (and thus on the matter/energy content) of the
universe, we will have different possible outcomes of its evolution.
The universe may expand for ever, recollapse in the future or approach
an asymptotic state in between.

\subsubsection{The expansion of the universe}

In 1929, Edwin P. Hubble observed a redshift in the spectra of distant
galaxies, which indicated that they were receding from us at a velocity
proportional to their distance to us~\cite{Hubble}. This was correctly
interpreted as mainly due to the expansion of the universe, that is, to
the fact that the scale factor today is larger than when the photons
were emitted by the observed galaxies. For simplicity, consider the
metric of a spatially flat universe, $ds^2=dt^2-a^2(t)\,d\vec x^2$ (the
generalization of the following argument to curved space is
straightforward). The scale factor $a(t)$ gives {\em physical size} to
the spatial coordinates $\vec x$, and the expansion is nothing but a
change of scale (of spatial units) with time. Except for {\em peculiar
velocities}, i.e. motion due to the local attraction of matter, galaxies
do not move in coordinate space, it is the space-time fabric which is
stretching between galaxies. Due to this continuous stretching, the
observed wavelength of photons coming from distant objects is greater
than when they were emitted by a factor precisely equal to the ratio of
scale factors,
\begin{equation}\label{Redshift}
{\lambda_{\rm obs}\over\lambda_{\rm em}} = {a_0\over a} \equiv 1 + z\,,
\end{equation}
where $a_0$ is the present value of the scale factor. Since the universe
today is larger than in the past, the observed wavelengths will be
shifted towards the red, or {\em redshifted}, by an amount characterized
by $z$, the redshift parameter.

In the context of a FRW metric, the universe expansion is characterized
by a quantity known as the Hubble rate of expansion, $H(t) = \dot
a(t)/a(t)$, whose value today is denoted by $H_0$. As I shall deduce
later, it is possible to compute the relation between the physical
distance $d_L$ and the present rate of expansion, in terms of the
redshift parameter,\footnote{The subscript $L$ refers to Luminosity,
which characterizes the amount of light emitted by an object. See
Eq.~(\ref{LuminosityDistance}).}
\begin{equation}\label{PhysicalDistance}
H_0\,d_L = z + {1\over2}(1-q_0)\,z^2 + {\cal O}(z^3)\,.
\end{equation}
At small distances from us, i.e. at $z\ll1$, we can safely keep only the
linear term, and thus the recession velocity becomes proportional to the
distance from us, $v = c\,z = H_0\,d_L$, the proportionality constant
being the Hubble rate, $H_0$. This expression constitutes the so-called
Hubble law, and is spectacularly confirmed by a huge range of data, up
to distances of hundreds of megaparsecs. In fact, only recently
measurements from very bright and distant supernovae, at $z\simeq1$,
were obtained, and are beginning to probe the second-order term,
proportional to the deceleration parameter $q_0$, see
Eq.~(\ref{DecelerationParameter}). I will come back to these
measurements in Section~3.

One may be puzzled as to {\em why} do we see such a stretching of
space-time. Indeed, if all spatial distances are scaled with a universal
scale factor, our local measuring units (our rulers) should also be
stretched, and therefore we should not see the difference when comparing
the two distances (e.g. the two wavelengths) at different times. The
reason we see the difference is because we live in a gravitationally
bound system, decoupled from the expansion of the universe: local
spatial units in these systems are {\em not} stretched by the
expansion.\footnote{The local space-time of a gravitationally bound
system is described by the Schwarzschild metric, which is
static~\cite{Weinberg}.} The wavelengths of photons are stretched along
their geodesic path from one galaxy to another. In this consistent world
picture, galaxies are like point particles, moving as a fluid in an
expanding universe.

\subsubsection{The matter and energy content of the universe}

So far I have only discussed the geometrical aspects of space-time. Let
us now consider the matter and energy content of such a universe. The
most general matter fluid consistent with the assumption of homogeneity
and isotropy is a perfect fluid, one in which an observer {\em comoving
with the fluid} would see the universe around it as isotropic. The
energy momentum tensor associated with such a fluid can be written 
as~\cite{Weinberg}
\begin{equation}\label{PerfectFluid}
T^{\mu\nu} = p\,g^{\mu\nu} + (p+\rho)\,U^\mu U^\nu\,,
\end{equation}
where $p(t)$ and $\rho(t)$ are the pressure and energy density of the
fluid at a given time in the expansion, and $U^\mu$ is the comoving
four-velocity, satisfying $U^\mu U_\mu=-1$.

Let us now write the equations of motion of such a fluid in an expanding
universe. According to general relativity, these equations can be
deduced from the Einstein equations (\ref{EinsteinEquations}), where we
substitute the FRW metric (\ref{FRWmetric}) and the perfect fluid tensor 
(\ref{PerfectFluid}). The $\mu=\nu=0$ component of the Einstein
equations constitutes the so-called Friedmann equation
\begin{equation}\label{FriedmannEquation}
H^2 = \left({\dot a\over a}\right)^2 = {8\pi G\over3}\,\rho +
{\Lambda\over3} - {K\over a^2}\,,
\end{equation}
where I have treated the cosmological constant $\Lambda$ as a different
component from matter. In fact, it can be associated with the vacuum
energy of quantum field theory, although we still do not understand why
should it have such a small value (120 orders of magnitude below that
predicted by quantum theory), if it is non-zero. This constitutes today
one of the most fundamental problems of physics, let alone cosmology.

The conservation of energy ($T^{\mu\nu}_{\hspace{3mm};\nu} = 0$), a
direct consequence of the general covariance of the theory
($G^{\mu\nu}_{\hspace{3mm};\nu} = 0$), can be written in terms of the
FRW metric and the perfect fluid tensor (\ref{PerfectFluid}) as
\begin{equation}\label{EnergyConservation}
{d \over dt}\Big(\rho\,a^3\Big) + p\,{d \over dt}\Big(a^3\Big) = 0\,,
\end{equation}
where the energy density and pressure can be split into its matter and
radiation components, $\rho=\rho_{\rm M}+\rho_{\rm R}, \\ p=p_{\rm
M}+p_{\rm R}$, with corresponding equations of state, $p_{\rm M}=0, \ \
p_{\rm R}=\rho_{\rm R}/3$. Together, the Friedmann and the
energy-conservation equation give the evolution equation for the scale
factor,
\begin{equation}\label{Evolution}
{\ddot a\over a} = -\,{4\pi G\over3}\,(\rho + 3p) + {\Lambda\over3}\,,
\end{equation}

I will now make a few useful definitions. We can write the Hubble
parameter today $H_0$ in units of 100 km\,s$^{-1}$Mpc$^{-1}$, in terms
of which one can estimate the order of magnitude for the present size
and age of the universe,
\begin{eqnarray}
H_0&=& 100\,h\ \ {\rm km\,s}^{-1}{\rm Mpc}^{-1}\,,\\
c\,H_0^{-1} &=& 3000\,h^{-1}\ {\rm Mpc}\,,\\
H_0^{-1} &=& 9.773\,h^{-1}\ {\rm Gyr}\,.
\end{eqnarray}
The parameter $h$ has been measured to be in the range $0.4 < h< 1$ for
decades, and only in the last few years has it been found to lie within
10\% of $h=0.65$.  I will discuss those recent measurements in the next
Section.

One can also define a {\em critical} density $\rho_c$, that which in the
absence of a cosmological constant would correspond to a flat universe,
\begin{eqnarray}\label{CriticalDensity}
\rho_c\equiv{3H_0^2\over8\pi G}&=& 1.88\,h^2\,10^{-29}\ {\rm g/cm}^3\\
&=& 2.77\,h^{-1}\,10^{11}\ M_\odot/(h^{-1}\,{\rm Mpc})^3\,,
\end{eqnarray}
where $M_\odot=1.989\times10^{33}$ g \ is a solar mass unit. The
critical density $\rho_c$ corresponds to approximately 4 protons per
cubic meter, certainly a very dilute fluid! In terms of the critical
density it is possible to define the ratios $\Omega_i \equiv
\rho_i/\rho_c$, for matter, radiation, cosmological constant and even
curvature, today, 
\begin{eqnarray}\label{Omega}
&&\Omega_{\rm M}={8\pi G\,\rho_{\rm M}\over3H_0^2}\hspace{3cm} 
\Omega_{\rm R}={8\pi G\,\rho_{\rm R}\over3H_0^2}\\
&&\,\Omega_\Lambda={\Lambda\over3H_0^2}\hspace{3.5cm}
\Omega_K=-\,{K\over a_0^2H_0^2}\,.\end{eqnarray} 

We can evaluate today the radiation component $\Omega_{\rm R}$,
corresponding to relativistic particles, from the density of microwave
background photons, $\rho_{_{\rm CMB}} = {\pi^2\over15}(kT_{_{\rm
CMB}})^4/(\hbar c)^3 = 4.5\times10^{-34}\ {\rm g/cm}^3$, which gives
$\Omega_{_{\rm CMB}} = 2.4\times 10^{-5}\ h^{-2}$. Three massless
neutrinos contribute an even smaller amount. Therefore, we can safely
neglect the contribution of relativistic particles to the total density
of the universe today, which is dominated either by non-relativistic
particles (baryons, dark matter or massive neutrinos) or by a
cosmological constant, and write the rate of expansion $H^2$ in terms of
its value today,
\begin{equation}\label{H2a}
H^2(a) = H_0^2\left(\Omega_{\rm R}\,{a_0^4\over a^4} + 
\Omega_{\rm M}\,{a_0^3\over a^3} + \Omega_\Lambda + 
\Omega_K\,{a_0^2\over a^2}\right)\,.
\end{equation}
An interesting consequence of these redefinitions is that I can now
write the Friedmann equation today, $a=a_0$, as a {\em cosmic sum rule}, 
\begin{equation}\label{CosmicSumRule}
1 = \Omega_{\rm M} + \Omega_\Lambda + \Omega_K\,,
\end{equation}
where we have neglected $\Omega_{\rm R}$ today. That is, in the context
of a FRW universe, the total fraction of matter density, cosmological
constant and spatial curvature today must add up to one.  For instance,
if we measure one of the three components, say the spatial curvature, we
can deduce the sum of the other two. Making use of the cosmic sum rule
today, we can write the matter and cosmological constant as a function
of the scale factor ($a_0\equiv1$)
\begin{eqnarray}\label{OmegaM}
&&\Omega_{\rm M}(a)={8\pi G\,\rho_{\rm M}\over3H^2(a)}=
{\Omega_{\rm M}\over a + \Omega_{\rm M}(1-a) + 
\Omega_\Lambda(a^3-a)}\hspace{5mm}\left\{
\begin{array}{lr}\stackrel{a\rightarrow0}{\longrightarrow}&1\\
\stackrel{a\rightarrow\infty}{\longrightarrow}&0\end{array}\right.\,,\\[2mm]
&&\,\Omega_\Lambda(a)={\Lambda\over3H^2(a)}={\Omega_\Lambda a^3\over
a + \Omega_{\rm M}(1-a) + \Omega_\Lambda(a^3-a)}\hspace{5mm}\left\{
\begin{array}{lr}\stackrel{a\rightarrow0}{\longrightarrow}&0\\
\label{OmegaL}
\stackrel{a\rightarrow\infty}{\longrightarrow}&1\end{array}\right.\,.
\end{eqnarray} 
This implies that for sufficiently early times, $a\ll1$, all
matter-dominated FRW universes can be described by Einstein-de Sitter
(EdS) models ($\Omega_K=0, \ \Omega_\Lambda=0$).\footnote{Note that in
the limit $a\to0$ the radiation component starts dominating, see
Eq.~(\ref{H2a}), but we still recover the EdS model.} On the other hand,
the vacuum energy will always dominate in the future.

Another relationship which becomes very useful is that of the
cosmological deceleration parameter today, $q_0$, in terms of the matter
and cosmological constant components of the universe, see 
Eq.~(\ref{Evolution}),
\begin{equation}\label{DecelerationParameter}
q_0 \equiv -\left.{\ddot a\over aH^2}\right|_0 = {1\over2}\Omega_{\rm M} - 
\Omega_\Lambda\,,
\end{equation}
which is independent of the spatial curvature. Uniform expansion
corresponds to $q_0=0$ and requires a precise cancellation: $\Omega_{\rm
M} = 2\Omega_\Lambda$. It represents spatial sections that are expanding
at a fixed rate, its scale factor growing by the same amount in
equally-spaced time intervals. Accelerated expansion corresponds to
$q_0<0$ and comes about whenever $\Omega_{\rm M} < 2\Omega_\Lambda$:
spatial sections expand at an increasing rate, their scale factor
growing at a greater speed with each time interval.  Decelerated
expansion corresponds to $q_0>0$ and occurs whenever $\Omega_{\rm M} >
2\Omega_\Lambda$: spatial sections expand at a decreasing rate, their
scale factor growing at a smaller speed with each time interval.

\subsubsection{Mechanical analogy}

\begin{figure}[htb]
\begin{center}
\includegraphics[width=7cm]{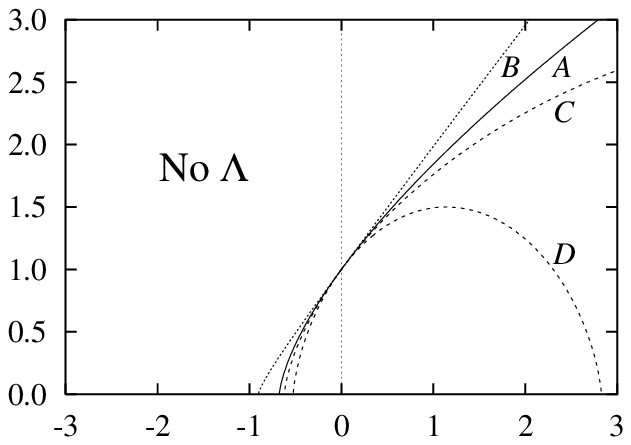}
\includegraphics[width=7cm]{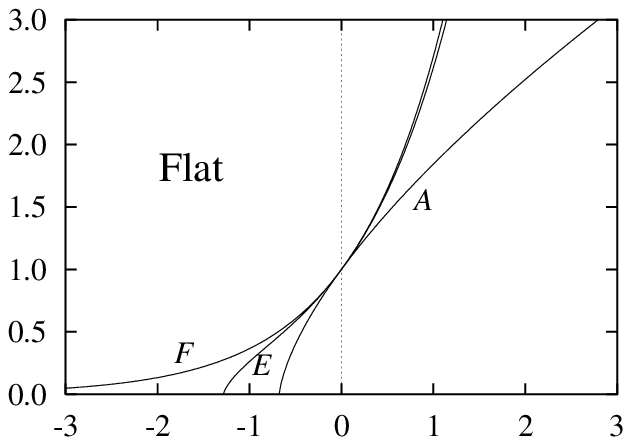}
\end{center}
\begin{center}
\includegraphics[width=7cm]{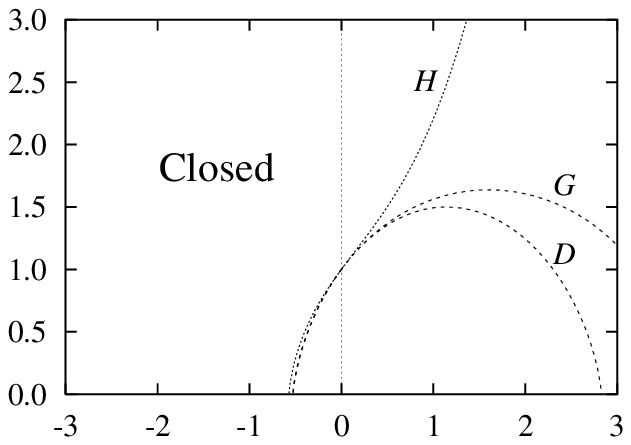}
\includegraphics[width=7cm]{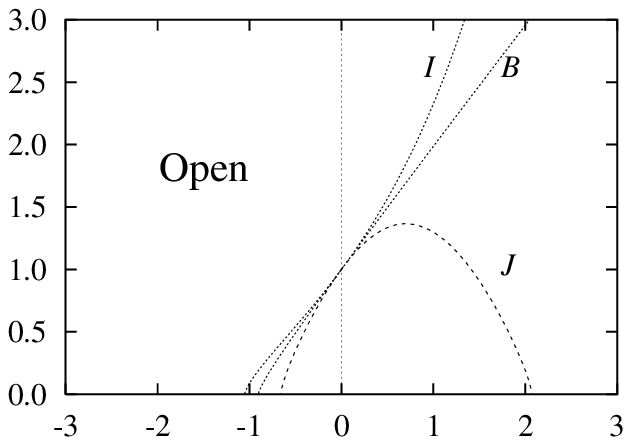}
\end{center}
\begin{center}
\includegraphics[width=7cm]{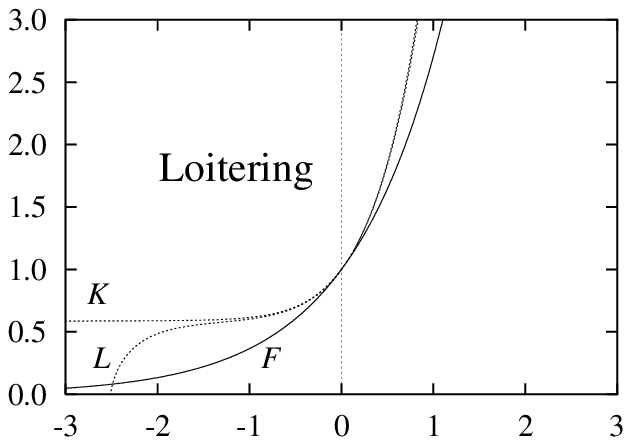}
\includegraphics[width=7cm]{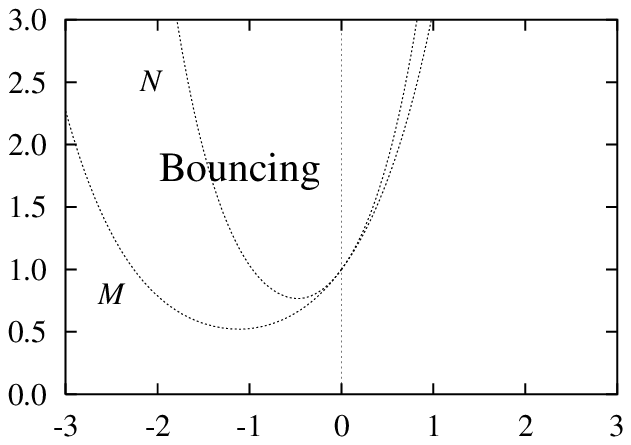}
\end{center}
\vspace{-5mm}
\caption{Evolution of the scale parameter with respect to time for
different values of matter density and cosmological parameter. The
horizontal axis represents $\tau=H_0(t-t_0)$, while the vertical axis is
$y=a/a_0$ in each case. The values of ($\Omega_{\rm M}$,
$\Omega_\Lambda$) for different plots are: A=(1,0), B=(0.1,0),
C=(1.5,0), D=(3,0), E=(0.1,0.9), F=(0,1), G=(3,.1), H=(3,1), I=(.1,.5),
J=(.5,$-1$), K=(1.1,2.707), L=(1,2.59), M=(0.1,1.5), N=(0.1,2.5). From
Ref.~\cite{Pal}. }
\label{fig1}
\end{figure}

It is enlightening to work with a mechanical analogy of the Friedmann
equation. Let us rewrite Eq.~(\ref{FriedmannEquation}) as
\begin{equation}\label{MechanicalAnalog}
{1\over2}\dot a^2 - {GM\over a} - {\Lambda\over6}\,a^2 = -{K\over2} =
{\rm constant}\,,
\end{equation}
where $M\equiv{4\pi\over3}\,\rho\,a^3$ is the equivalent of mass for the
whole volume of the universe. Equation (\ref{MechanicalAnalog}) can be
understood as the energy conservation law $E=T+V$ for a test particle
of unit mass in the central potential
\begin{equation}\label{EscapePotential}
V(r)= - {GM\over r} + {1\over2}\,k\,r^2 \,,
\end{equation}
corresponding to a Newtonian potential plus a harmonic oscillator
potential with a {\em negative} spring constant $k\equiv-\Lambda/3$.
Note that, in the absence of a cosmological constant ($\Lambda=0$), a
critical universe, defined as the borderline between indefinite
expansion and recollapse, corresponds, through the Friedmann equations
of motion, precisely with a flat universe ($K=0$). In that case, and
{\em only} in that case, a spatially open universe ($K=-1$) corresponds
to an eternally expanding universe, and a spatially closed universe
($K=+1$) to a recollapsing universe in the future. Such a well known
(textbook) correspondence is incorrect when $\Omega_\Lambda\neq0$:
spatially open universes may recollapse while closed universes can
expand forever. One can see in Fig.~\ref{fig1} a range of possible
evolutions of the scale factor, for various pairs of values of
$(\Omega_{\rm M}, \Omega_\Lambda)$.

One can show that, for $\Omega_\Lambda\neq0$, a critical universe ($H =
\dot H = 0$) corresponds to those points $x\equiv a_0/a>0$, for which
$f(x)\equiv H^2(a)$ and $f'(x)$ vanish, while $f''(x)>0$,
\begin{eqnarray}\label{FCritical}
&&f(x) = x^3\Omega_{\rm M} + x^2\Omega_K + \Omega_\Lambda = 0\,,\\[2mm]
&&f'(x) = 3x^2\Omega_{\rm M} + 2x\Omega_K = 0\hspace{5mm}\left\{
\begin{array}{l}x = 0\\x=-2\Omega_K/3\Omega_{\rm M}>0\end{array}\right.\,,\\
&&f''(x) = 6x\Omega_{\rm M} + 2\Omega_K = \hspace{5mm}\left\{
\begin{array}{cl}+2\Omega_K > 0&\hspace{5mm}x = 0\\
-2\Omega_K > 0&\hspace{5mm}x=2|\Omega_K|/3\Omega_{\rm M}\end{array}\right.\,.
\end{eqnarray} 
Using the cosmic sum rule (\ref{CosmicSumRule}), we can write the
solutions as
\begin{equation}\label{Critical}
\Omega_\Lambda = \hspace{1mm}\left\{
\begin{array}{ll}0&\hspace{2cm}\Omega_{\rm M} \leq1\\[2mm]
4\Omega_{\rm M}\,\sin^3\!\Big[{1\over3}{\rm arcsin}
(1-\Omega_{\rm M}^{-1})\Big]&\hspace{2cm}\Omega_{\rm M} \geq1
\end{array}\right.\,.
\end{equation}
The first solution corresponds to the critical point $x=0$
($a=\infty$), and $\Omega_K > 0$, while the second one to
$x=2|\Omega_K|/3\Omega_{\rm M}$, and $\Omega_K < 0$. Expanding around
$\Omega_{\rm M}=1$, we find $\Omega_\Lambda \simeq {4\over27}\,
(\Omega_{\rm M}-1)^3/\Omega_{\rm M}^2$, for $\Omega_{\rm M}
\geq1$. These critical solutions are asymptotic to the Einstein-de
Sitter model ($\Omega_{\rm M}=1, \ \Omega_\Lambda=0$), see Fig.~2.

\begin{figure}[htb]
\vspace*{-4cm}
\begin{center}
\includegraphics[width=11cm]{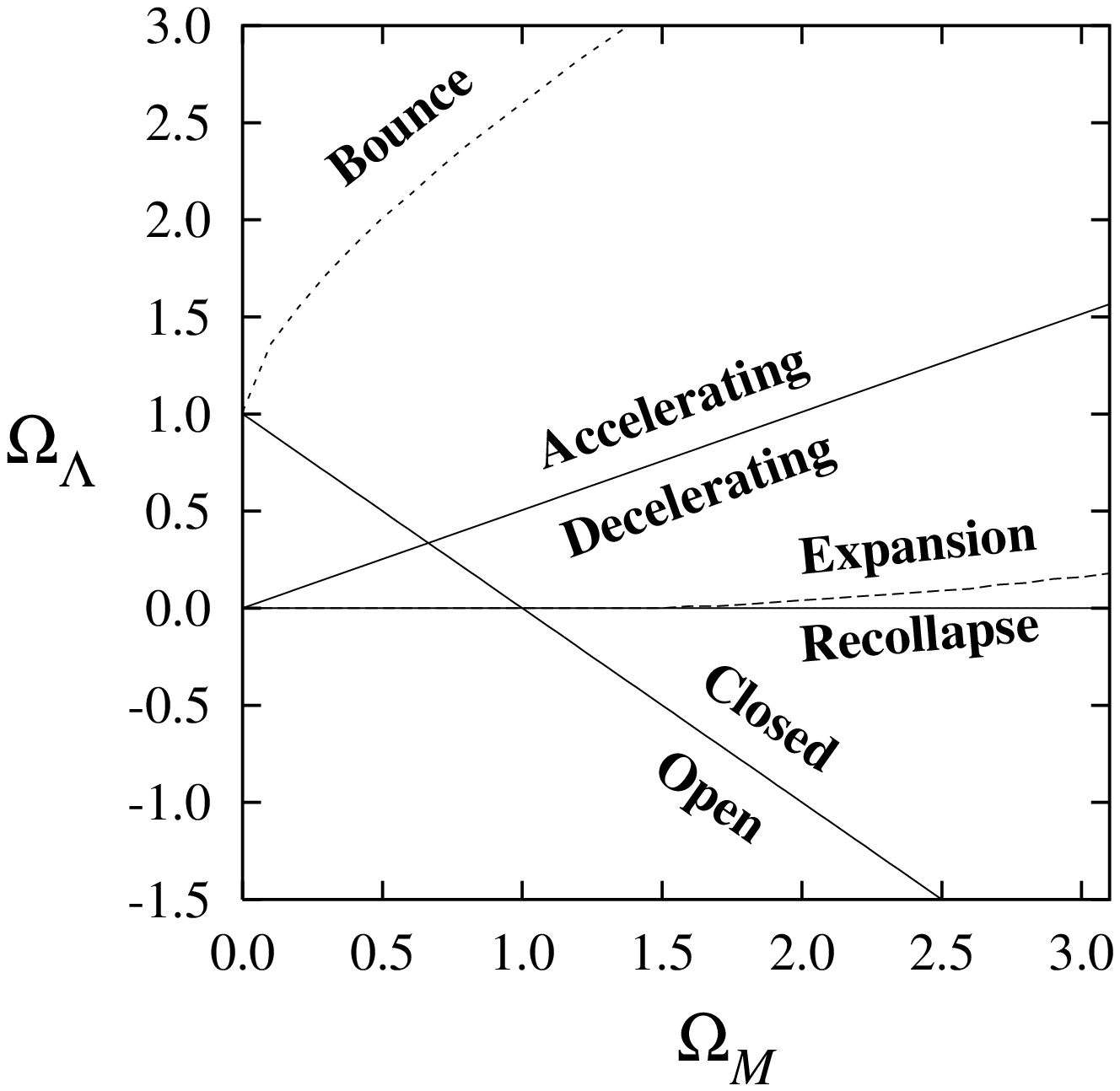}
\caption{Parameter space $(\Omega_{\rm M}, \Omega_\Lambda)$.  The line
$\Omega_\Lambda=1-\Omega_{\rm M}$ corresponds to a flat universe,
$\Omega_K=0$, separating open from closed universes. The line
$\Omega_\Lambda=\Omega_{\rm M}/2$ corresponds to uniform expansion,
$q_0 = 0$, separating accelerating from decelerating universes. The
dashed line corresponds to critical universes, separating eternal
expansion from recollapse in the future. Finally, the dotted line
corresponds to $t_0H_0=\infty$, beyond which the universe has a bounce. }
\end{center}
\label{fig2}
\end{figure}

\subsubsection{Thermodynamical analogy}

It is also enlightening to find an analogy between the energy
conservation equation (\ref{EnergyConservation}) and the second law of
Thermodynamics,
\begin{equation}\label{SecondLaw}
TdS = dU + pdV\,,
\end{equation}
where $U=\rho V$ is the total energy of the closed system and $V=a^3$ is
its physical volume. Equation (\ref{EnergyConservation}) implies that
the expansion of the universe is adiabatic or isoentropic ($dS=0$),
corresponding to a fluid in thermal equilibrium at a temperature T. For
a barotropic fluid, satisfying the equation of state $p=\omega\rho$, we
can write the energy density evolution as
\begin{equation}\label{DensityEvolution}
{d\over dt}(\rho a^3) = - p\,{d \over dt}(a^3) = 
-3H \omega\,(\rho a^3)\,.
\end{equation}
For relativistic particles in thermal equilibrium, the trace of the
energy-momentum tensor vanishes (because of conformal invariance) and
thus $p_{\rm R} = \rho_{\rm R}/3 \,\Rightarrow\,\omega=1/3$. In that
case, the energy density of radiation in thermal equilibrium can be
written as~\cite{KT}
\begin{eqnarray}\label{RhoRadiation}
\rho_{\rm R} &=& {\pi^2\over30} g_* T^4\,,\\
g_* &=& \sum_{i={\rm bosons}} g_i\left({T_i\over T}\right)^4 +
{7\over8}\sum_{i={\rm fermions}} g_i\left({T_i\over T}\right)^4 \,,
\end{eqnarray}
where $g_*$ is the number of relativistic degrees of freedom, coming
from both bosons and fermions.  Using the equilibrium expressions for
the pressure and density, we can write $dp=(\rho+p)dT/T$, and therefore
\begin{equation}\label{ComovingEntropy}
dS = {1\over T}d[(\rho+p)V] - (\rho+p)V{dT\over T^2} =
d\left[{(\rho+p)V\over T} + {\rm const.}\right]
\end{equation}
That is, up to an additive constant, the entropy per comoving volume is
$S=a^3(\rho+p)V/T$, which is conserved. The entropy per comoving volume
is dominated by the contribution of relativistic particles, so that, to
very good approximation,
\begin{eqnarray}\label{EntropyRadiation}
S &=& {2\pi^2\over45} g_{*s} (aT)^3 = \ {\rm constant}\,,\\
g_{*s} &=& \sum_{i={\rm bosons}} g_i\left({T_i\over T}\right)^3 +
{7\over8}\sum_{i={\rm fermions}} g_i\left({T_i\over T}\right)^3 \,.
\end{eqnarray}
A consequence of Eq.~(\ref{EntropyRadiation}) is that, during the
adiabatic expansion of the universe, the scale factor grows inversely
proportional to the temperature of the universe, $a\propto1/T$.
Therefore, the observational fact that the universe is expanding today
implies that in the past the universe must have been much hotter and
denser, and that in the future it will become much colder and
dilute. Since the ratio of scale factors can be described in terms of
the redshift parameter $z$, see Eq.~(\ref{Redshift}), we can find the
temperature of the universe at an earlier epoch by
\begin{equation}\label{TemperatureRedshift}
T = T_0\,(1+z)\,.
\end{equation}
Such a relation has been spectacularly confirmed with observations of
absorption spectra from quasars at large distances, which showed that,
indeed, the temperature of the radiation background scaled with redshift
in the way predicted by the hot Big Bang model.

\subsection{Brief thermal history of the universe}

In this Section, I will briefly summarize the thermal history of the
universe, from the Planck era to the present. As we go back in time, the
universe becomes hotter and hotter and thus the amount of energy
available for particle interactions increases. As a consequence, the
nature of interactions goes from those described at low energy by long
range gravitational and electromagnetic physics, to atomic physics,
nuclear physics, all the way to high energy physics at the electroweak
scale, gran unification (perhaps), and finally quantum gravity. The last
two are still uncertain since we do not have any experimental evidence
for those ultra high energy phenomena, and perhaps Nature has followed a
different path.~\footnote{See the recent theoretical developments on
large extra dimensions and quantum gravity at the TeV~\cite{ADV}.}

The way we know about the high energy interactions of matter is via
particle accelerators, which are unravelling the details of those
fundamental interactions as we increase in energy. However, one should
bear in mind that the physical conditions that take place in our high
energy colliders are very different from those that occurred in the
early universe. These machines could never reproduce the conditions of
density and pressure in the rapidly expanding thermal plasma of the
early universe. Nevertheless, those experiments are crucial in
understanding the nature and {\em rate} of the local fundamental
interactions available at those energies. What interests cosmologists is
the statistical and thermal properties that such a plasma should have,
and the role that causal horizons play in the final outcome of the early
universe expansion. For instance, of crucial importance is the time at
which certain particles {\em decoupled} from the plasma, i.e. when their
interactions were not quick enough compared with the expansion of the
universe, and they were left out of equilibrium with the plasma.

One can trace the evolution of the universe from its origin till today.
There is still some speculation about the physics that took place in the
universe above the energy scales probed by present colliders.
Nevertheless, the overall layout presented here is a plausible and
hopefully testable proposal. According to the best accepted view, the
universe must have originated at the Planck era ($10^{19}$ GeV,
$10^{-43}$ s) from a quantum gravity fluctuation. Needless to say, we
don't have any experimental evidence for such a statement: Quantum
gravity phenomena are still in the realm of physical speculation.
However, it is plausible that a primordial era of cosmological {\em
inflation} originated then. Its consequences will be discussed below.
Soon after, the universe may have reached the Grand Unified Theories
(GUT) era ($10^{16}$ GeV, $10^{-35}$ s). Quantum fluctuations of the
inflaton field most probably left their imprint then as tiny
perturbations in an otherwise very homogenous patch of the universe. At
the end of inflation, the huge energy density of the inflaton field was
converted into particles, which soon thermalized and became the origin
of the hot Big Bang as we know it. Such a process is called {\em
reheating} of the universe. Since then, the universe became radiation
dominated. It is probable (although by no means certain) that the
asymmetry between matter and antimatter originated at the same time as
the rest of the energy of the universe, from the decay of the
inflaton. This process is known under the name of {\em baryogenesis}
since baryons (mostly quarks at that time) must have originated then,
from the leftovers of their annihilation with antibaryons. It is a
matter of speculation whether baryogenesis could have occurred at
energies as low as the electroweak scale ($100$ GeV, $10^{-10}$ s).
Note that although particle physics experiments have reached energies as
high as 100 GeV, we still do not have observational evidence that the
universe actually went through the EW phase transition. If confirmed,
baryogenesis would constitute another ``window'' into the early
universe. As the universe cooled down, it may have gone through the
quark-gluon phase transition ($10^2$ MeV, $10^{-5}$ s), when baryons
(mainly protons and neutrons) formed from their constituent quarks.

The furthest window we have on the early universe at the moment is that
of {\em primordial nucleosynthesis} ($1 - 0.1$ MeV, 1 s -- 3 min), when
protons and neutrons were cold enough that bound systems could form,
giving rise to the lightest elements, soon after {\em neutrino
decoupling}: It is the realm of nuclear physics. The observed relative
abundances of light elements are in agreement with the predictions of
the hot Big Bang theory. Immediately afterwards, electron-positron
annihilation occurs (0.5 MeV, 1 min) and all their energy goes into
photons. Much later, at about (1 eV, $\sim 10^5$ yr), matter and
radiation have equal energy densities. Soon after, electrons become
bound to nuclei to form atoms (0.3 eV, $3\times10^5$ yr), in a process
known as {\em recombination}: It is the realm of atomic physics.
Immediately after, photons decouple from the plasma, travelling freely
since then. Those are the photons we observe as the cosmic microwave
background. Much later ($\sim 1-10$ Gyr), the small inhomogeneities
generated during inflation have grown, via gravitational collapse, to
become galaxies, clusters of galaxies, and superclusters, characterizing
the epoch of {\em structure formation}. It is the realm of long range
gravitational physics, perhaps dominated by a vacuum energy in the form
of a cosmological constant. Finally (3K, 13 Gyr), the Sun, the Earth,
and biological life originated from previous generations of stars, and
from a primordial soup of organic compounds, respectively.

I will now review some of the more robust features of the Hot Big Bang
theory of which we have precise observational evidence.

\subsubsection{Primordial nucleosynthesis and light element abundance}

In this subsection I will briefly review Big Bang nucleosynthesis and
give the present observational constraints on the amount of baryons in
the universe. In 1920 Eddington suggested that the sun might derive its
energy from the fusion of hydrogen into helium. The detailed reactions
by which stars burn hydrogen were first laid out by Hans Bethe in 1939.
Soon afterwards, in 1946, George Gamow realized that similar processes
might have occurred also in the hot and dense early universe and gave
rise to the first light elements~\cite{Gamow}. These processes could
take place when the universe had a temperature of around $T_{_{\rm NS}}
\sim 1-0.1$ MeV, which is about 100 times the temperature in the core of
the Sun, while the density is $\rho_{_{\rm NS}} =
{\pi^2\over30}g_*T_{_{\rm NS}}^4\sim 82$ g\,cm$^{-3}$, about the same
density as the core of the Sun. Note, however, that although both
processes are driven by identical thermonuclear reactions, the physical
conditions in star and Big Bang nucleosynthesis are very different. In
the former, gravitational collapse heats up the core of the star and
reactions last for billions of years (except in supernova explosions,
which last a few minutes and creates all the heavier elements beyond
iron), while in the latter the universe expansion cools the hot and
dense plasma in just a few minutes. Nevertheless, Gamow reasoned that,
although the early period of cosmic expansion was much shorter than the
lifetime of a star, there was a large number of free neutrons at that
time, so that the lighter elements could be built up quickly by
succesive neutron captures, starting with the reaction \ $n + p
\rightarrow D + \gamma$. The abundances of the light elements would then
be correlated with their neutron capture cross sections, in rough
agreement with observations~\cite{Weinberg,Burles}.

\begin{figure}[htb]
\begin{center}
\includegraphics[width=8.5cm]{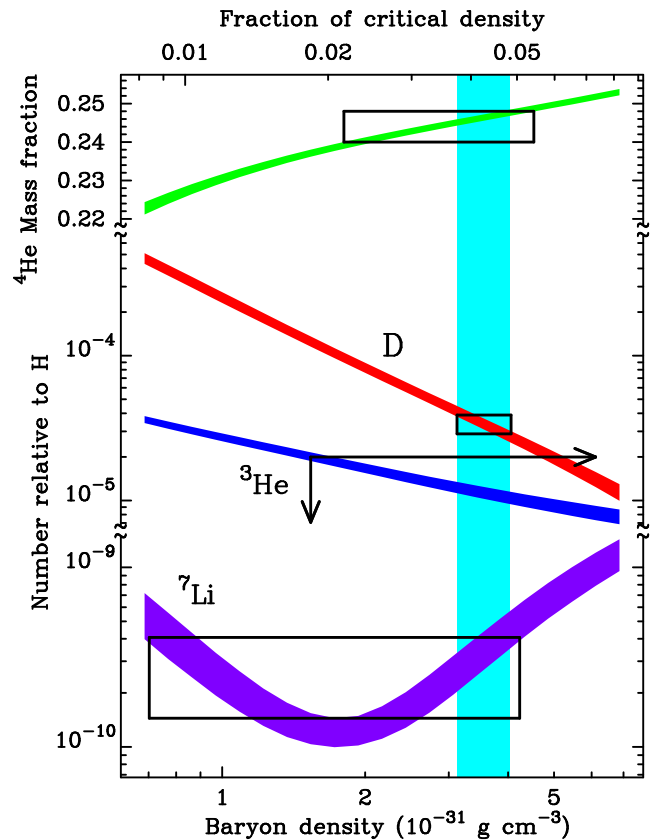} 
\caption{The relative abundance of light elements to Hidrogen. Note the
large range of scales involved. From Ref.~\cite{Burles}.}
\label{fig3}
\end{center}
\end{figure}

Nowadays, Big Bang nucleosynthesis (BBN) codes compute a chain of around
30 coupled nuclear reactions, to produce all the light elements up to
beryllium-7.~\footnote{The rest of nuclei, up to iron (Fe), are produced
in heavy stars, and beyond Fe in novae and supernovae explosions.} Only
the first four or five elements can be computed with accuracy better
than 1\% and compared with cosmological observations. These light
elements are $H, {}^4\!He, D, {}^3\!He, {}^7\!Li$, and perhaps also
${}^6\!Li$. Their observed relative abundance to hydrogen is
$[1:0.25:3\cdot10^{-5}:2\cdot10^{-5}:2\cdot10^{-10}]$ with various
errors, mainly systematic. The BBN codes calculate these abundances
using the laboratory measured nuclear reaction rates, the decay rate of
the neutron, the number of light neutrinos and the homogeneous FRW
expansion of the universe, as a function of {\em only} one variable, the
number density fraction of baryons to photons, $\eta\equiv n_{\rm
B}/n_\gamma$.  In fact, the present observations are only consistent,
see Fig.~\ref{fig3} and Ref.~\cite{BBN,Burles}, with a very narrow range
of values of
\begin{equation}\label{EtaBaryon}
\eta_{10} \equiv 10^{10}\,\eta = 4.6 - 5.9 \,.
\end{equation}
Such a small value of $\eta$ indicates that there is about one baryon
per $10^9$ photons in the universe today. Any acceptable theory of
baryogenesis should account for such a small number. Furthermore, the
present baryon fraction of the critical density can be calculated from
$\eta_{10}$ as~\cite{Burles}
\begin{equation}\label{OmegaBaryon}
\Omega_{\rm B}h^2 = 3.6271\times 10^{-3}\,\eta_{10} = 
0.0190 \pm 0.0024 \hspace{5mm} (95\%\ {\rm c.l.})
\end{equation}
Clearly, this number is well below closure density, so baryons cannot
account for all the matter in the universe, as I shall discuss below.

\subsubsection{Neutrino decoupling}

Just before the nucleosynthesis of the lightest elements in the early
universe, weak interactions were too slow to keep neutrinos in thermal
equilibrium with the plasma, so they decoupled. We can estimate the
temperature at which decoupling occurred from the weak interaction
cross section, $\sigma_{\rm w} \simeq G_F^2 T^2$ at finite temperature
$T$, where $G_F=1.2\times10^{-5}$ GeV$^{-2}$ is the Fermi constant.  The
neutrino interaction rate, via W boson exchange in \ $n+\nu
\leftrightarrow p+e^-$ and \ $p+\bar\nu\leftrightarrow n+e^+$, can
be written as~\cite{KT}
\begin{equation}\label{NeutrinoInteraction}
\Gamma_\nu = n_\nu\langle\sigma_{\rm w}|v|\rangle\simeq G_F^2T^5\,,
\end{equation}
while the rate of expansion of the universe at that time ($g_*=10.75$)
was $H\simeq5.4\ T^2/M_{\rm P}$, where $M_{\rm P} = 1.22\times10^{19}$
GeV is the Planck mass. Neutrinos decouple when their interaction rate
is slower than the universe expansion, $\Gamma_\nu \leq H$ or,
equivalently, at $T_{\nu-{\rm dec}} \simeq 0.8$ MeV. Below this
temperature, neutrinos are no longer in thermal equilibrium with the
rest of the plasma, and their temperature continues to decay inversely
proportional to the scale factor of the universe. Since neutrinos
decoupled before $e^+e^-$ annihilation, the cosmic background of
neutrinos has a temperature today lower than that of the microwave
background of photons. Let us compute the difference. At temperatures
above the the mass of the electron, $T>m_e = 0.511$ MeV, and below 0.8
MeV, the only particle species contributing to the entropy of the
universe are the photons ($g_*=2$) and the electron-positron pairs
($g_*=4\times {7\over8}$); total number of degrees of freedom
$g_*={11\over2}$. At temperatures $T\simeq m_e$, electrons and positrons
annihilate into photons, heating up the plasma (but not the neutrinos,
which had decoupled already). At temperatures $T< m_e$, only photons
contribute to the entropy of the universe, with $g_*=2$ degrees of
freedom. Therefore, from the conservation of entropy, we find that the
ratio of $T_\gamma$ and $T_\nu$ today must be
\begin{equation}\label{NeutrinoTemperature}
{T_\gamma\over T_\nu} = \Big({11\over4}\Big)^{1/3} = 1.401
\hspace{5mm} \Rightarrow \hspace{5mm} T_\nu = 1.945\ {\rm K}\,,
\end{equation}
where I have used $T_{_{\rm CMB}} = 2.725\pm0.002$ K. We still have not
measured such a relic background of neutrinos, and probably will remain
undetected for a long time, since they have an average energy of order
$10^{-4}$ eV, much below that required for detection by present
experiments (of order GeV), precisely because of the relative
weakness of the weak interactions. Nevertheless, it would be fascinating
if, in the future, ingenious experiments were devised to detect such a
background, since it would confirm one of the most robust features of
Big Bang cosmology. 

\subsubsection{Matter-radiation equality}

Relativistic species have energy densities proportional to the quartic
power of temperature and therefore scale as $\rho_{\rm R}\propto
a^{-4}$, while non-relativistic particles have essentially zero pressure
and scale as $\rho_{\rm M}\propto a^{-3}$, see
Eq.~(\ref{DensityEvolution}). Therefore, there will be a time in the
evolution of the universe in which both energy densities are equal
$\rho_{\rm R}(t_{\rm eq})=\rho_{\rm M}(t_{\rm eq})$.  Since then both
decay differently, and thus
\begin{equation}\label{Equality}
1+z_{\rm eq} = {a_0\over a_{\rm eq}} = {\Omega_{\rm M}\over\Omega_{\rm
R}} = 3.1\times10^4\ \Omega_{\rm M} h^2\,,
\end{equation}
where I have used $\Omega_{\rm R} h^2= \Omega_{_{\rm CMB}} h^2 +
\Omega_\nu h^2 = 3.24\times10^{-5}$ for three massless neutrinos at
$T=T_\nu$.  As I will show later, the matter content of the universe
today is below critical, $\Omega_{\rm M} \simeq 0.3$, while
$h\simeq0.65$, and therefore $(1+z_{\rm eq}) \simeq 3900$, or about
$t_{\rm eq} = 1.2\times10^3\, (\Omega_{\rm M} h^2)^{-2} \simeq
7\times10^4$ years after the origin of the universe. Around the time of
matter-radiation equality, the rate of expansion (\ref{H2a}) can be
written as ($a_0\equiv1$)
\begin{equation}\label{RateEquality}
H(a) = H_0\,\Big(\Omega_{\rm R}\,a^{-4} + \Omega_{\rm M}\,a^{-3}\Big)^{1/2} =
H_0\,\Omega_{\rm M}^{1/2}\,a^{-3/2}\Big(1+{a_{\rm eq}\over a}\Big)^{1/2}\,.
\end{equation}
The {\em horizon size} is the coordinate distance travelled by a
photon since the beginning of the universe, $d_H\sim H^{-1}$, i.e.
the size of causally connected regions in the universe. The
{\em comoving} horizon size is then given by
\begin{equation}\label{HorizonSize}
d_H={c\over aH(a)}=c\,H_0^{-1}\Omega_{\rm M}^{-1/2}\,a^{1/2}
\Big(1+{a_{\rm eq}\over a}\Big)^{-1/2}\,.
\end{equation}
Thus the horizon size at matter-radiation equality ($a=a_{\rm eq}$) is
\begin{equation}\label{HorizonSizeEquality}
d_H(a_{\rm eq})={c\,H_0^{-1}\over\sqrt2}\,\Omega_{\rm M}^{-1/2}\,
a_{\rm eq}^{1/2} \simeq 12\,(\Omega_{\rm M} h)^{-1}\,h^{-1}{\rm Mpc}\,.
\end{equation}
This scale plays a very important role in theories of structure
formation.

\begin{figure}[htb]
\begin{center}
\includegraphics[width=6cm,angle=-90]{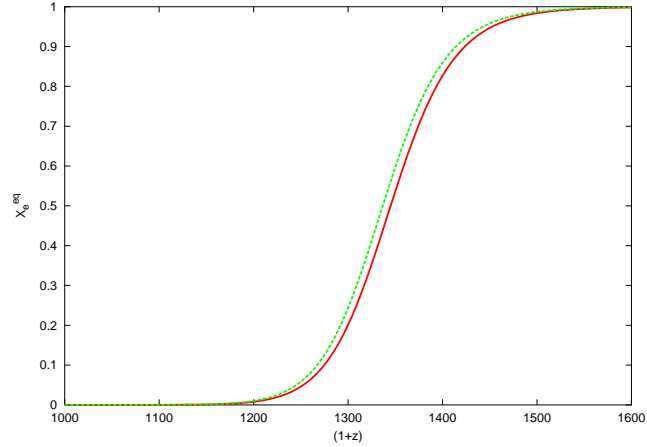} 
\caption{The equilibrium ionization fraction $X_e^{\rm eq}$
as a function of redshift. The two lines show the 
range of $\eta_{10}=4.6-5.9$. }
\label{fig4}
\end{center}
\end{figure}

\subsubsection{Recombination and photon decoupling}

As the temperature of the universe decreased, electrons could eventually
become bound to protons to form neutral hydrogen. Nevertheless, there is
always a non-zero probability that a rare energetic photon ionizes
hydrogen and produces a free electron. The {\em ionization fraction} of
electrons in equilibrium with the plasma at a given temperature is given
by~\cite{KT}
\begin{equation}\label{IonizationFraction}
{1-X_e^{\rm eq}\over X_e^{\rm eq}} = {4\sqrt2\zeta(3)\over\sqrt\pi}\,
\eta\,\left({T\over m_e}\right)^{3/2}\,e^{E_{\rm ion}/T}\,,
\end{equation}
where $E_{\rm ion} = 13.6$ eV is the ionization energy of hydrogen, and
$\eta$ is the baryon-to-photon ratio (\ref{EtaBaryon}). If we now use
Eq.~(\ref{TemperatureRedshift}), we can compute the ionization fraction
$X_e^{\rm eq}$ as a function of redshift $z$, see Fig.~\ref{fig4}. Note
that the huge number of photons with respect to electrons (in the ratio
$^{4}\!He: H: \gamma \simeq 1:4:10^{10}$) implies that even at a very
low temperature, the photon distribution will contain a sufficiently
large number of high-energy photons to ionize a significant fraction of
hydrogen. In fact, {\em defining} recombination as the time at which
$X_e^{\rm eq}\equiv0.1$, one finds that the recombination temperature is
$T_{\rm rec} = 0.3\ {\rm eV} \ll E_{\rm ion}$, for $\eta_{10}\simeq
5.2$. Comparing with the present temperature of the microwave
background, we deduce the corresponding redshift at recombination,
$(1+z_{\rm rec}) \simeq 1270$.

Photons remain in thermal equilibrium with the plasma of baryons and
electrons through elastic Thomson scattering, with cross section
\begin{equation}\label{ThomsonCrossSection}
\sigma_{_T} = {8\pi\alpha^2\over3m_e^2} = 6.65\times 10^{-25} \
{\rm cm}^2 =0.665\ {\rm barn}\,,
\end{equation}
where $\alpha=1/137.036$ \ is the dimensionless electromagnetic coupling
constant. The mean free path of photons $\lambda_\gamma$ in such a
plasma can be estimated from the photon interaction rate,
$\lambda_\gamma^{-1}\sim \Gamma_\gamma = n_e\sigma_{_T}$. For
temperatures above a few eV, the mean free path is much smaller that the
causal horizon at that time and
photons suffer multiple scattering: the plasma is like a dense
fog. Photons will decouple from the plasma when their interaction rate
cannot keep up with the expansion of the universe and the mean free path
becomes larger than the horizon size: the universe becomes
transparent. We can estimate this moment by evaluating $\Gamma_\gamma =
H$ at photon decoupling.  Using $n_e=X_e\,\eta\,n_\gamma$, one can
compute the decoupling temperature as $T_{\rm dec} = 0.26$ eV, and the
corresponding redshift as $(1+z_{\rm dec}) \simeq 1100$. This redshift
defines the so called {\em last scattering surface}, when photons last
scattered off protons and electrons and travelled freely ever
since. This decoupling occurred when the universe was approximately
$t_{\rm dec} = 1.8\times 10^5\,(\Omega_{\rm M} h^2)^{-1/2}
\simeq 5\times 10^5$ years old.

\begin{figure}[htb]
\begin{center}\hspace{-1.8cm}
\includegraphics[width=6cm,angle=90]{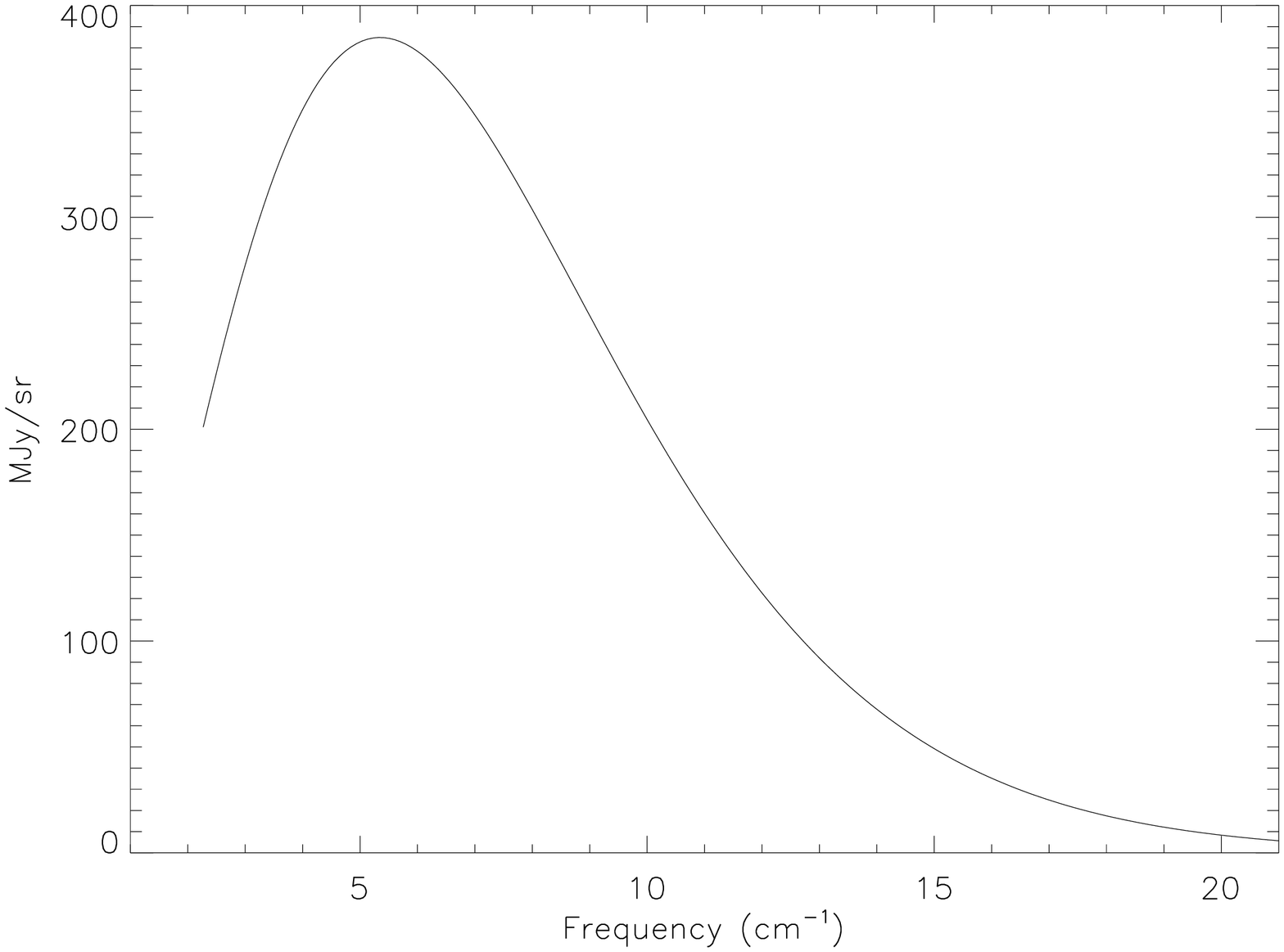}
\includegraphics[width=6cm,angle=90]{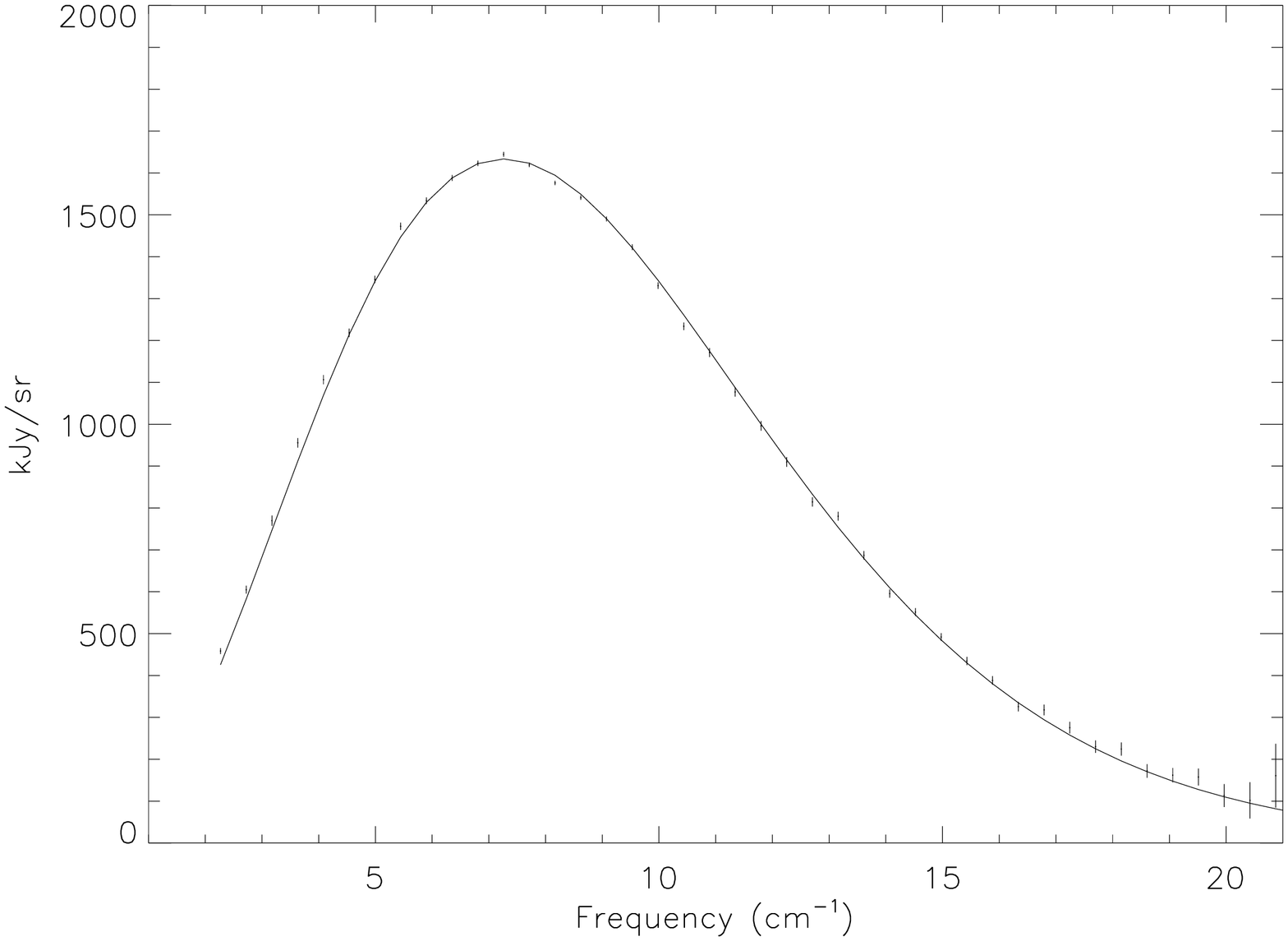}
\end{center}\vspace*{-5mm}
\caption{The Cosmic Microwave Background Spectrum seen by the FIRAS
instrument on COBE. The left panel corresponds to the monopole spectrum,
$T_0 = 2.725\pm0.002$ K, where the error bars are smaller than the 
line width. The right panel shows the dipole spectrum, $\delta T_1 =
3.372\pm0.014$ mK. From Ref.~\cite{FIRAS}.}
\label{fig5}
\end{figure}

\subsubsection{The microwave background}

One of the most remarkable observations ever made my mankind is the
detection of the relic background of photons from the Big Bang. This
background was predicted by George Gamow and collaborators in the 1940s,
based on the consistency of primordial nucleosynthesis with the observed
helium abundance. They estimated a value of about 10 K, although a
somewhat more detailed analysis by Alpher and Herman in 1950 predicted
$T_\gamma \approx 5$ K. Unfortunately, they had doubts whether the
radiation would have survived until the present, and this remarkable
prediction slipped into obscurity, until Dicke, Peebles, Roll and
Wilkinson~\cite{Dicke} studied the problem again in 1965. Before they
could measure the photon background, they learned that Penzias and
Wilson had observed a weak isotropic background signal at a radio
wavelength of 7.35 cm, corresponding to a blackbody temperature of
$T_\gamma=3.5\pm1$ K. They published their two papers back to back, with
that of Dicke et al.  explaining the fundamental significance of their
measurement~\cite{Weinberg}.

Since then many different experiments have confirmed the existence of
the microwave background. The most outstanding one has been the Cosmic
Background Explorer (COBE) satellite, whose FIRAS instrument measured
the photon background with great accuracy over a wide range of
frequencies ($\nu = 1-97$ cm$^{-1}$), see Ref.~\cite{FIRAS}, with a
spectral resolution ${\Delta\nu\over\nu}=0.0035$. Nowadays,
the photon spectrum is confirmed to be a blackbody spectrum with a
temperature given by~\cite{FIRAS}
\begin{equation}\label{T0}
T_{_{\rm CMB}} = 2.725 \pm 0.002 \ {\rm K} \ 
({\rm systematic}, \ 95\%\ {\rm c.l.}) \ 
\pm 7 \ \mu\!{\rm K} \ (1\sigma\ {\rm statistical})
\end{equation}
In fact, this is the best blackbody spectrum ever measured, see
Fig.~\ref{fig5}, with spectral distortions below the level of 10 parts
per million (ppm).

\begin{figure}[htb]
\vspace*{-5mm}
\begin{center}
\includegraphics[width=8cm,angle=0]{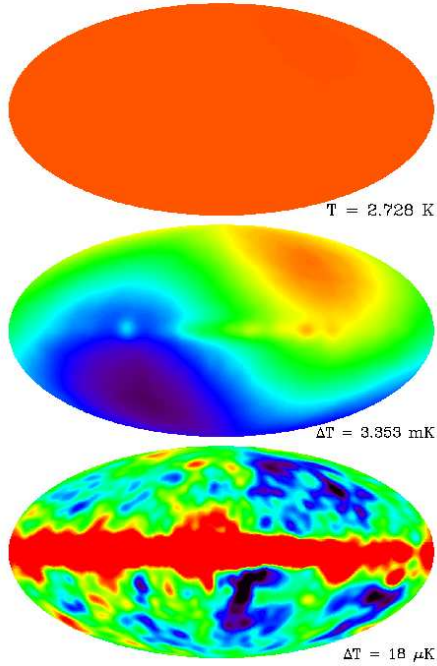}
\end{center}
\vspace*{-1cm}
\caption{The Cosmic Microwave Background Spectrum seen by the DMR
instrument on COBE. The top figure corresponds to the monopole, $T_0 =
2.725\pm0.002$ K. The middle figure shows the dipole, $\delta T_1 =
3.372\pm0.014$ mK, and the lower figure shows the quadrupole and higher
multipoles, $\delta T_2 = 18\pm2 \ \mu$K. The central region corresponds
to foreground by the galaxy. From Ref.~\cite{DMR}.}
\label{fig6}
\end{figure}

Moreover, the differential microwave radiometer (DMR) instrument on
COBE, with a resolution of about $7^\circ$ in the sky, has also
confirmed that it is an extraordinarily isotropic background. The
deviations from isotropy, i.e. differences in the temperature of the
blackbody spectrum measured in different directions in the sky, are of
the order of 20\,$\mu$K on large scales, or one part in $10^5$, see
Ref.~\cite{DMR}.  There is, in fact, a dipole anisotropy of one part in
$10^3$, $\delta T_1 = 3.372\pm0.007$ mK (95\% c.l.), in the direction of
the Virgo cluster, $(l,b) = (264.14^\circ \pm 0.30, 48.26^\circ \pm
0.30)$ (95\% c.l.). Under the assumption that a Doppler effect is
responsible for the entire CMB dipole, the velocity of the Sun with
respect to the CMB rest frame is $v_\odot=371\pm0.5$ km/s, see
Ref.~\cite{FIRAS}.\footnote{COBE even determined the annual variation
due to the Earth's motion around the Sun -- the ultimate proof of
Copernicus' hypothesis.} When subtracted, we are left with a whole
spectrum of anisotropies in the higher multipoles (quadrupole, octupole,
etc.), $\delta T_2 = 18\pm2 \ \mu$K (95\% c.l.), see Ref.~\cite{DMR} and
Fig.~\ref{fig6}.

Soon after COBE, other groups quickly confirmed the detection of
temperature anisotropies at around 30\,$\mu$K and above, at higher
multipole numbers or smaller angular scales. As I shall discuss below,
these anisotropies play a crucial role in the understanding of the
origin of structure in the universe.

\subsection{Large-scale structure formation}

Although the isotropic microwave background indicates that the universe
in the {\em past} was extraordinarily homogeneous, we know that the
universe {\em today} is not exactly homogeneous: we observe galaxies,
clusters and superclusters on large scales. These structures are
expected to arise from very small primordial inhomogeneities that grow
in time via gravitational instability, and that may have originated from
tiny ripples in the metric, as matter fell into their troughs. Those
ripples must have left some trace as temperature anisotropies in the
microwave background, and indeed such anisotropies were finally
discovered by the COBE satellite in 1992. The reason why they took so
long to be discovered was that they appear as perturbations in
temperature of only one part in $10^5$.

While the predicted anisotropies have finally been seen in the CMB, not
all kinds of matter and/or evolution of the universe can give rise to
the structure we observe today. If we define the density contrast
as~\cite{Peebles}
\begin{equation}\label{DensityContrast}
\delta(\vec x,a) \equiv {\rho(\vec x,a)-\bar\rho(a)\over\bar\rho(a)}
= \int d^3\vec{k}\ \delta_k(a)\ e^{i\vec{k}\cdot\vec{x}}\,,
\end{equation}
where $\bar\rho(a)=\rho_0\,a^{-3}$ is the average cosmic density, we need a
theory that will grow a density contrast with amplitude $\delta \sim
10^{-5}$ at the last scattering surface ($z=1100$) up to density
contrasts of the order of $\delta \sim 10^2$ for galaxies at redshifts
$z\ll1$, i.e. today. This is a {\em necessary} requirement for any
consistent theory of structure formation~\cite{Padmanabhan}.

Furthermore, the anisotropies observed by the COBE satellite correspond
to a small-amplitude scale-invariant primordial power spectrum of
inhomogeneities
\begin{equation}\label{HarrisonZeldovich}
P(k) = \langle|\delta_k|^2\rangle \propto k^n\,, \hspace{5mm} {\rm with}
\hspace{5mm} n=1\,,
\end{equation}
where the brackets $\langle\cdot\rangle$ represent integration over an
ensemble of different universe realizations. These inhomogeneities are
like waves in the space-time metric. When matter fell in the troughs of
those waves, it created density perturbations that collapsed
gravitationally to form galaxies and clusters of galaxies, with a
spectrum that is also scale invariant.  Such a type of spectrum was
proposed in the early 1970s by Edward R. Harrison, and independently by
the Russian cosmologist Yakov B. Zel'dovich, see Ref.~\cite{HZ}, to
explain the distribution of galaxies and clusters of galaxies on very
large scales in our observable universe.

\begin{figure}[htb]
\vspace*{-1cm}
\begin{center}
\includegraphics[width=10cm,angle=0]{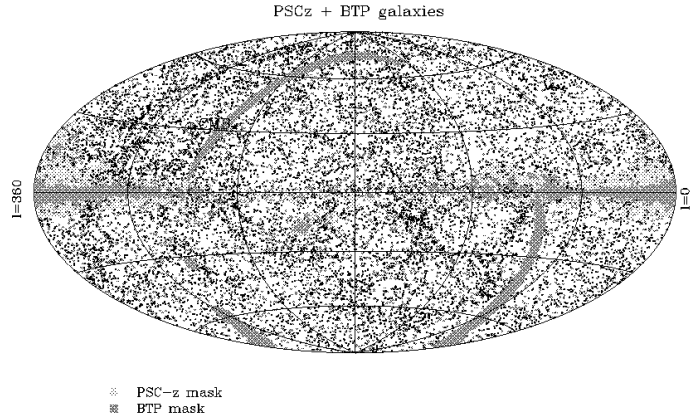}
\end{center}\vspace*{-1.2cm}
\caption{The IRAS Point Source Catalog redshift survey contains some
15,000 galaxies, covering over 83\% of the sky up to redshifts of
$z\leq0.05$. We show here the projection of the galaxy distribution in
galactic coordinates. From Ref.~\cite{PSCz}.}
\label{fig7}
\end{figure}

\begin{figure}[htb]
\begin{center}\hspace{2mm}
\includegraphics[width=7.8cm]{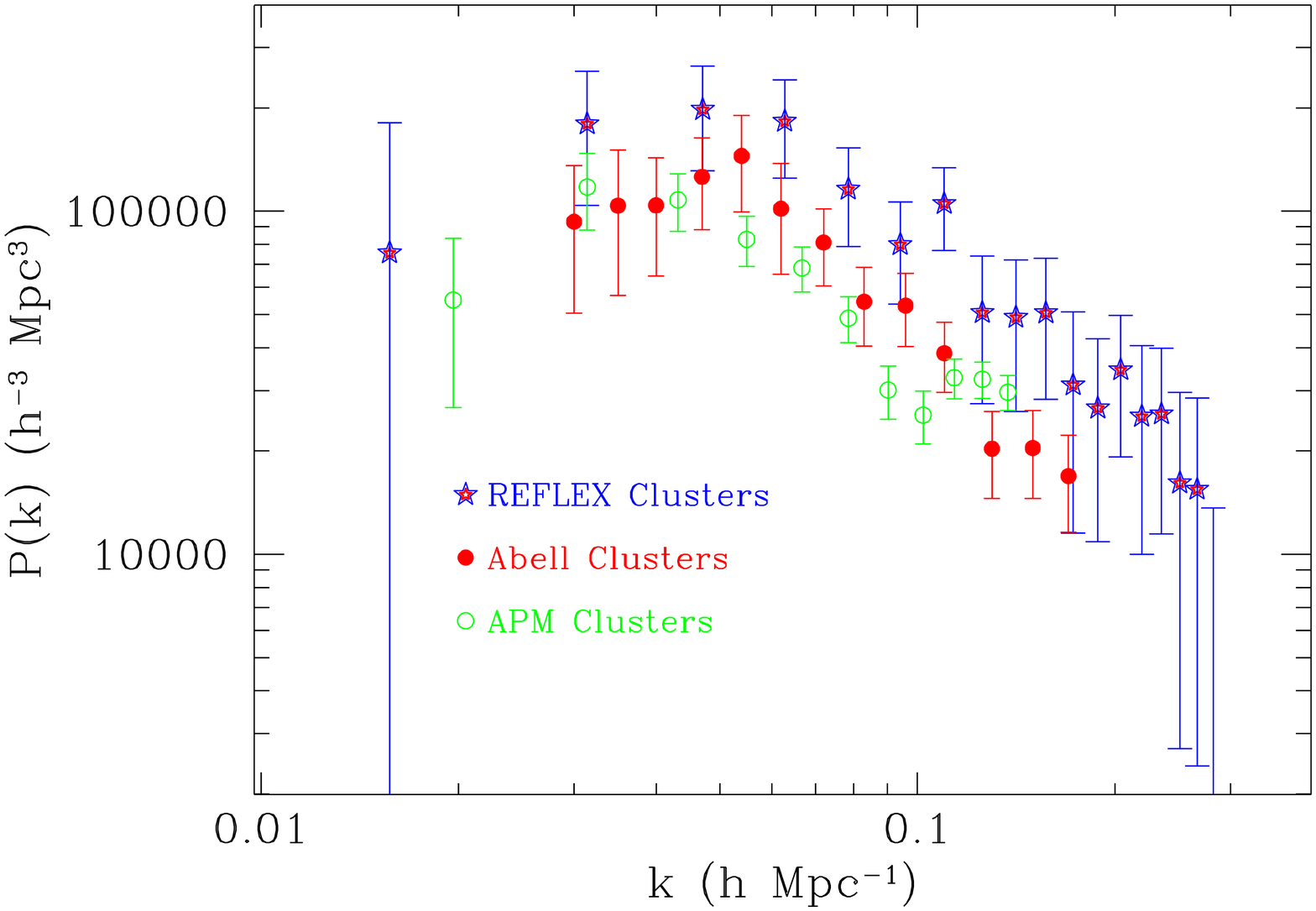} 
\includegraphics[width=7.8cm]{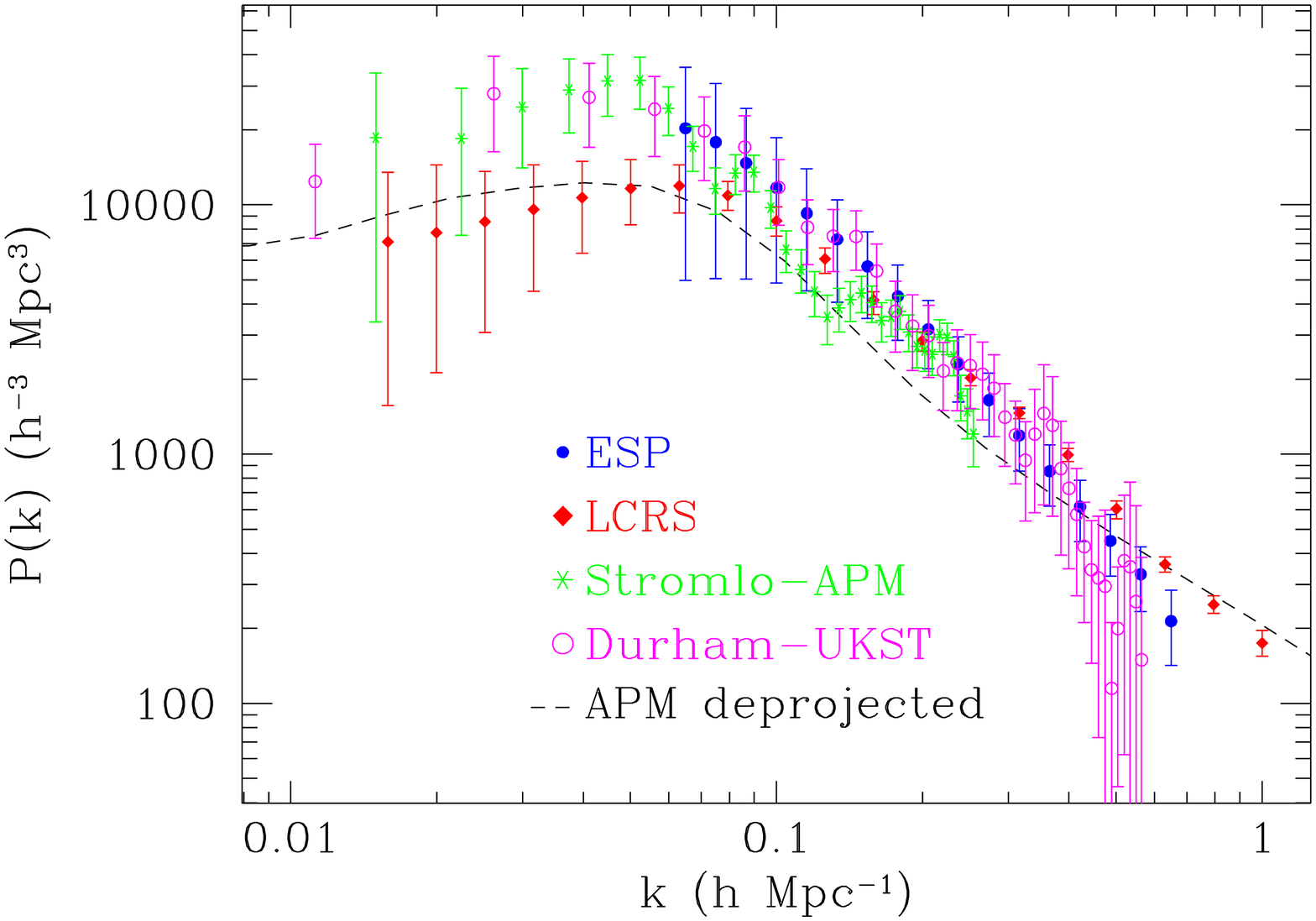} 
\end{center}\vspace*{-1cm}
\caption{The left panel shows the matter power spectrum for clusters of
galaxies, from three different cluster surveys. The right panel shows a
compilation of the most recent estimates of the power spectrum of galaxy
clustering, from four of the largest available redshift surveys of
optically-selected galaxies, compared to the deprojected spectrum of the
2D APM galaxy survey. From Ref.~\cite{Guzzo}.}
\label{fig8}
\end{figure}

Today various telescopes -- like the Hubble Space Telescope, the twin
Keck telescopes in Hawaii and the European Southern Observatory
telescopes in Chile -- are exploring the most distant regions of the
universe and discovering the first galaxies at large distances. The
furthest galaxies observed so far are at redshifts of $z\simeq5$, or 12
billion light years from the Earth, whose light was emitted when the
universe had only about 5\% of its present age.  Only a few galaxies are
known at those redshifts, but there are at present various catalogs like
the CfA and APM galaxy catalogs, and more recently the IRAS Point Source
redshift Catalog, see Fig.~\ref{fig7}, and Las Campanas redshift
surveys, that study the spatial distribution of hundreds of thousands of
galaxies up to distances of a billion light years, or $z<0.1$, that
recede from us at speeds of tens of thousands of kilometres per
second. These catalogs are telling us about the evolution of clusters of
galaxies in the universe, and already put constraints on the theory of
structure formation. From these observations one can infer that most
galaxies formed at redshifts of the order of $2 - 6$; clusters of
galaxies formed at redshifts of order 1, and superclusters are forming
now. That is, cosmic structure formed from the bottom up: from galaxies
to clusters to superclusters, and not the other way around. This
fundamental difference is an indication of the type of matter that gave
rise to structure. The observed power spectrum of the galaxy matter
distribution from a selection of deep redshift catalogs can be seen
in Fig.~\ref{fig8}.

We know from Big Bang nucleosynthesis that all the baryons in the
universe cannot account for the observed amount of matter, so there must
be some extra matter (dark since we don't see it) to account for its
gravitational pull. Whether it is relativistic (hot) or non-relativistic
(cold) could be inferred from observations: relativistic particles tend
to diffuse from one concentration of matter to another, thus
transferring energy among them and preventing the growth of structure on
small scales. This is excluded by observations, so we conclude that most
of the matter responsible for structure formation must be cold. How much
there is is a matter of debate at the moment. Some recent analyses
suggest that there is not enough cold dark matter to reach the critical
density required to make the universe flat. If we want to make sense of
the present observations, we must conclude that some other form of
energy permeates the universe. In order to resolve this issue, even
deeper galaxy redshift catalogs are underway, looking at millions of
galaxies, like the Sloan Digital Sky Survey (SDSS) and the
Anglo-Australian two degree field (2dF) Galaxy Redshift Survey, which
are at this moment taking data, up to redshifts of $z\lsim0.5$, over a
large region of the sky. These important observations will help
astronomers determine the nature of the dark matter and test the
validity of the models of structure formation.

Before COBE discovered the anisotropies of the microwave background
there were serious doubts whether gravity alone could be responsible for
the formation of the structure we observe in the universe today. It
seemed that a new force was required to do the job. Fortunately, the
anisotropies were found with the right amplitude for structure to be
accounted for by gravitational collapse of primordial inhomogeneities
under the attraction of a large component of non-relativistic dark
matter. Nowadays, the standard theory of structure formation is a cold
dark matter model with a non vanishing cosmological constant in a
spatially flat universe. Gravitational collapse amplifies the density
contrast initially through linear growth and later on via non-linear
collapse. In the process, overdense regions decouple from the Hubble
expansion to become bound systems, which start attracting eachother to
form larger bound structures. In fact, the largest structures,
superclusters, have not yet gone non-linear.

The primordial spectrum (\ref{HarrisonZeldovich}) is reprocessed by
gravitational instability after the universe becomes matter dominated
and inhomogeneities can grow. Linear perturbation theory shows that the
growing mode~\footnote{The decaying modes go like $\delta(t)\sim
t^{-1}$, for all $\omega$.} of small density contrasts go
like~\cite{Peebles,Padmanabhan}
\begin{equation}\label{LinearGrowth}
\delta(a) \propto a^{1+3\omega} = \left\{\begin{array}{ll}
a^2\,,&\hspace{1cm}a<a_{\rm eq}\\
a\,,&\hspace{1cm}a>a_{\rm eq}\end{array}\right.
\end{equation}
in the Einstein-de Sitter limit ($\omega = p/\rho = 1/3$ and 0, for
radiation and matter, respectively). There are slight deviations for
$a\gg a_{\rm eq}$, if $\Omega_{\rm M}\neq1$ or $\Omega_\Lambda\neq0$,
but we will not be concerned with them here. The important observation
is that, since the density contrast at last scattering is of order
$\delta \sim 10^{-5}$, and the scale factor has grown since then only a
factor $z_{\rm dec} \sim 10^3$, one would expect a density contrast
today of order $\delta_0 \sim 10^{-2}$. Instead, we observe structures
like galaxies, where $\delta\sim10^2$. So how can this be possible? The
microwave background shows anisotropies due to fluctuations in the
baryonic matter component only (to which photons couple,
electromagnetically). If there is an additional matter component that
only couples through very weak interactions, fluctuations in that
component could grow as soon as it decoupled from the plasma, well
before photons decoupled from baryons. The reason why baryonic
inhomogeneities cannot grow is because of photon pressure: as baryons
collapse towards denser regions, radiation pressure eventually halts the
contraction and sets up acoustic oscillations in the plasma that prevent
the growth of perturbations, until photon decoupling. On the other hand,
a weakly interacting cold dark matter component could start
gravitational collapse much earlier, even before matter-radiation
equality, and thus reach the density contrast amplitudes observed today.
The resolution of this mismatch is one of the strongest arguments for
the existence of a weakly interacting cold dark matter component of the
universe.

\begin{figure}[htb]
\vspace*{-2mm}
\begin{center}
\includegraphics[width=9cm]{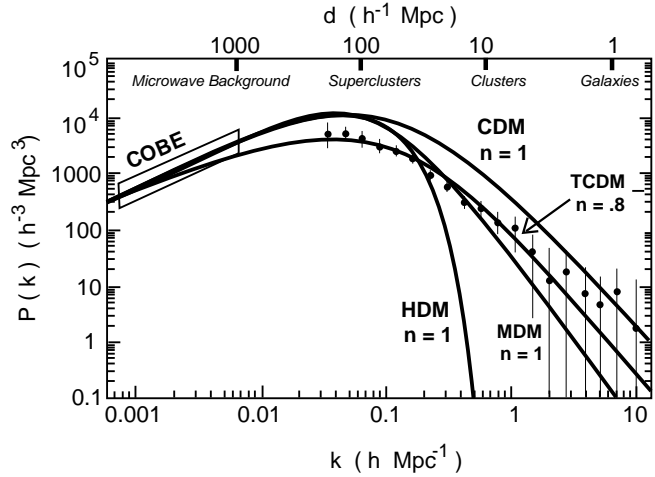}
\end{center}\vspace*{-5mm}
\caption{The power spectrum for cold dark matter (CDM), tilted cold dark
matter (TCDM), hot dark matter (HDM), and mixed hot plus cold dark
matter (MDM), normalized to COBE, for large-scale structure formation. 
From Ref.~\cite{Steinhardt}.}
\label{fig9}
\end{figure}

How much dark matter there is in the universe can be deduced from the
actual power spectrum (the Fourier transform of the two-point
correlation function of density perturbations) of the observed large
scale structure. One can decompose the density contrast in Fourier
components, see Eq.~(\ref{DensityContrast}). This is very convenient
since in linear perturbation theory individual Fourier components evolve
independently.  A comoving wavenumber $k$ is said to ``enter the
horizon'' when $k=d_H^{-1}(a)=aH(a)$. If a certain perturbation, of
wavelength $\lambda =k^{-1}<d_H(a_{\rm eq})$, enters the horizon before
matter-radiation equality, the fast radiation-driven expansion prevents
dark-matter perturbations from collapsing. Since light can only cross
regions that are smaller than the horizon, the suppression of growth due
to radiation is restricted to scales smaller than the horizon, while
large-scale perturbations remain unaffected. This is the reason why the
horizon size at equality, Eq.~(\ref{HorizonSizeEquality}), sets an
important scale for structure growth,
\begin{equation}\label{EqualityScale}
k_{\rm eq} = d_H^{-1}(a_{\rm eq})\simeq 0.083\,(\Omega_{\rm M} h)\,h\ 
{\rm Mpc}^{-1}\,.
\end{equation}
The suppression factor can be easily computed from (\ref{LinearGrowth})
as $f_{\rm sup} = (a_{\rm enter}/a_{\rm eq})^2 = (k_{\rm eq}/k)^2$.
In other words, the processed power spectrum $P(k)$ will have the form:
\begin{equation}\label{PowerSpectrum}
P(k) \propto \left\{\begin{array}{ll}
k\,,&\hspace{1cm}k\ll k_{\rm eq}\\[2mm]
k^{-3}\,,&\hspace{1cm}k\gg k_{\rm eq}\end{array}\right.
\end{equation}
This is precisely the shape that large-scale galaxy catalogs are bound
to test in the near future, see Fig.~\ref{fig9}. Furthermore, since
relativistic Hot Dark Matter (HDM) transfer energy between clumps of
matter, they will wipe out small scale perturbations, and this should be
seen as a distinctive signature in the matter power spectra of future
galaxy catalogs. On the other hand, non-relativistic Cold Dark Matter
(CDM) allow structure to form on {\em all} scales via gravitational
collapse. The dark matter will then pull in the baryons, which will
later shine and thus allow us to see the galaxies.

Naturally, when baryons start to collapse onto dark matter potential
wells, they will convert a large fraction of their potential energy into
kinetic energy of protons and electrons, ionizing the medium. As a
consequence, we expect to see a large fraction of those baryons
constituting a hot ionized gas surrounding large clusters of galaxies.
This is indeed what is observed, and confirms the general picture of
structure formation.

\section{DETERMINATION OF COSMOLOGICAL PARAMETERS}

In this Section, I will restrict myself to those recent measurements of
the cosmological parameters by means of standard cosmological
techniques, together with a few instances of new results from recently
applied techniques. We will see that a large host of observations are
determining the cosmological parameters with some reliability of the
order of 10\%. However, the majority of these measurements are dominated
by large systematic errors. Most of the recent work in observational
cosmology has been the search for virtually systematic-free observables,
like those obtained from the microwave background anisotropies, and
discussed in Section 4.4. I will devote, however, this Section to the
more `classical' measurements of the following cosmological parameters:
The rate of expansion $H_0$; the matter content $\Omega_{\rm M}$; the
cosmological constant $\Omega_\Lambda$; the spatial curvature
$\Omega_K$, and the age of the universe $t_0$.\footnote{We will take the
baryon fraction as given by observations of light element abundances, in
accordance with Big Bang nucleosynthesis, see Eq.~(\ref{OmegaBaryon}).}

These five basic cosmological parameters are not mutually independent.
Using the homogeneity and isotropy on large scales observed by COBE, we
can infer relationships between the different cosmological parameters
through the Einstein-Friedmann equations. In particular, we can deduce
the value of the spatial curvature from the Cosmic Sum Rule,
\begin{equation}
1 = \Omega_{\rm M} + \Omega_\Lambda + \Omega_K\,,
\end{equation}
or viceversa, if we determine that the universe is spatially flat from
observations of the microwave background, we can be sure that the sum of
the matter content plus the cosmological constant must be one.

\begin{figure}[htb]
\vspace*{-2mm}\hspace{4cm}
\begin{center}
\includegraphics[width=11cm]{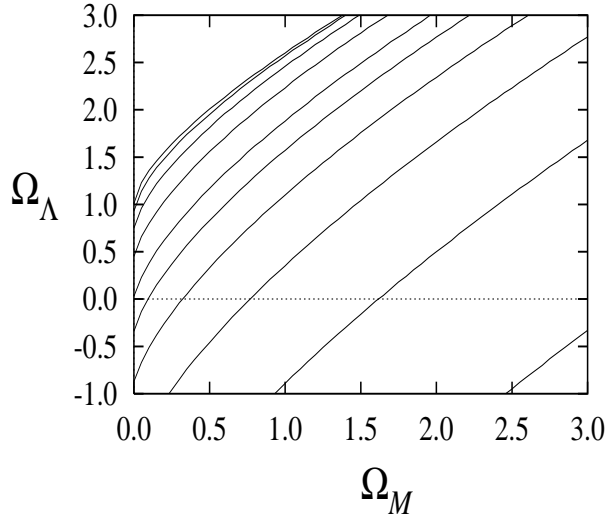}
\vspace*{-2mm}
\caption{The contour lines correspond to equal \ $t_0H_0 = 0.5 - 1.0,
1.2, 1.5, 2.0$ and 5.0, from bottom to top, in parameter space
$(\Omega_{\rm M}, \Omega_\Lambda)$. The line $t_0H_0 = \infty$ would
be indistinguishable from that of \ $t_0H_0 = 5$. From Ref.~\cite{Pal}.}
\end{center}
\label{fig10}
\end{figure}

Another relationship between parameters appears for the age of the
universe. In a FRW cosmology, the cosmic expansion is determined by the
Friedmann equation (\ref{FriedmannEquation}). Defining a new time and
normalized scale factor,
\begin{equation}\label{Redefinitions}
y \equiv {a\over a_0} = {1\over1+z}\,, \hspace{2cm}
\tau \equiv H_0(t-t_0)\,,
\end{equation}
we can write the Friedmann equation with the help of the Cosmic Sum Rule
(\ref{CosmicSumRule}) as
\begin{equation}\label{ReFriedmann}
y'(\tau) = \Big[1 + (y^{-1} - 1)\Omega_{\rm M} + (y^2-1)\Omega_\Lambda
\Big]^{1/2}\,,
\end{equation}
with initial condition $y(0)=1, \ y'(0)=1$. Therefore, the present age
$t_0$ is a function of the other parameters, $t_0=f(H_0, \Omega_{\rm M}, 
\Omega_\Lambda)$, determined from
\begin{equation}\label{PresentAge}
t_0H_0 = \int_0^1 dy\,\Big[1 + (y^{-1} - 1)\Omega_{\rm M} + 
(y^2-1)\Omega_\Lambda \Big]^{-1/2}\,.
\end{equation}
We show in Fig.~\ref{fig10} the contour lines for constant $t_0H_0$ in
parameter space $(\Omega_{\rm M}, \Omega_\Lambda)$.

\begin{figure}[htb]
\vspace*{2mm}
\begin{center}
\includegraphics[width=9cm]{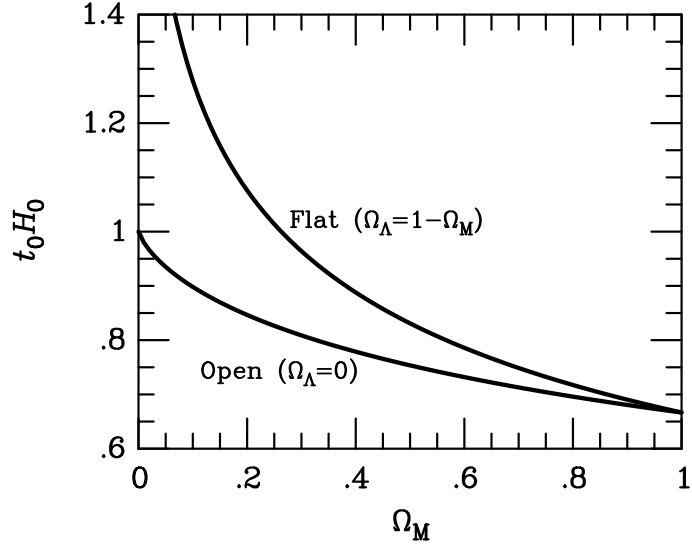}
\end{center}\vspace*{-2cm}
\caption{The age of the universe as a function of the matter content,
for an open and a flat universe. From Ref.~\cite{Raffelt}.}
\label{fig11}
\end{figure}

\noindent
There are two specific limits of interest: an {\em open universe} with
$\Omega_\Lambda=0$, for which the age is given by
\begin{equation}\label{OpenAge}
t_0H_0 = {1\over1-\Omega_{\rm M}}-{\Omega_{\rm M}\over
(1-\Omega_{\rm M})^{3/2}}\,
\ln\left[{1+(1-\Omega_{\rm M})^{1/2}\over\Omega_{\rm M}^{1/2}}\right] =
2\,\sum_{n=0}^\infty{(1-\Omega_{\rm M})^n\over(2n+1)(2n+3)}\,,
\end{equation}
and a {\em flat universe} with $\Omega_\Lambda=1-\Omega_{\rm M}$, for which
the age can also be expressed in compact form,
\begin{equation}\label{FlatAge}
t_0H_0 = {2\over3(1-\Omega_{\rm M})^{1/2}}\,\ln
\left[{1+(1-\Omega_{\rm M})^{1/2}\over\Omega_{\rm M}^{1/2}}\right]=
{2\over3}\,\sum_{n=0}^\infty{(1-\Omega_{\rm M})^n\over2n+1}\,.
\end{equation}
We have plotted these functions in Fig.~\ref{fig11}. It is clear that in
both cases $t_0H_0\to2/3$ as $\Omega_{\rm M}\to1$. We can now use these
relations as a consistency check between the cosmological observations
of $H_0$, $\Omega_{\rm M}$, $\Omega_\Lambda$ and $t_0$. Of course, we
{\em cannot} measure the age of the universe directly, but only the age
of its constituents: stars, galaxies, globular clusters, etc. Thus we
can only find a lower bound on the age of the universe, $t_0\gsim t_{\rm
gal} + 1.5$ Gyr. As we will see, this is not a trivial bound and, in
several occasions, during the progress towards better determinations of
the cosmological parameters, the universe {\em seemed} to be younger
than its constituents, a logical inconsistency, of course, only due to
an incorrect assessment of systematic errors~\cite{Freedman}.

In order to understand those recent measurements, one should also define
what is known as the {\em luminosity distance} to an object in the
universe. Imagine a source that is emitting light at a distance $d_L$
from a detector of area $dA$. The {\em absolute luminosity} ${\cal L}$
of such a source is nothing but the energy emitted per unit time. A {\em
standard candle} is a luminous object that can be calibrated with some
accuracy and therefore whose absolute luminosity is known, within
certain errors.  For example, Cepheid variable stars and type Ia
supernovae are considered to be reasonable standard candles, i.e. their
calibration errors are within bounds. The energy flux ${\cal F}$
received at the detector is the {\em measured} energy per unit time per
unit area of the detector coming from that source. The luminosity
distance $d_L$ is then defined as the radius of the sphere centered on
the source for which the absolute luminosity would give the observed
flux, $\ {\cal F} \equiv {\cal L}/4\pi d_L^2$. In a
Friedmann-Robertson-Walker universe, light travels along null geodesics,
$ds^2=0$, or, see Eq.~(\ref{FRWmetric}),
\begin{equation}\label{NullGeodesics}
{dr\over\sqrt{1+a_0^2H_0^2\,r^2\,\Omega_K}} = {1\over a_0^2H_0^2}\,
{dz\over\sqrt{(1+z)^2(1+z\Omega_{\rm M}) - z(2+z)\Omega_\Lambda}}\,,
\end{equation}
which determines the coordinate distance $r = r(z, H_0, \Omega_{\rm M},
\Omega_\Lambda)$, as a function of redshift $z$ and the other
cosmological parameters. Now let us consider the effect of the universe
expansion on the observed flux coming from a source at a certain
redshift $z$ from us.  First, the photon energy on its way here will be
redshifted, and thus the observed energy $E_0 = E/(1+z)$. Second, the
rate of photon arrival will be time-delayed with respect to that emitted
by the source, $dt_0 = (1+z)dt$. Finally, the fraction of the area of
the 2-sphere centered on the source that is covered by the detector is
$dA/4\pi a_0^2\,r^2(z)$. Therefore, the total flux detected is
\begin{equation}\label{TotalFlux}
{\cal F} = {{\cal L}\over4\pi a_0^2\,r^2(z)} \equiv 
{{\cal L}\over4\pi d_L^2}\,.
\end{equation}
The final expression for the luminosity distance $d_L$ as a function of
redshift is thus given by~\cite{KT}
\begin{equation}\label{LuminosityDistance}
H_0\,d_L = (1+z)\,|\Omega_K|^{-1/2}\,{\rm sinn}\left[|\Omega_K|^{1/2}
\int_0^z {dz'\over\sqrt{(1+z')^2(1+z'\Omega_{\rm M}) - 
z'(2+z')\Omega_\Lambda}}\right]\,,
\end{equation}
where ${\rm sinn}(x)= x\ {\rm if}\ K=0;\ \sin(x)\ {\rm if}\ K=+1\ {\rm
and}\ \sinh(x)\ {\rm if}\ K=-1$. Expanding to second order around $z=0$,
we obtain Eq.~(\ref{PhysicalDistance}), 
\begin{equation}
H_0\,d_L = z + {1\over2}\Big(1-{\Omega_{\rm M}\over2}+\Omega_\Lambda\Big)\,z^2 + 
{\cal O}(z^3)\,.
\end{equation}
This expression goes beyond the leading linear term, corresponding to the
Hubble law, into the second order term, which is sensitive to the
cosmological parameters $\Omega_{\rm M}$ and $\Omega_\Lambda$. It is only
recently that cosmological observations have gone far enough back into
the early universe that we can begin to probe the second term, as I will
discuss shortly. Higher order terms are not yet probed by cosmological
observations, but they would contribute as important consistency checks.

Let us now pursue the analysis of the recent determinations of the most
important cosmological parameters: the rate of expansion $H_0$, the
matter content $\Omega_{\rm M}$, the cosmological constant $\Omega_\Lambda$,
the spatial curvature $\Omega_K$, and the age of the universe $t_0$.

\subsection{The rate of expansion $H_0$}

Over most of last century the value of $H_0$ has been a constant source
of disagreement~\cite{Freedman}. Around 1929, Hubble measured the rate
of expansion to be $H_0 = 500$ km\,s$^{-1}$Mpc$^{-1}$, which implied an
age of the universe of order $t_0\sim2$ Gyr, in clear conflict with
geology.  Hubble's data was based on Cepheid standard candles that were
incorrectly calibrated with those in the Large Magellanic Cloud. Later
on, in 1954 Baade recalibrated the Cepheid distance and obtained a lower
value, $H_0 = 250$ km\,s$^{-1}$Mpc$^{-1}$, still in conflict with ratios
of certain unstable isotopes. Finally, in 1958 Sandage realized that the
brightest stars in galaxies were ionized HII regions, and the Hubble
rate dropped down to $H_0 = 60$ km\,s$^{-1}$ Mpc$^{-1}$, still with
large (factor of two) systematic errors. Fortunately, in the past 15
years there has been significant progress towards the determination of
$H_0$, with systematic errors approaching the 10\% level. These
improvements come from two directions. First, technological, through the
replacement of photographic plates (almost exclusively the source of
data from the 1920s to 1980s) with charged couple devices (CCDs), i.e.
solid state detectors with excellent flux sensitivity per pixel, which
were previously used successfully in particle physics detectors. Second,
by the refinement of existing methods for measuring extragalactic
distances (e.g. parallax, Cepheids, supernovae, etc.). Finally, with the
development of completely new methods to determine $H_0$, which fall
into totally independent and very broad categories: a) Gravitational
lensing; b) Sunyaev-Zel'dovich effect; c) Extragalactic distance scale,
mainly Cepheid variability and type Ia Supernovae; d) Microwave
background anisotropies. I will review here the first three, and leave
the last method for Section~4.4, since it involves knowledge about the
primordial spectrum of inhomogeneities.

\subsubsection{Gravitational lensing}

Imagine a quasi-stellar object (QSO) at large redshift ($z\gg1$) whose
light is lensed by an intervening galaxy at redshift $z\sim1$ and
arrives to an observer at $z=0$. There will be at least two different
images of the same background {\em variable} point source. The arrival
times of photons from two different gravitationally lensed images of the
quasar depend on the different path lengths and the gravitational
potential traversed. Therefore, a measurement of the time delay and the
angular separation of the different images of a variable quasar can be
used to determine $H_0$ with great accuracy. This method, proposed in
1964 by Refsdael~\cite{Refsdael}, offers tremendous potential because it
can be applied at great distances and it is based on very solid physical
principles~\cite{Blandford}.

Unfortunately, there are very few systems with both a favourable
geometry (i.e. a known mass distribution of the intervening galaxy) and
a variable background source with a measurable time delay. That is the
reason why it has taken so much time since the original proposal for the
first results to come out. Fortunately, there are now very powerful
telescopes that can be used for these purposes. The best candidate
to-date is the QSO\ $0957+561$, observed with the 10m Keck telescope,
for which there is a model of the lensing mass distribution that is
consistent with the measured velocity dispersion. Assuming a flat space
with $\Omega_{\rm M}=0.25$, one can determine~\cite{Grogin}
\begin{equation}
H_0 = 72 \pm 7 \ (1\sigma\ {\rm statistical})\ 
\pm 15\%\ ({\rm systematic})\ \ {\rm km\,s}^{-1}{\rm Mpc}^{-1}\,.
\end{equation}
The main source of systematic error is the degeneracy between the mass
distribution of the lens and the value of $H_0$. Knowledge of the
velocity dispersion within the lens as a function of position helps
constrain the mass distribution, but those measurements are very
difficult and, in the case of lensing by a cluster of galaxies, the dark
matter distribution in those systems is usually unknown, associated with
a complicated cluster potential. Nevertheless, the method is just
starting to give promising results and, in the near future, with the
recent discovery of several systems with optimum properties, the
prospects for measuring $H_0$ and lowering its uncertainty with this
technique are excellent.

\subsubsection{Sunyaev-Zel'dovich effect}

\begin{figure}[htb]
\vspace*{1mm}
\begin{center}
\includegraphics[width=5cm,angle=0]{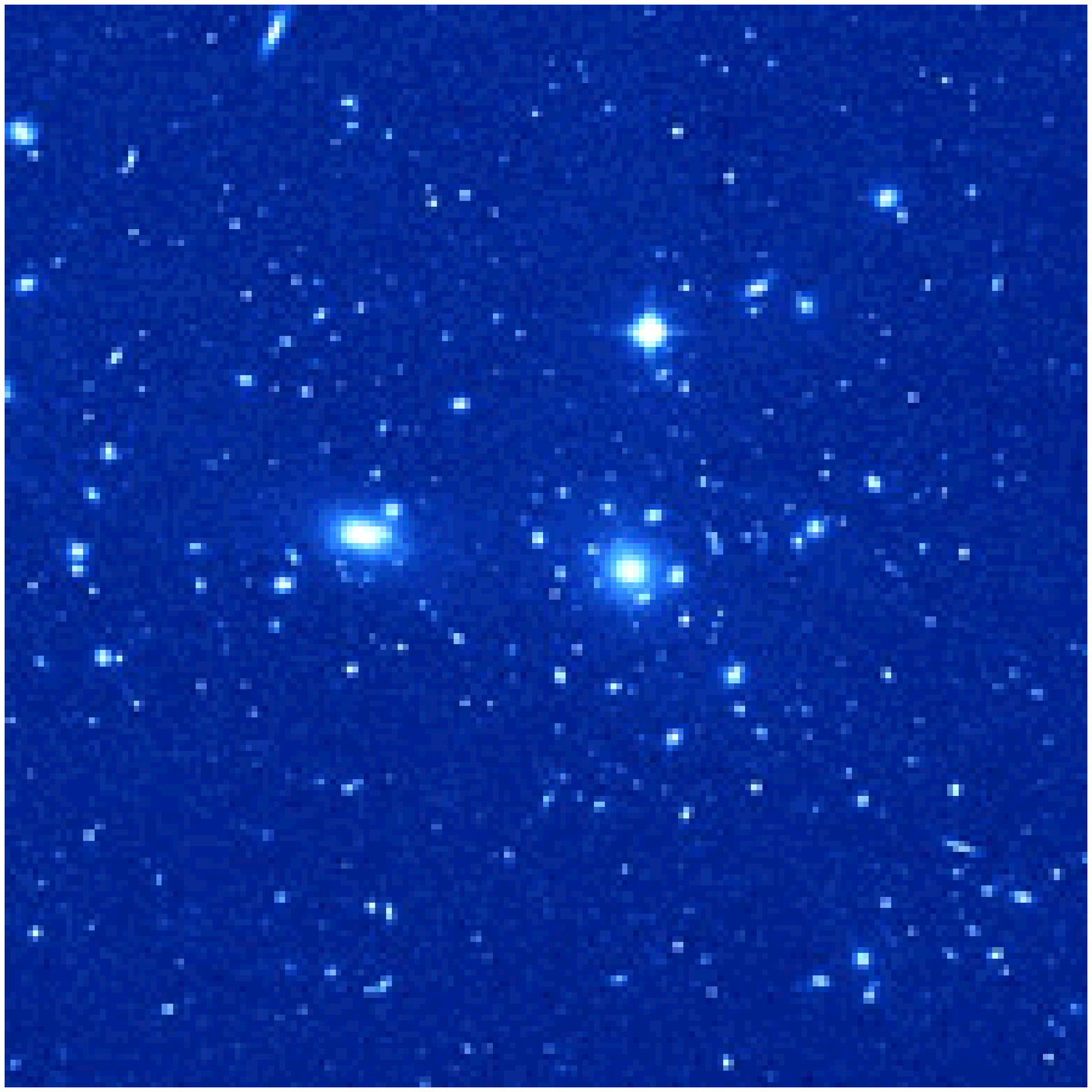}
\includegraphics[width=5cm,angle=0]{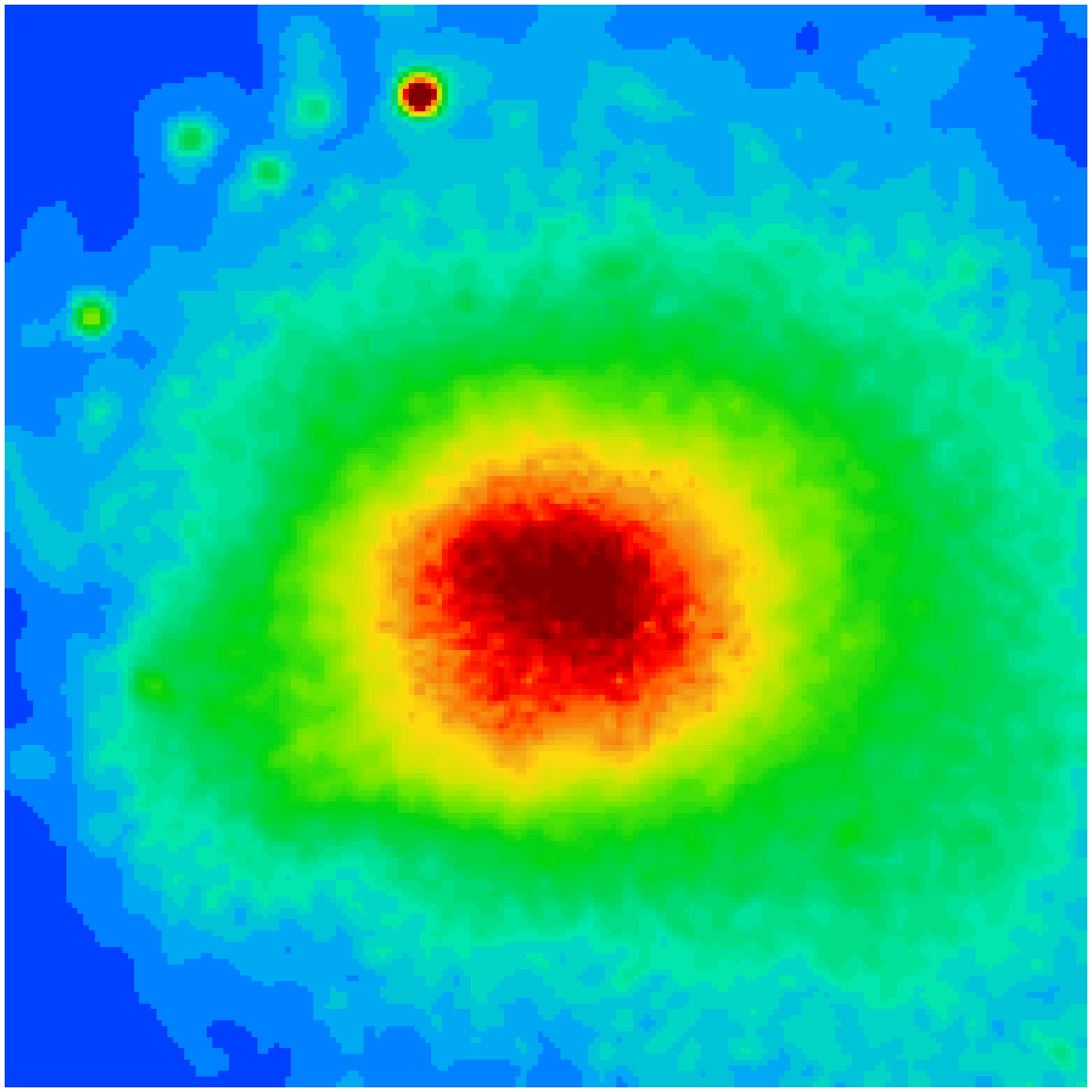}
\end{center}
\vspace*{-1mm}
\caption{The Coma cluster of galaxies, seen here in an optical image
(left) and an X-ray image (right), taken by the recently launched 
Chandra X-ray Observatory. From Ref.~\cite{Chandra}.}
\label{fig12}
\end{figure}

As discussed in the previous Section, the gravitational collapse of
baryons onto the potential wells generated by dark matter gave rise to
the reionization of the plasma, generating an X-ray halo around rich
clusters of galaxies, see Fig.~\ref{fig12}. The inverse-Compton
scattering of microwave background photons off the hot electrons in the
X-ray gas results in a measurable distortion of the blackbody spectrum
of the microwave background, known as the Sunyaev-Zel'dovich (SZ)
effect. Since photons acquire extra energy from the X-ray electrons, we
expect a shift towards higher frequencies of the spectrum,
$(\Delta\nu/\nu) \simeq (k_{\rm B}T_{\rm gas}/m_e c^2) \sim
10^{-2}$. This corresponds to a {\em decrement} of the microwave
background temperature at low frequencies (Rayleigh-Jeans region) and an
increment at high frequencies, see Ref.~\cite{Birkinshaw}.  

Measuring the {\em spatial} distribution of the SZ effect (3 K
spectrum), together with a high resolution X-ray map ($10^8$ K spectrum)
of the cluster, one can determine the density and temperature
distribution of the hot gas. Since the X-ray flux is distance-dependent
(${\cal F}={\cal L}/4\pi d_L^2$), while the SZ decrement is not (because
the energy of the CMB photons increases as we go back in redshift,
$\nu=\nu_0(1+z)$, and exactly compensates the redshift in energy of the
photons that reach us), one can determine from there the distance to the
cluster, and thus the Hubble rate $H_0$.

The advantages of this method are that it can be applied to large
distances and it is based on clear physical principles. The main
systematics come from possible clumpiness of the gas (which would reduce
$H_0$), projection effects (if the clusters are prolate, $H_0$ could be
larger), the assumption of hydrostatic equilibrium of the X-ray gas,
details of models for the gas and electron densities, and possible
contaminations from point sources. Present measurements give the
value~\cite{Birkinshaw}
\begin{equation}
H_0 = 60 \pm 10 \ (1\sigma\ {\rm statistical})\ 
\pm 20\%\ ({\rm systematic})\ \ {\rm km\,s}^{-1}{\rm Mpc}^{-1}\,,
\end{equation}
compatible with other determinations. A great advantage of this
completely new and independent method is that nowadays more and more
clusters are observed in the X-ray, and soon we will have
high-resolution 2D maps of the SZ decrement from several balloon
flights, as well as from future microwave background satellites,
together with precise X-ray maps and spectra from the Chandra X-ray
observatory recently launched by NASA, as well as from the European
X-ray satellite XMM launched a few months ago by ESA, which will deliver
orders of magnitude better resolution than the existing Einstein X-ray
satellite.

\subsubsection{Cepheid variability}

Cepheids are low-mass variable stars with a period-luminosity relation
based on the helium ionization cycles inside the star, as it contracts
and expands. This time variability can be measured, and the star's
absolute luminosity determined from the calibrated relationship. From
the observed flux one can then deduce the luminosity distance, see
Eq.~(\ref{LuminosityDistance}), and thus the Hubble rate $H_0$. The
Hubble Space Telescope (HST) was launched by NASA in 1990 (and repaired
in 1993) with the specific project of calibrating
the extragalactic distance scale and thus determining the Hubble rate
with 10\% accuracy. The most recent results from HST are the
following~\cite{HST}
\begin{equation}
H_0 = 71 \pm 4 \ ({\rm random})\ \pm 7\ ({\rm systematic})\ \ 
{\rm km\,s}^{-1}{\rm Mpc}^{-1}\,.
\end{equation}
The main source of systematic error is the distance to the Large
Magellanic Cloud, which provides the fiducial comparison for Cepheids in
more distant galaxies. Other systematic uncertainties that affect the
value of $H_0$ are the internal extinction correction method used, a
possible metallicity dependence of the Cepheid period-luminosity
relation and cluster population incompleteness bias, for a set of 21
galaxies within 25 Mpc, and 23 clusters within $z\lsim0.03$.

With better telescopes coming up soon, like the Very Large Telescope
(VLT) interferometer of the European Southern Observatory (ESO) in the
Chilean Atacama desert, with 4 synchronized telescopes by the year 2005,
and the Next Generation Space Telescope (NGST) proposed by NASA for
2008, it is expected that much better resolution and therefore accuracy
can be obtained for the determination of $H_0$.

\subsection{The matter content $\Omega_{\rm M}$}

In the 1920s Hubble realized that the so called nebulae were actually
distant galaxies very similar to our own. Soon afterwards, in 1933,
Zwicky found dynamical evidence that there is possibly ten to a hundred
times more mass in the Coma cluster than contributed by the luminous
matter in galaxies~\cite{Zwicky}. However, it was not until the 1970s
that the existence of dark matter began to be taken more seriously. At
that time there was evidence that rotation curves of galaxies did not
fall off with radius and that the dynamical mass was increasing with
scale from that of individual galaxies up to clusters of galaxies. Since
then, new possible extra sources to the matter content of the universe
have been accumulating:
\begin{eqnarray}\label{MatterContent}
\Omega_{\rm M} &=& \Omega_{\rm B,\ lum} \hspace{1.8cm} 
({\rm stars\ in\ galaxies})\\
&+& \Omega_{\rm B,\ dark} \hspace{1.7cm} ({\rm MACHOs?})\\
&+& \Omega_{\rm CDM} \hspace{2cm} ({\rm weakly\ interacting:\ 
axion,\ neutralino?})\\
&+& \Omega_{\rm HDM} \hspace{2cm} ({\rm massive\ neutrinos?})
\end{eqnarray}

The empirical route to the determination of $\Omega_{\rm M}$ is nowadays one
of the most diversified of all cosmological parameters. The matter
content of the universe can be deduced from the mass-to-light ratio of
various objects in the universe; from the rotation curves of galaxies;
from microlensing and the direct search of Massive Compact Halo Objects
(MACHOs); from the cluster velocity dispersion with the use of the Virial
theorem; from the baryon fraction in the X-ray gas of clusters; from
weak gravitational lensing; from the observed matter distribution of the
universe via its power spectrum; from the cluster abundance and its
evolution; from direct detection of massive neutrinos at
SuperKamiokande; from direct detection of Weakly Interacting Massive
Particles (WIMPs) at DAMA and UKDMC, and finally from microwave background
anisotropies. I will review here just a few of them.

\subsubsection{Luminous matter}

The most straight forward method of estimating $\Omega_{\rm M}$ is to measure
the luminosity of stars in galaxies and then estimate the mass-to-light
ratio, defined as the mass per luminosity density observed from an
object, $\Upsilon={\cal M}/{\cal L}$. This ratio is usually expressed in
solar units, ${\cal M}_\odot/{\cal L}_\odot$, so that for the sun
$\Upsilon_\odot=1$. The luminosity of stars depends very sensitively on
their mass and stage of evolution. The mass-to-light ratio of stars in
the solar neighbourhood is of order $\Upsilon\approx3$. For globular
clusters and spiral galaxies we can determine their mass and luminosity
independently and this gives $\Upsilon\approx$ few. For our galaxy,
\begin{equation}\label{LuminosityGalaxy}
{\cal L}_{\rm gal} = (1.0\pm0.3)\times10^8\,h\,L_\odot\,{\rm Mpc}^{-3}
\hspace{1cm} {\rm and} \hspace{1cm} \Upsilon_{\rm gal} = 6\pm3\,.
\end{equation}
The contribution of galaxies to the luminosity density of the universe
(in the visible-V spectral band, centered at $\sim 5500$ \AA)
is~\cite{Huchra}
\begin{equation}\label{LuminosityUniverse}
{\cal L}_V = (1.7\pm0.6)\times10^8\,h\,L_\odot\,{\rm Mpc}^{-3}\,,
\end{equation}
which can be translated into a mass density by multiplying by the observed
$\Upsilon$ in that band,
\begin{equation}\label{MassDensity}
\Omega_{\rm M}\,h = (6.1\pm2.2)\times10^{-4}\,\Upsilon_V\,.
\end{equation}
All the luminous matter in the universe, from galaxies, clusters of
galaxies, etc., account for $\Upsilon\approx10$, and thus~\cite{Copi}
\begin{equation}\label{OmegaLuminous}
0.002 \leq \Omega_{\rm lum}\,h \leq 0.006\,.
\end{equation}
As a consequence, the luminous matter alone is far from the critical
density. Moreover, comparing with the amount of baryons from Big Bang
nucleosynthesis (\ref{OmegaBaryon}), we conclude that $\Omega_{\rm
lum} \ll \Omega_{\rm B}$, so there must be a large fraction of baryons
that are dark, perhaps in the form of very dim stars.

\subsubsection{Rotation curves of spiral galaxies}

\begin{figure}[htb]
\begin{center}
\includegraphics[width=7cm,angle=0]{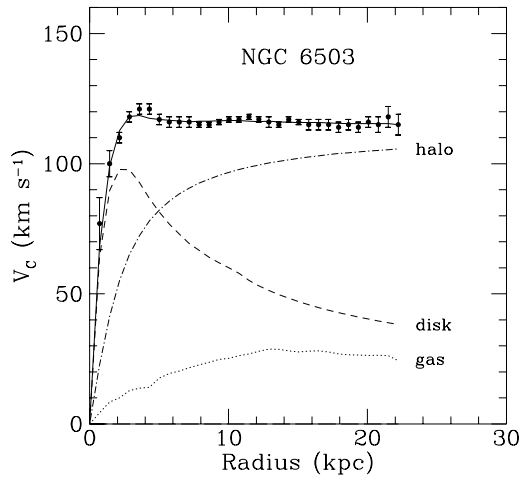}
\end{center}\vspace*{-5mm}
\caption{The rotation curve of the spiral galaxy NGC 6503, determined by
radio observations of hydrogen gas in the disk~\cite{Begeman}. The
dashed line shows the rotation curve expected from the disk material
alone, the dot-dashed line is from the dark matter halo alone.}
\label{fig13}
\end{figure}

The flat rotation curves of spiral galaxies provide the most direct
evidence for the existence of large amounts of dark matter. Spiral
galaxies consist of a central bulge and a very thin disk, stabilized
against gravitational collapse by angular momentum conservation, and
surrounded by an approximately spherical halo of dark matter. One can
measure the orbital velocities of objects orbiting around the disk as a
function of radius from the Doppler shifts of their spectral lines. The
rotation curve of the Andromeda galaxy was first measured by Babcock in
1938, from the stars in the disk. Later it became possible to measure
galactic rotation curves far out into the disk, and a trend was
found~\cite{Freeman}. The orbital velocity rose linearly from the center
outward until it reached a typical value of 200 km/s, and then remained
flat out to the largest measured radii. This was completely unexpected
since the observed surface luminosity of the disk falls off
exponentially with radius, $I(r) = I_0 \exp(-r/r_D)$, see
Ref.~\cite{Freeman}. Therefore, one would expect that most of the
galactic mass is concentrated within a few disk lengths $r_D$, such that
the rotation velocity is determined as in a Keplerian orbit, $v_{\rm
rot} = (GM/r)^{1/2} \propto r^{-1/2}$. No such behaviour is observed. In
fact, the most convincing observations come from radio emission (from
the 21 cm line) of neutral hydrogen in the disk, which has been measured
to much larger galactic radii than optical tracers. A typical case is
that of the spiral galaxy NGC 6503, where $r_D = 1.73$ kpc, while the
furthest measured hydrogen line is at $r=22.22$ kpc, about 13 disk
lengths away. The measured rotation curve is shown in Fig.~\ref{fig13}
together with the relative components associated with the disk, the halo
and the gas.

Nowadays, thousands of galactic rotation curves are known, and all
suggest the existence of about ten times more mass in the halos of
spiral galaxies than in the stars of the disk. Recent numerical
simulations of galaxy formation in a CDM cosmology~\cite{Frenk} suggest
that galaxies probably formed by the infall of material in an overdense
region of the universe that had decoupled from the overall expansion.
The dark matter is supposed to undergo violent relaxation and create a
virialized system, i.e. in hydrostatic equilibrium. This picture has led
to a simple model of dark-matter halos as isothermal spheres, with
density profile $\rho(r) = \rho_c/(r_c^2 + r^2)$, where $r_c$ is a core
radius and $\rho_c = v_\infty^2/4\pi G$, with $v_\infty$ equal to the
plateau value of the flat rotation curve. This model is consistent with
the universal rotation curve seen in Fig.~\ref{fig13}. At large radii
the dark matter distribution leads to a flat rotation curve. Adding up
all the matter in galactic halos up to maximum radii, one finds
$\Upsilon_{\rm halo} \geq 30\,h$, and therefore
\begin{equation}\label{OmegaHalo}
\Omega_{\rm halo} \geq 0.03 - 0.05\,.
\end{equation}
Of course, it would be extraordinary if we could confirm, through direct
detection, the existence of dark matter in our own galaxy. For that
purpose, one should measure its rotation curve, which is much more
difficult because of obscuration by dust in the disk, as well as
problems with the determination of reliable galactocentric distances for
the tracers. Nevertheless, the rotation curve of the Milky Way has been
measured and conforms to the usual picture, with a plateau value of the
rotation velocity of 220 km/s, see Ref.~\cite{Fich}. For dark matter
searches, the crucial quantity is the dark matter density in the solar
neighbourhood, which turns out to be (within a factor of two uncertainty
depending on the halo model) $\rho_{\rm DM} = 0.3$ GeV/cm$^3$. We will
come back to direct searched of dark matter in a later subsection.

\subsubsection{Microlensing}

The existence of large amounts of dark matter in the universe, and in our
own galaxy in particular, is now established beyond any reasonable
doubt, but its nature remains a mystery. We have seen that baryons
cannot account for the whole matter content of the universe; however,
since the contribution of the halo (\ref{OmegaHalo}) is comparable in
magnitude to the baryon fraction of the universe (\ref{OmegaBaryon}),
one may ask whether the galactic halo could be made of purely baryonic
material in some non-luminous form, and if so, how one should search for
it. In other words, are MACHOs the non-luminous baryons filling the gap
between $\Omega_{\rm lum}$ and $\Omega_{\rm B}$? If not, what are they?

Let us start a systematic search for possibilities. They cannot be
normal stars since they would be luminous; neither hot gas since it
would shine; nor cold gas since it would absorb light and reemit in the
infrared. Could they be burnt-out stellar remnants? This seems
implausible since they would arise from a population of normal stars of
which there is no trace in the halo. Neutron stars or black holes would
typically arise from Supernova explosions and thus eject heavy elements
into the galaxy, while the overproduction of helium in the halo is
strongly constrained. They could be white dwarfs, i.e. stars not massive
enough to reach supernova phase. Despite some recent arguments, a halo
composed by white dwarfs is not rigorously excluded. Are they stars too
small to shine? Perhaps M-dwarfs, stars with a mass $M\leq 0.1\,M_\odot$
which are intrinsically dim; however, very long exposure images of the
Hubble Space Telescope restrict the possible M-dwarf contribution to the
galaxy to be below 6\%. The most plausible alternative is a halo
composed of brown dwarfs with mass $M\leq 0.08\, M_\odot$, which never
ignite hydrogen and thus shine only from the residual energy due to
gravitational contraction.\footnote{A sometimes discussed alternative,
planet-size Jupiters, can be classified as low-mass brown dwarfs.} In
fact, the extrapolation of the stellar mass function to small masses
predicts a large number of brown dwarfs within normal stellar
populations. A final possibility is primordial black holes (PBH), which
could have been created in the early universe from early phase
transitions~\cite{LGBW}, even before baryons were formed, and thus may
be classified as non-baryonic. They could make a large contribution
towards the total $\Omega_{\rm M}$, and still be compatible with Big
Bang nucleosynthesis.

\begin{figure}[htb]
\begin{center}
\includegraphics[width=8.5cm]{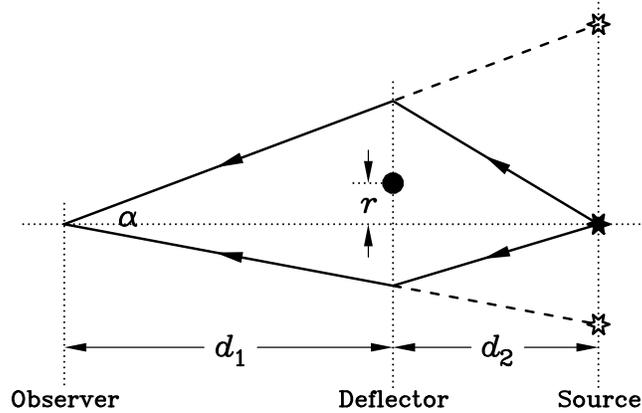} 
\caption{Geometry of the light deflection by a pointlike mass which
gives two images of a source viewed by an observer. From 
Ref.~\cite{Raffelt}.}
\label{fig14}
\end{center}
\end{figure}

\begin{figure}[htb]
\begin{center}
\includegraphics[width=7.5cm]{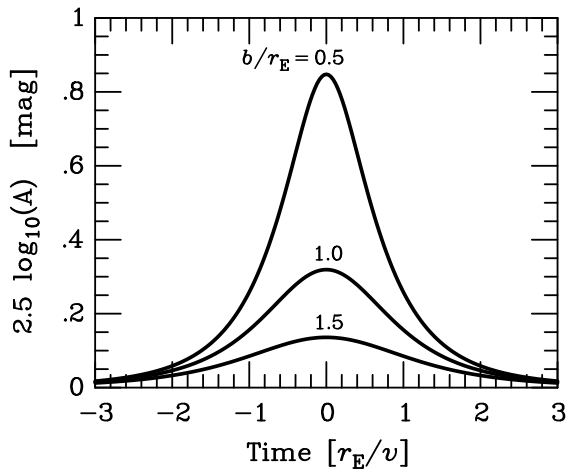} 
\caption{The apparent lightcurve of a source if a pointlike MACHO
passes through the line of sight with a transverse velocity $v$ and an
impact parameter $b$. The amplification factor $A$ is shown in
logarithmic scale to give the usual astronomical magnitude of an object.
From Ref.~\cite{Raffelt}.}
\label{fig15}
\end{center}
\end{figure}

\begin{figure}[htb]
\begin{center}
\includegraphics[width=7.5cm]{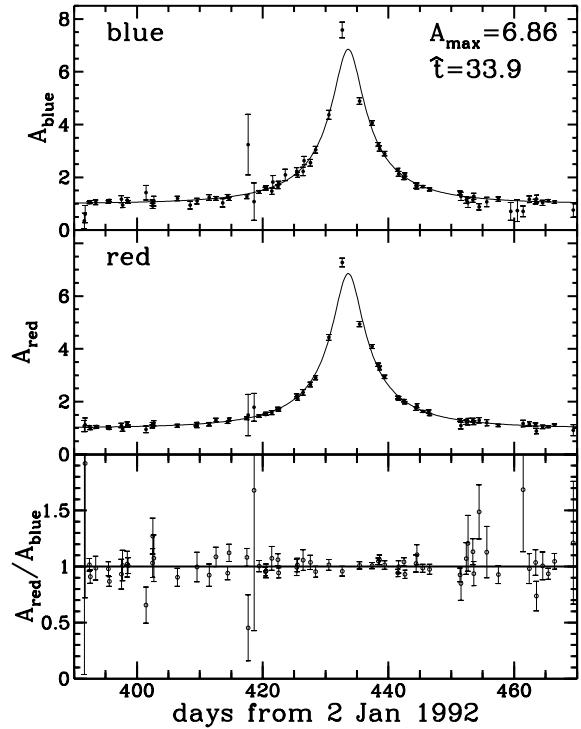} 
\caption{The best candidate (LMC-1) for microlensing from the MACHO
Collaboration in the direction of the Large Mage\-llanic Cloud. A recent
reanalysis of this event suggested an amplification factor \ $A_{\rm
max}=7.20\pm0.09$, with achromaticity $A_{\rm red}/A_{\rm blue} =
1.00\pm0.05$, and a duration of \ $\hat t = 34.8\pm0.2$. From
Ref.~\cite{microlensing}.}
\label{fig16}
\end{center}
\end{figure}

\begin{figure}[htb]
\begin{center}
\includegraphics[width=6cm,angle=-90]{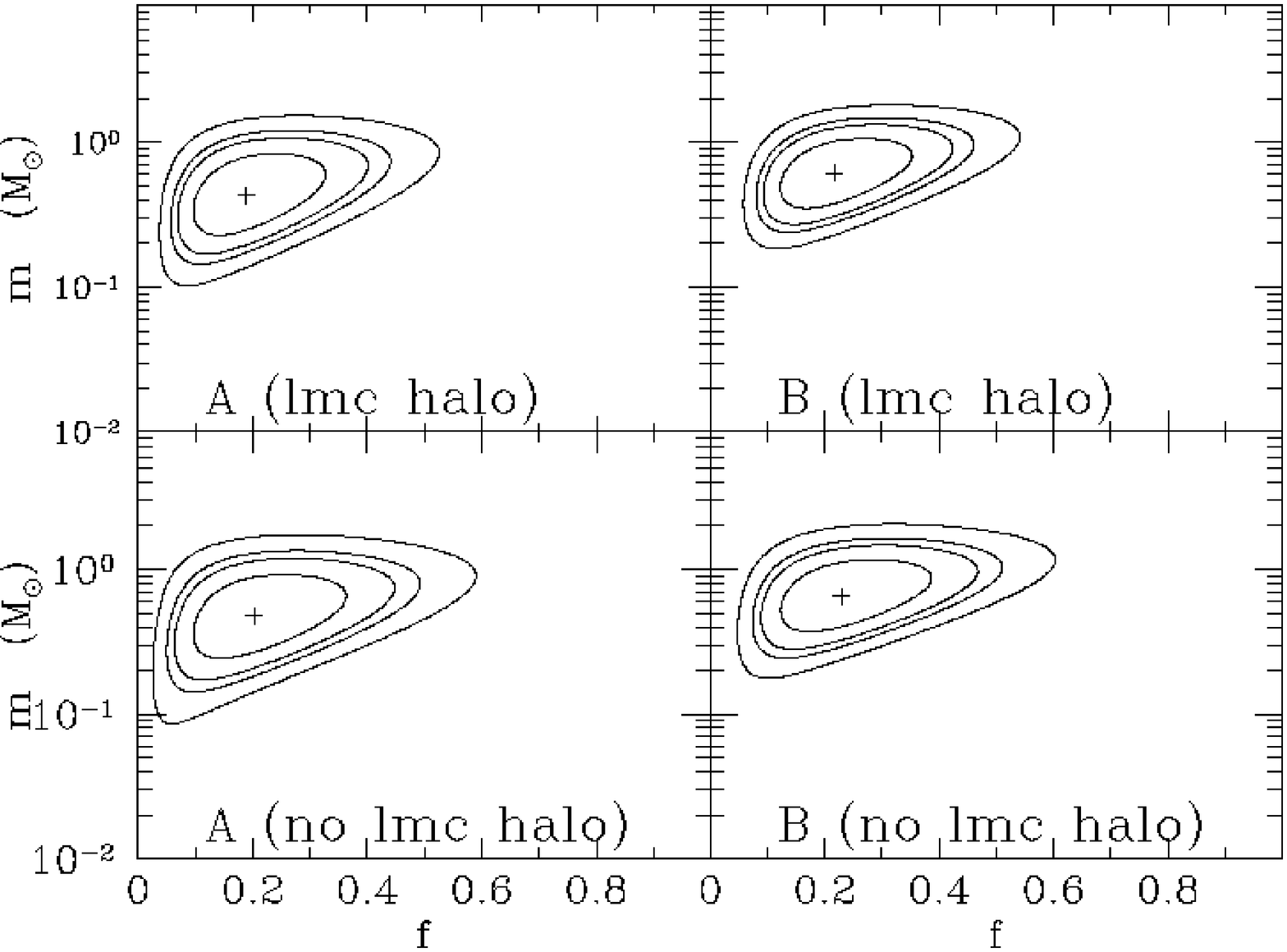} 
\caption{Likelihood contours for MACHO mass $m$ (in units of solar mass)
and halo fraction $f$ for a typical size halo. The plus sign shows the
maximum likelihood estimate and the contours enclose regions of 68\%,
90\%, 95\% and 99\% probability. The panels are labeled according to
different sets of selection criteria (A or B), and whether or not an
LMC halo with MACHO fraction $f$ is included. From Ref.~\cite{LMC}.}
\label{fig17}
\end{center}
\end{figure}

Whatever the arguments for or against baryonic objects as galactic dark
matter, nothing would be more convincing than a direct detection of the
various candidates, or their exclusion, in a direct search experiment.
Fortunately, in 1986 Paczy\'nski proposed a method for detecting faint
stars in the halo of our galaxy~\cite{Paczynski}. The idea is based on
the well known effect that a point-like mass deflector placed between an
observer and a light source creates two different images, as shown in
Fig.~\ref{fig14}. When the source is exactly aligned with the deflector
of mass $M_D$, the image would be an annulus, an {\em Einstein ring},
with radius
\begin{equation}\label{EinsteinRadius}
r_{\rm E}^2 = 4 G M_D\,d\,, \hspace{1cm} {\rm where} \hspace{1cm} 
d={d_1d_2\over d_1+d_2}
\end{equation}
is the {\em reduced} distance to the source, see Fig.~\ref{fig14}. If
the two images cannot be separated because their angular distance
$\alpha$ is below the resolving power of the observer's telescope, the
only effect will be an apparent brightening of the star, an effect known
as {\em gravitational microlensing}. The amplification factor 
is~\cite{Paczynski}
\begin{equation}\label{Amplification}
A={2+u^2\over u\sqrt{4+u^2}}\,, \hspace{1cm} {\rm where} \hspace{1cm} 
u\equiv{r\over r_{\rm E}}\,,
\end{equation}
with $r$ the distance from the line of sight to the deflector. Imagine
an observer on Earth watching a distant star in the Large Magellanic
Cloud (LMC), 50 kpc away. If the galactic halo is filled with MACHOs,
one of them will occasionally pass near the line of sight and thus cause
the image of the background star to brighten. If the MACHO moves with
velocity $v$ transverse to the line of sight, and if its {\em impact
parameter}, i.e. the minimal distance to the line of sight, is $b$, then
one expects an apparent lightcurve as shown in Fig.~\ref{fig15} for
different values of $b/r_{\rm E}$. The natural time unit is $\Delta t =
r_{\rm E}/v$, and the origin corresponds to the time of closest approach
to the line of sight.

The probability for a target star to be lensed is independent of the
mass of the dark matter object~\cite{Paczynski,Raffelt}. For stars in
the LMC one finds a probability, i.e. an {\em optical depth} for
microlensing of the galactic halo, of approximately $\tau\sim
10^{-6}$. Thus, if one looks simultaneously at several millions of stars
in the LMC during extended periods of time, one has a good chance of
seeing at least a few of them brightened by a dark halo object. In order
to be sure one has seen a microlensing event one has to monitor a large
sample of stars long enough to identify the characteristic light curve
shown in Fig.~\ref{fig15}.  The unequivocal signatures of such an event
are the following: it must be a) unique (non-repetitive in time); b)
time-symmetric; and c) achromatic (because of general covariance). These
signatures allow one to discriminate against variable stars which
constitute the background. The typical duration of the light curve is
the time it takes a MACHO to cross an Einstein radius, $\Delta t =
r_{\rm E}/v$. If the deflector mass is $1\ M_\odot$, the average
microlensing time will be 3~months, for $10^{-2}\ M_\odot$ it is 9~days,
for $10^{-4}\ M_\odot$ it is 1~day, and for $10^{-6}\ M_\odot$ it is
2~hours. A characteristic event, of duration 34 days,
is shown in Fig.~\ref{fig16}.

The first microlensing events towards the LMC were reported by the MACHO
and EROS collaborations in 1993~\cite{MACHO,EROS}. Nowadays, there are
12 candidates towards the LMC, 2 towards the SMC, around 40 towards the
bulge of our own galaxy, and about 2 towards Andromeda, seen by
AGAPE~\cite{AGAPE}, with a slightly different technique based on pixel
brightening rather than individual stars. Thus, microlensing is a well
established technique with a rather robust future. In particular, it has
allowed the MACHO and EROS collaboration to draw exclusion plots for
various mass ranges in terms of their maximum allowed halo fraction, see
Fig.~\ref{fig17}. The MACHO Collaboration conclude in their 5-year
analysis, see Ref.~\cite{LMC}, that the spatial distribution of events
is consistent with an extended lens distribution such as Milky Way or
LMC halo, consisting partially of compact objects. A maximum likelihood
analysis gives a MACHO halo fraction of 20\% for a typical halo model
with a 95\% confidence interval of 8\% to 50\%. A 100\% MACHO halo is
ruled out at 95\% c.l. for all except their most extreme halo model.
The most likely MACHO mass is between 0.15 $M_\odot$ and 0.9 $M_\odot$,
depending on the halo model. The lower mass is characteristic of white
dwarfs, but a galactic halo composed primarily of white dwarfs is barely
compatible with a range of observational constraints. On the other hand,
if one wanted to attribute the observed events to brown dwarfs, one
needs to appeal to a very non-standard density and/or velocity
distribution of these objects. It is still unclear what sort of objects
the microlensing experiments are seeing towards the LMC and where the
lenses are. Nevertheless, the field is expanding, with several new
experiments already underway, to search for clear signals of parallax,
or binary systems, where the degeneracy between mass and distance can be
resolved. For a discussion of those new results, see
Ref.~\cite{microlensing}.

\subsubsection{Virial theorem and large scale motion}

\begin{figure}[htb]
\begin{center}
\includegraphics[width=8.3cm,angle=0]{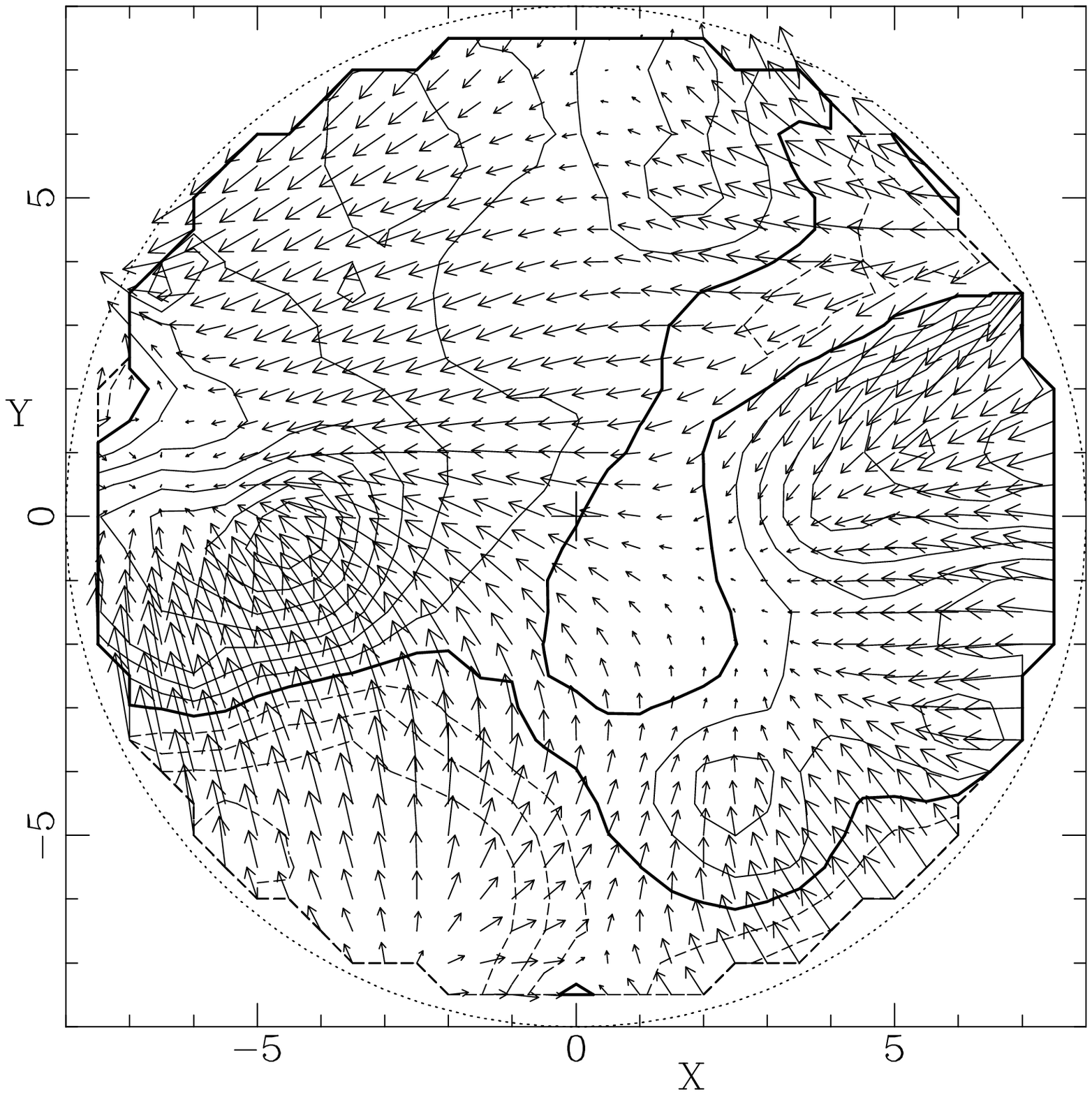} 
\includegraphics[width=5cm,angle=-90]{fig18b.ps} 
\end{center}
\caption{The velocity and density fluctuation fields in the
Supergalactic Plane as recovered by the POTENT method from the Mark III
velocities of about 3,000 galaxies with 12 $h^{-1}$ smoothing. The
vectors are projections of the 3D velocity field in the frame of the
CMB. Coordinates are in units of 10 $h^{-1}$ Mpc. The marked structures
are the Local Group (LG), the ``Great Attractor'' (GA), the Coma 
cluster ``Great Wall'' (GW), the Perseus-Pisces (PP) region and the 
``Southern Wall'' (SW). From Ref.~\cite{Dekel}.}
\label{fig18}
\end{figure}

Clusters of galaxies are the largest gravitationally bound systems in
the universe (superclusters are not yet in equilibrium). We know today
several thousand clusters; they have typical radii of $1 - 5$ Mpc and
typical masses of $2 - 9$ $\times10^{14}\ M_\odot$. Zwicky noted in 1933
that these systems appear to have large amounts of dark
matter~\cite{Zwicky}. He used the virial theorem (for a gravitationally
bound system in equilibrium), $2\langle E_{\rm kin}\rangle = - \langle
E_{\rm grav}\rangle$, where $\langle E_{\rm
kin}\rangle={1\over2}m\langle v^2\rangle$ is the average kinetic energy
of one of the bound objects (galaxies) of mass $m$ and $\langle E_{\rm
grav}\rangle = - m\langle GM/r\rangle$ is the average gravitational
potential energy caused by the attraction of the other
galaxies. Measuring the velocity dispersion $\langle v^2\rangle$ from
the Doppler shifts of the spectral lines and estimating the geometrical
size of the system gives an estimate of its total mass~$M$. As Zwicky
noted, this {\em virial mass} of clusters far exceeds their luminous
mass, typically leading to a mass-to-light ratio $\Upsilon_{\rm cluster}
= 200 \pm 70$. Assuming that the average cluster $\Upsilon$ is
representative of the entire universe~\footnote{Recent observations
indicate that $\Upsilon$ is independent of scale up to supercluster
scales $\sim 100\ h^{-1}$ Mpc.} one finds for the cosmic matter
density~\cite{Carlberg}
\begin{equation}\label{OmegaCluster}
\Omega_{\rm M} = 0.24 \pm 0.05\ (1\sigma\ {\rm statistical})\ 
\pm 0.09\ ({\rm systematic})\,.
\end{equation}

On scales larger than clusters the motion of galaxies is dominated by
the overall cosmic expansion. Nevertheless, galaxies exhibit {\em
peculiar velocities} with respect to the global cosmic flow. For
example, our Local Group of galaxies is moving with a speed of
$627\pm22$ km/s relative to the cosmic microwave background reference
frame, towards the Great Attractor.

In the context of the standard gravitational instability theory of
structure formation, the peculiar motions of galaxies are attributed to
the action of gravity during the universe evolution, caused by the
matter density inhomogeneities which give rise to the formation of
structure. The observed large-scale velocity fields, together with the
observed galaxy distributions, can then be translated into a measure for
the mass-to-light ratio required to explain the large-scale flows. An
example of the reconstruction of the matter density field in our
cosmological vicinity from the observed velocity field is shown in
Fig.~\ref{fig18}. The cosmic matter density inferred from such analyses 
is~\cite{Dekel,DBW}
\begin{equation}\label{OmegaFlows}
\Omega_{\rm M} > 0.3 \hspace{1cm} 95\%\ {\rm c.l.}
\end{equation}
Related methods that are more model-dependent give even larger
estimates.

\subsubsection{Baryon fraction in clusters}

Since large clusters of galaxies form through gravitational collapse,
they scoop up mass over a large volume of space, and therefore the ratio
of baryons over the total matter in the cluster should be representative
of the entire universe, at least within a 20\% systematic error. Since
the 1960s, when X-ray telescopes became available, it is known that
galaxy clusters are the most powerful X-ray sources in the
sky~\cite{Sarazin}. The emission extends over the whole cluster and
reveals the existence of a hot plasma with temperature $T\sim 10^7 -
10^8$ K, where X-rays are produced by electron bremsstrahlung. Assuming
the gas to be in hydrostatic equilibrium and applying the virial theorem
one can estimate the total mass in the cluster, giving general agreement
(within a factor of 2) with the virial mass estimates. From these
estimates one can calculate the baryon fraction of clusters
\begin{equation}\label{BaryonFraction}
f_{\rm B}h^{3/2} = 0.03 - 0.08 \hspace{1cm} \Rightarrow
\hspace{1cm} {\Omega_{\rm B}\over\Omega_{\rm M}} \approx 0.15\,,
\hspace{5mm} {\rm for} \hspace{3mm} h=0.65\,,
\end{equation}
which together with (\ref{OmegaLuminous}) indicates that clusters
contain far more baryonic matter in the form of hot gas than in the form
of stars in galaxies. Assuming this fraction to be representative of the
entire universe, and using the Big Bang nucleosynthesis value of
$\Omega_{\rm B} = 0.05 \pm 0.01$, for $h=0.65$, we find
\begin{equation}\label{OmegaXray}
\Omega_{\rm M} = 0.3 \pm 0.1\ ({\rm statistical})\ 
\pm 20\%\ ({\rm systematic})\,.
\end{equation}
This value is consistent with previous determinations of $\Omega_{\rm M}$.
If some baryons are ejected from the cluster during gravitational
collapse, or some are actually bound in nonluminous objects like planets,
then the actual value of $\Omega_{\rm M}$ is smaller than this estimate.

\subsubsection{Weak gravitational lensing}

\begin{figure}[htb]
\begin{center}
\includegraphics[width=5.5cm,angle=-90]{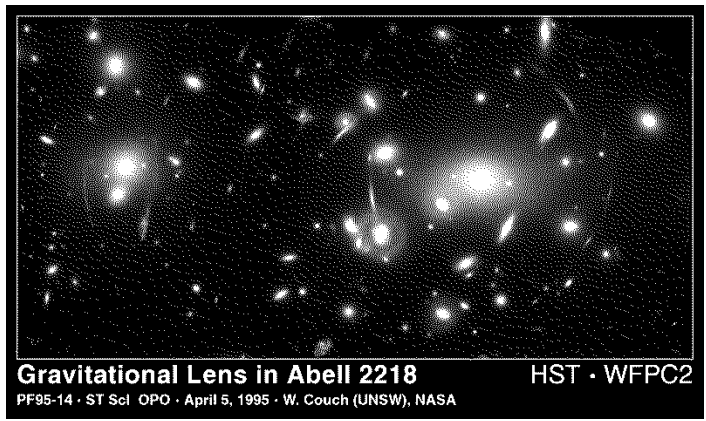} 
\end{center}\vspace{-2mm}
\caption{The most famous image of weak gravitational lensing around
the Abell 2218 cluster, made by the Hubble Space Telescope. From
Ref.~\cite{DeepField}.}
\label{fig19}
\end{figure}

Since the mid 1980s, deep surveys with powerful telescopes have observed
huge arc-like features in galaxy clusters, see for instance
Fig.~\ref{fig19}. The spectroscopic analysis showed that the cluster and
the giant arcs were at very different redshifts. The usual
interpretation is that the arc is the image of a distant background
galaxy which is in the same line of sight as the cluster so that it
appears distorted and magnified by the gravitational lens effect: the
giant arcs are essentially partial Einstein rings. From a systematic
study of the cluster mass distribution one can reconstruct the shear
field responsible for the gravitational distortion, see 
Ref.~\cite{Bartelmann}. 

This analysis shows that there are large amounts of dark matter in the
clusters, in rough agreement with the virial mass estimates, although
the lensing masses tend to be systematically larger. At present, the
estimates indicate $\Omega_{\rm M} = 0.2 - 0.3$ on scales $\lsim
6\,h^{-1}$ Mpc, while $\Omega_{\rm M} = 0.4$ for the Corona Borealis
supercluster, on scales of order 20 Mpc.

\subsubsection{Structure formation and the matter power spectrum}

One the most important constraints on the amount of matter in the
universe comes from the present distribution of galaxies. As we
mentioned in the Section~2.3, gravitational instability increases the
primordial density contrast, seen at the last scattering surface as
temperature anisotropies, into the present density field responsible for
the large and the small scale structure.

Since the primordial spectrum is very approximately represented by a
scale-invariant {\em Gaussian random field}, the best way to present the
results of structure formation is by working with the 2-point
correlation function in Fourier space (the equivalent to the Green's
function in QFT), the so-called {\em power spectrum}. If the reprocessed
spectrum of inhomogeneities remains Gaussian, the power spectrum is all
we need to describe the galaxy distribution. Non-Gaussian effects are
expected to arise from the non-linear gravitational collapse of
structure, and may be important at small scales~\cite{Peebles}.

The power spectrum measures the degree of inhomogeneity in the mass
distribution on different scales. It depends upon a few basic
ingredientes: a) the primordial spectrum of inhomogeneities, whether
they are Gaussian or non-Gaussian, whether {\em adiabatic}
(perturbations in the energy density) or {\em isocurvature}
(perturbations in the entropy density), whether the primordial spectrum
has {\em tilt} (deviations from scale-invariance), etc.; b) the recent
creation of inhomogeneities, whether {\em cosmic strings} or some other
topological defect from an early phase transition are responsible for
the formation of structure today; and c) the cosmic evolution of the
inhomogeneity, whether the universe has been dominated by cold or hot
dark matter or by a cosmological constant since the beginning of
structure formation, and also depending on the rate of expansion of the
universe.

The working tools used for the comparison between the observed power
spectrum and the predicted one are very precise N-body numerical
simulations and theoretical models that predict the {\em shape} but not
the {\em amplitude} of the present power spectrum. Even though a large
amount of work has gone into those analyses, we still have large
uncertainties about the nature and amount of matter necessary for
structure formation.  A model that has become a working paradigm is a
flat cold dark matter model with a cosmological constant and
$\Omega_{\rm M} = 0.3 - 0.4$. This model will soon be confronted with
very precise measurements from SDSS, 2dF, and several other large
redshift catalogs, that are already taking data, see Section~4.5.

The observational constraints on the power spectrum have a huge lever
arm of measurements at very different scales, mainly from the observed
cluster abundance, on 10 Mpc scales, to the CMB fluctuations, on 1000
Mpc scales, which determines the normalization of the spectrum. At
present, deep redshift surveys are probing scales between 100 and 1000
Mpc, which should begin to see the turnover corresponding to the peak of
the power spectrum at $k_{\rm eq}$, see Figs.~\ref{fig8} and \ref{fig9}.
The standard CDM model with $\Omega_{\rm M} = 1$, normalized to the CMB
fluctuations on large scales, is inconsistent with the cluster 
abundance. The power spectra of both a flat model with a cosmological
constant or an open universe with $\Omega_{\rm M} = 0.3$ (defined as
$\Lambda$CDM and OCDM, respectively) can be normalized so that they
agree with both the CMB and cluster observations. In the near future,
galaxy survey observations will greatly improve the power spectrum
constraints and will allow a measurement of $\Omega_{\rm M}$ from the
shape of the spectrum. At present, these measurements suggest a low
value of $\Omega_{\rm M}$, but with large uncertainties.

\subsubsection{Cluster abundance and evolution}

\begin{figure}[htb]
\begin{center}
\includegraphics[width=10cm,angle=0]{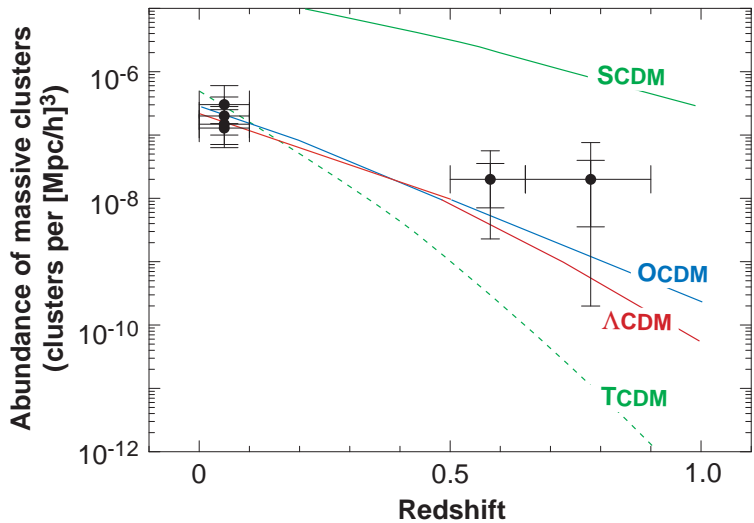} 
\end{center}\vspace{-1cm}
\caption{The evolution of the cluster abundance as a function of
redshift, compared with observations from massive clusters. The four
models are normalized to COBE. From Ref.~\cite{BOPS}.}
\label{fig20}
\end{figure}

Rich clusters are the most recently formed gravitationally bound systems
in the universe. Their number density as a function of time (or
redshift) helps determine the amount of dark matter. The observed
present ($z\sim0$) cluster abundance provides a strong constraint on the
normalization of the power spectrum of density perturbations on cluster
scales. Both $\Lambda$CDM and OCDM are consistent with the observed
cluster abundance at $z\sim0$, see Fig.~\ref{fig20}, while Standard CDM
(Einstein-De Sitter model, with $\Omega_{\rm M} = 1$), when normalized
at COBE scales, produces too many clusters at all redshifts.

The {\em evolution} of the cluster abundance with redshift breaks the 
degeneracy among the models at $z\sim0$. The low-mass models (Open and
$\Lambda$-CDM) predict a relatively small change in the number density of
rich clusters as a function of redshift because, due to the low density,
hardly any structure growth occurs since $z\sim1$. The high-mass models
(Tilted and Standard CDM) predict that structure has grown steadily and
rich clusters only formed recently: the number density of rich clusters
at $z\sim1$ is predicted to be exponentially smaller than today. The
observation of a single massive cluster is enough to rule out the 
$\Omega_{\rm M} = 1$ model. In fact, three clusters have been seen,
suggesting a low density universe~\cite{Bahcall},
\begin{equation}\label{ClusterEvolution}
\Omega_{\rm M} = 0.25^{\ \ +0.15}_{\ \ -0.10}\ (1\sigma\ 
{\rm statistical})\ \pm 20\%\ ({\rm systematic})\,.
\end{equation}
But one should be cautious. There is the caveat that for this constraint
it is assumed that the initial spectrum of density perturbations is
Gaussian, as predicted in the simplest models of inflation, but that has
not yet been confirmed observationally on cluster scales.

\begin{figure}[htb]
\begin{center}
\includegraphics[width=7.5cm]{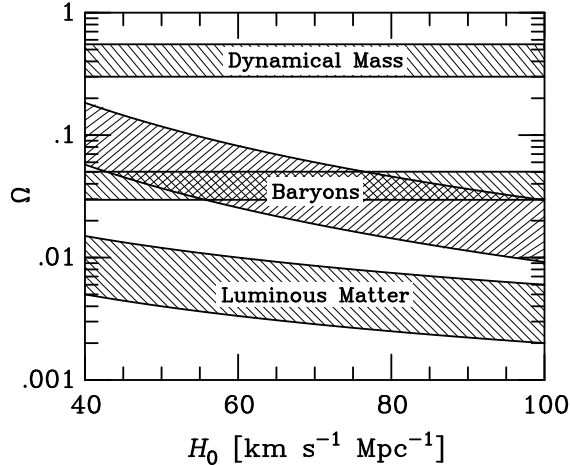} 
\caption{The observed cosmic matter components as functions of the
Hubble expansion parameter. The luminous matter component is given by
Eq.~(\ref{OmegaLuminous}); the galactic halo component is the horizontal
band, Eq.~(\ref{OmegaHalo}), crossing the baryonic component from BBN,
Eq.~(\ref{OmegaBaryon}); and the dynamical mass component from large
scale structure analysis is given by Eq.~(\ref{OmegaXray}). Note that
in the range $H_0 = 70\pm7$ km/s/Mpc, there are {\em three} dark matter
problems, see the text.}
\label{fig21}
\end{center}
\end{figure}

\subsubsection{Summary of the matter content}

We can summarize the present situation with Fig.~\ref{fig21}, for
$\Omega_{\rm M}$ as a function of $H_0$. There are four bands, the
luminous matter $\Omega_{\rm lum}$; the baryon content $\Omega_{\rm B}$,
from BBN; the galactic halo component $\Omega_{\rm halo}$, and the
dynamical mass from clusters, $\Omega_{\rm M}$. From this figure it is
clear that there are in fact {\em three} dark matter problems: The first
one is where are 90\% of the baryons. Between the fraction predicted by
BBN and that seen in stars and diffuse gas there is a huge fraction
which is in the form of dark baryons. They could be in small clumps of
hydrogen that have not started thermonuclear reactions and perhaps
constitute the dark matter of spiral galaxies' halos. Note that although
$\Omega_{\rm B}$ and $\Omega_{\rm halo}$ coincide at $H_0\simeq70$
km/s/Mpc, this could be just a coincidence.  The second problem is what
constitutes 90\% of matter, from BBN baryons to the mass inferred from
cluster dynamics. This is the standard dark matter problem and could be
solved by direct detection of a weakly interacting massive particle in
the laboratory.  And finally, since we know from observations of the
CMB, see Section~4.4, that the universe is flat, what constitutes around
60\% of the energy density, from dynamical mass to critical density,
$\Omega_0=1$? One possibility could be that the universe is dominated by
a diffuse vacuum energy, i.e. a cosmological constant, which only
affects the very large scales.  Alternatively, the theory of gravity
(general relativity) may need to be modified on large scales, e.g. due
to quantum gravity effects. The need to introduce an effective
cosmological constant on large scales is nowadays the only reason why
gravity may need to be modified at the quantum level. Since we still do
not have a quantum theory of gravity, such a proposal is still very
speculative, and most of the approaches simply consider the inclusion of
a cosmological constant as a phenomenological parameter.

\subsubsection{Massive neutrinos}

\begin{figure}[htb]
\begin{center}\vspace*{-1cm}
\includegraphics[height=14cm,width=12cm]{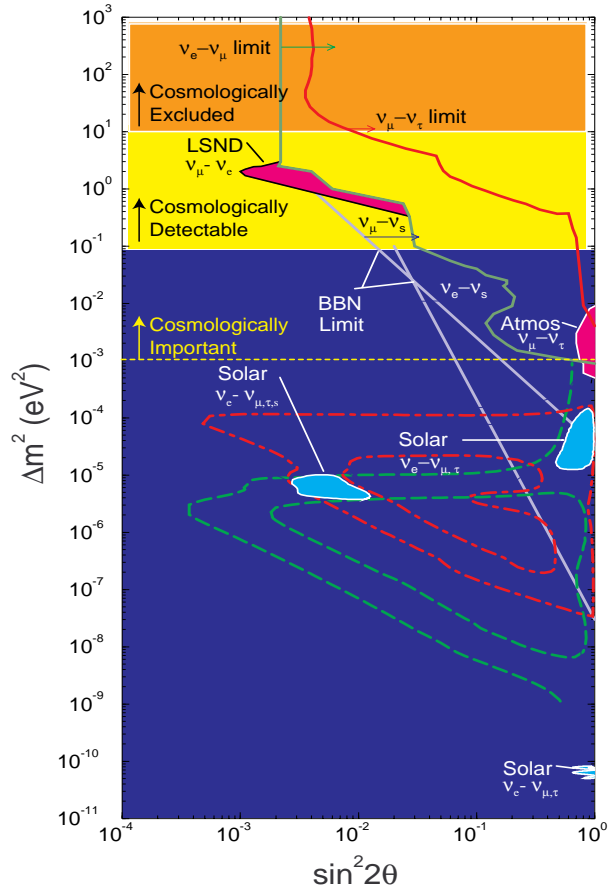} 
\vspace*{-1cm}
\caption{The neutrino parameter space, mixing angle against $\Delta
m^2$, including the results from the different solar and atmospheric
neutrino oscillation experiments. Note the threshold of cosmologically
important masses, cosmologically detectable neutrinos (by CMB and LSS
observations), and cosmologically excluded range of masses. From
Ref.~\cite{HET}.}
\label{fig22}
\end{center}
\end{figure}

One of the `usual suspects' when addressing the problem of dark matter
are neutrinos. They are the only candidates known to exist. If neutrinos
have a mass, could they constitute the missing matter? We know from the
Big Bang theory, see Section~2.2.2, that there is a cosmic neutrino
background at a temperature of approximately 2K. This allows one to
compute the present number density in the form of neutrinos, which turns
out to be, for massless neutrinos, $n_\nu(T_\nu) = {3\over11}\,
n_\gamma(T_\gamma) = 112\ {\rm cm}^{-3}$, per species of neutrino. If
neutrinos have mass, as recent experiments seem to suggest, see
Fig.~\ref{fig22}, the cosmic energy density in massive neutrinos would be
$\rho_\nu = \sum n_\nu m_\nu = {3\over11}\,n_\gamma\,\sum m_\nu$, and
therefore its contribution today,
\begin{equation}\label{OmegaNeutrinos}
\Omega_\nu h^2 = {\sum m_\nu\over94\ {\rm eV}}\,.
\end{equation}
The discussion in the previous Sections suggest that $\Omega_{\rm M}
\leq 0.4$, and thus, for any of the three families of neutrinos, $m_\nu
\leq 40$ eV. Note that this limit improves by six orders of magnitude
the present bound on the tau-neutrino mass~\cite{PDG}. Supposing that
the missing mass in non-baryonic cold dark matter arises from a single
particle dark matter (PDM) component, its contribution to the critical
density is bounded by \ $0.05 \leq \Omega_{\rm PDM}h^2 \leq 0.4$, see
Fig.~\ref{fig21}.

I will now go through the various logical arguments that exclude
neutrinos as the {\em dominant} component of the missing dark matter in
the universe. Is it possible that neutrinos with a mass \ $4\ {\rm eV}
\leq m_\nu \leq 40$ eV be the non-baryonic PDM component? For instance,
could massive neutrinos constitute the dark matter halos of galaxies?
For neutrinos to be gravitationally bound to galaxies it is necessary
that their velocity be less that the escape velocity $v_{\rm esc}$, and
thus their maximum momentum is $p_{\rm max} = m_\nu\,v_{\rm esc}$. How
many neutrinos can be packed in the halo of a galaxy? Due to the Pauli
exclusion principle, the maximum number density is given by that of a
completely degenerate Fermi gas with momentum $p_{\rm F} = p_{\rm max}$,
i.e. $n_{\rm max} = p_{\rm max}^3/3\pi^2$. Therefore, the maximum local
density in dark matter neutrinos is $\rho_{\rm max} = n_{\rm max}m_\nu =
m_\nu^4\, v_{\rm esc}^3/3\pi^2$, which must be greater than the typical
halo density $\rho_{\rm halo} = 0.3$ GeV\,cm$^{-3}$. For a typical
spiral galaxy, this constraint, known as the Tremaine-Gunn limit, gives
$m_\nu \geq 40$ eV, see Ref.~\cite{TG}. However, this mass, even for a
single species, say the tau-neutrino, gives a value for $\Omega_\nu
h^2=0.5$, which is far too high for structure formation. Neutrinos of
such a low mass would constitute a relativistic hot dark matter
component, which would wash-out structure below the supercluster scale,
against evidence from present observations, see Fig.~\ref{fig22}.
Furthermore, applying the same phase-space argument to the neutrinos as
dark matter in the halo of dwarf galaxies gives $m_\nu \geq 100$ eV,
beyond closure density (\ref{OmegaNeutrinos}). We must conclude that the
simple idea that light neutrinos could constitute the particle dark
matter on all scales is ruled out. They could, however, still play a
role as a sub-dominant hot dark matter component in a flat CDM model. In
that case, a neutrino mass of order 1 eV is not cosmological excluded,
see Fig.~\ref{fig22}.

Another possibility is that neutrinos have a large mass, of order a few
GeV. In that case, their number density at decoupling, see Section
2.2.2, is suppressed by a Boltzmann factor, $\sim \exp(-m_\nu/T_{\rm
dec})$.  For masses $m_\nu > T_{\rm dec} \simeq 0.8$ MeV, the present
energy density has to be computed as a solution of the corresponding
Boltzmann equation. Apart from a logarithmic correction, one finds
$\Omega_\nu h^2 \simeq 0.1 (10\ {\rm GeV}/m_\nu)^2$ for Majorana
neutrinos and slightly smaller for Dirac neutrinos. In either case,
neutrinos could be the dark matter only if their mass was a few
GeV. Laboratory limits for $\nu_\tau$ of around 18 MeV~\cite{PDG}, and
much more stringent ones for $\nu_\mu$ and $\nu_e$, exclude the known
light neutrinos. However, there is always the possibility of a fourth
unknown heavy and stable (perhaps sterile) neutrino. If it couples to
the Z boson and has a mass below 45 GeV for Dirac neutrinos (39.5 GeV
for Majorana neutrinos), then it is ruled out by measurements at LEP of
the invisible width of the Z. There are two logical alternatives, either
it is a sterile neutrino (it does not couple to the Z), or it does
couple but has a larger mass. In the case of a Majorana neutrino (its
own antiparticle), their abundance, for this mass range, is too small
for being cosmologically relevant, $\Omega_\nu h^2 \leq 0.005$. If it
were a Dirac neutrino there could be a lepton asymmetry, which may
provide a higher abundance (similar to the case of
baryogenesis). However, neutrinos scatter on nucleons via the weak
axial-vector current (spin-dependent) interaction. For the small
momentum transfers imparted by galactic WIMPs, such collisions are
essentially coherent over an entire nucleus, leading to an enhancement
of the effective cross section. The relatively large detection rate in
this case allowes one to exclude fourth-generation Dirac neutrinos for
the galactic dark matter~\cite{WIMP}. Anyway, it would be very
implausible to have such a massive neutrino today, since it would have
to be stable, with a life-time greater than the age of the universe, and
there is no theoretical reason to expect a massive sterile neutrino that
does not oscillate into the other neutrinos.

Of course, the definitive test to the possible contribution of neutrinos
to the overall density of the universe would be to measure {\em
directly} their mass in laboratory experiments.\footnote{For a review of
Neutrinos, see Bilenky's contribution to these
Proceedings~\cite{Bilenky}.} There are at present two types of
experiments: neutrino oscillation experiments, which measure only {\em
differences} in squared masses, and direct mass-searches experiments,
like the tritium $\beta$-spectrum and the neutrinoless double-$\beta$
decay experiments, which measure directly the mass of the electron
neutrino and give a bound $m_{\nu_e} \lsim $ 2 eV. Neutrinos with such
a mass could very well constitute the HDM component of the universe,
$\Omega_{\rm HDM} \lsim 0.15$. The oscillation experiments give a
variety of possibilities for $\Delta\, m_\nu^2 = 0.3 - 3\ {\rm eV}^2$
from LSND (not yet confirmed), to the atmospheric neutrino oscillations
from SuperKamiokande ($\Delta\, m_\nu^2 \simeq 3\times 10^{-3}\ {\rm
eV}^2$) and the solar neutrino oscillations ($\Delta\, m_\nu^2 \simeq
10^{-5}\ {\rm eV}^2$). Only the first two possibilities would be
cosmologically relevant, see Fig.~\ref{fig22}.

\subsubsection{Weakly Interacting Massive Particles}

Unless we drastically change the theory of gravity on large scales,
baryons cannot make up the bulk of the dark matter. Massive neutrinos
are the only alternative among the known particles, but they are
essentially ruled out as a universal dark matter candidate, even if they
may play a subdominant role as a hot dark matter component. There
remains the mystery of what is the physical nature of the dominant
cold dark matter component.

Something like a heavy stable neutrino, a generic Weakly Interacting
Massive Particle (WIMP), could be a reasonable candidate because its
present abundance could fall within the expected range,
\begin{equation}\label{OmegaPDM}
\Omega_{\rm PDM} h^2 \sim {G^{3/2}T_0^3h^2\over H_0^2\langle
\sigma_{\rm ann}v_{\rm rel}\rangle} =
{3\times10^{-27}\ {\rm cm}^3{\rm s}^{-1}\over \langle
\sigma_{\rm ann}v_{\rm rel}\rangle}\,.
\end{equation}
Here $v_{\rm rel}$ is the relative velocity of the two incoming dark
matter particles and the brackets $\langle\dots\rangle$ denote a thermal
ave\-rage at the freeze-out temperature, $T_{\rm f}\simeq m_{\rm
PDM}/20$, when the dark matter particles go out of equilibrium with
radiation. The value of $\langle\sigma_{\rm ann}v_{\rm rel}\rangle$
needed for $\Omega_{\rm PDM} \approx 1$ is remarkably close to what one
would expect for a WIMP with a mass $m_{\rm PDM}=100$ GeV, \
$\langle\sigma_{\rm ann}v_{\rm rel} \rangle\sim \alpha^2/8\pi\,m_{\rm
PDM}\sim 3\times10^{-27}\ {\rm cm}^3 {\rm s}^{-1}$. We still do not know
whether this is just a coincidence or an important hint on the nature of
dark matter.

\begin{figure}[htb]
\begin{center}\vspace*{1cm}
\includegraphics[width=10cm]{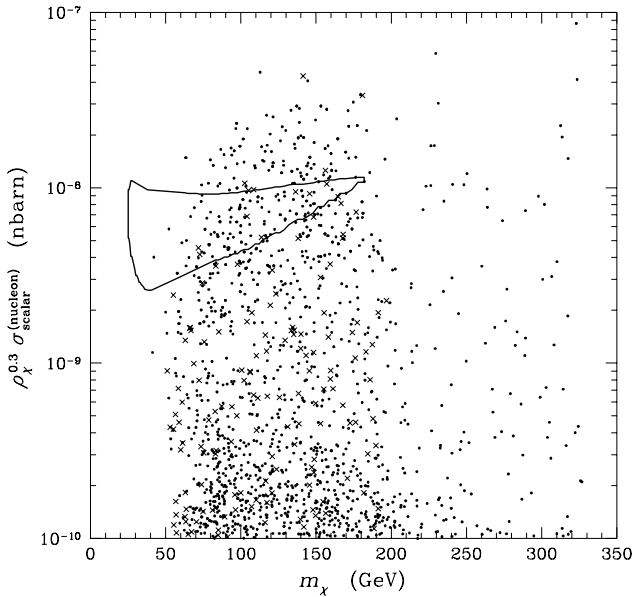}
\vspace*{-2.8cm}
\caption{The maximum likelihood region from the annual-modulation signal
consistent with a neutralino of mass $\ m_\chi = 59^{\ +17}_{\ -14}$ GeV
and a proton cross section of $\ \xi\sigma_p = 7.0 ^{\ +0.4}_{\ -1.2}
\times10^{-6}$ pb, see the text. The scatter plot represents the
theoretical predictions of a generic MSSM. From Ref.~\cite{DAMA}.}
\label{fig23}
\end{center}
\end{figure}

There are a few theoretical candidates for WIMPs, like the neutralino,
coming from supersymme\-tric extensions of the standard model of particle
physics,\footnote{For a review of Supersymmetry (SUSY), see Carena's
contribution to these Proceedings.} but at present there is no empirical
evidence that such extensions are indeed realized in nature. In fact,
the non-observation of supersymmetric particles at current accelerators
places stringent limits on the neutralino mass and interaction cross
section~\cite{neutralino}.

If WIMPs constitute the dominant component of the halo of our galaxy, it
is expected that some may cross the Earth at a reasonable rate to be
detected. The direct experimental search for them rely on elastic WIMP
collisions with the nuclei of a suitable target. Dark matter WIMPs move
at a typical galactic virial velocity of around $200-300$ km/s,
depending on the model. If their mass is in the range $10-100$ GeV, the
recoil energy of the nuclei in the elastic collision would be of order
10 keV. Therefore, one should be able to identify such energy
depositions in a macroscopic sample of the target. There are at present
three different methods: First, one could search for scintillation light
in NaI crystals or in liquid xenon; second, search for an ionization
signal in a semiconductor, typically a very pure germanium crystal; and
third, use a cryogenic detector at 10 mK and search for a measurable
temperature increase of the sample. The main problem with such a type of
experiment is the low expected signal rate, with a typical number below
1 event/kg/day. To reduce natural radioactive contamination one must
use extremely pure substances, and to reduce the background caused by
cosmic rays requires that these experiments be located deeply
underground.

\begin{figure}[htb]
\begin{center}\vspace*{-8mm}
\includegraphics[width=12cm]{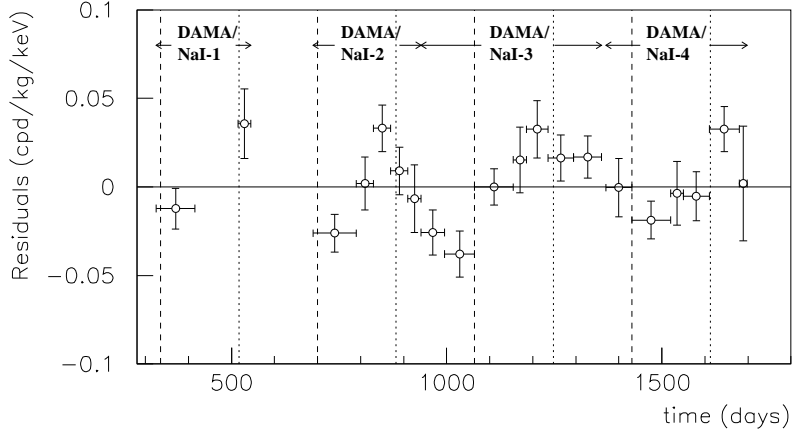}
\vspace*{-1mm}
\caption{The DAMA experiment sees an annual variation, of order 7\%, in
the WIMP flux due to the Earth's motion around the Sun. The model
independent residual rate in the lowest ($2-6$ keV) cumulative energy
interval (in counts per day/kg/keV) is shown as a function of time since
1 January of the first year of data taking. The expected behaviour of a
WIMP signal is a cosine function with a minimum (maximum) roughly at the 
dashed (dotted) vertical lines. From Ref.~\cite{DAMA}.}
\label{fig24}
\end{center}
\end{figure}

The best limits on WIMP scattering cross sections come from some
germanium experiments~\cite{GE}, as well as from the NaI scintillation
detectors of the UK dark matter collaboration (UKDMC) in the Boulby salt
mine in England~\cite{UKDMC}, and the DAMA experiment in the Gran Sasso
laboratory in Italy~\cite{DAMA}. Current experiments already touch the
parameter space expected from supersymmetric particles, see
Fig.~\ref{fig23}, and therefore there is a chance that they actually
discover the nature of the missing dark matter. The problem, of course,
is to attribute a tentative signal unambiguously to galactic WIMPs
rather than to some unidentified radioactive background.

\begin{figure}[htb]
\begin{center}
\includegraphics[width=10cm]{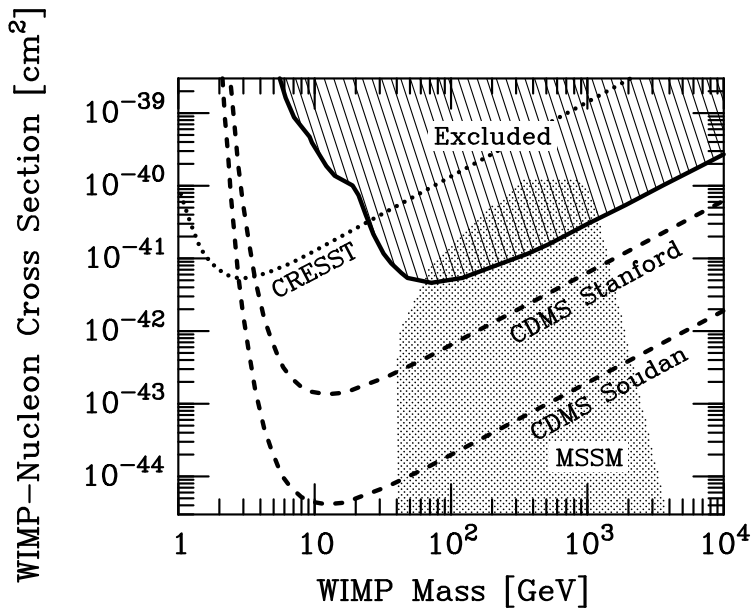}
\caption{Exclusion range for the spin-independent WIMP scattering cross
section per nucleon from the NaI experiments and the Ge detectors. Also
shown is the range of expected counting rates for neutralinos in the
MSSM. The search goals for the upcoming large-scale cryogenic detectors
CRESST and CDMS are also shown. From Ref.~\cite{Raffelt}.}
\label{fig25}
\end{center}
\end{figure}

One specific signature is the annual modulation which arises as the
Earth moves around the Sun.\footnote{The time scale of the Sun's orbit
around the center of the galaxy is too large to be relevant in the
analysis.} Therefore, the net speed of the Earth relative to the
galactic dark matter halo varies, causing a modulation of the expected
counting rate. The DAMA/NaI experiment has actually reported such a
modulation signal, see Fig.~\ref{fig24}, from the combined analysis of
their 4-year data~\cite{DAMA}, which provides a confidence level of
99.6\% for a neutralino mass of $\ m_\chi = 52^{\ +10}_{\ -8}$ GeV and
a proton cross section of $\ \xi\sigma_p = 7.2 ^{\ +0.4}_{\ -0.9}
\times10^{-6}$ pb, where $\xi = \rho_\chi/0.3$ GeV\,cm$^{-3}$ is
the local neutralino energy density in units of the galactic halo
density. There has been no confirmation yet of this result from other
dark matter search groups, but hopefully in the near future we will have
much better sensitivity at low masses from the Cryogenic Rare Event
Search with Superconducting Thermometers (CRESST) experiment at Gran
Sasso as well as at weaker cross sections from the CDMS experiment at
Stanford and the Soudan mine, see Fig.~\ref{fig25}. The CRESST
experiment~\cite{CRESST} uses sapphire crystals as targets and a new
method to simultaneously measure the phonons and the scintillating light
from particle interactions inside the crystal, which allows excellent
background discrimination.  Very recently there has been the interesting
proposal of a completely new method based on a Superheated Droplet
Detector (SDD), which claims to have already a similar sensitivity as
the more standard methods described above, see Ref.~\cite{SDD}.

There exist other {\em indirect} methods to search for galactic
WIMPs~\cite{Griest}. Such particles could self-annihilate at a certain
rate in the galactic halo, producing a potentially detectable background
of high energy photons or antiprotons. The absence of such a background
in both gamma ray satellites and the Alpha Matter
Spectrometer~\cite{AMS} imposes bounds on their density in the
halo. Alternatively, WIMPs traversing the solar system may interact with
the matter that makes up the Earth or the Sun so that a small fraction
of them will lose energy and be trapped in their cores, building up over
the age of the universe. Their annihilation in the core would thus
produce high energy neutrinos from the center of the Earth or from the
Sun which are detectable by neutrino telescopes. In fact,
SuperKamiokande already covers a large part of SUSY parameter space. In
other words, neutrino telescopes are already competitive with direct
search experiments. In particular, the AMANDA experiment at the South
Pole~\cite{AMANDA}, which is expected to have $10^3$ Cherenkov detectors
2.3 km deep in very clear ice, over a volume $\sim 1$ km$^3$, is
competitive with the best direct searches proposed. The advantages of
AMANDA are also directional, since the arrays of Cherenkov detectors
will allow one to reconstruct the neutrino trajectory and thus its
source, whether it comes from the Earth or the Sun.

\subsection{The cosmological constant $\Omega_\Lambda$}

A cosmological constant is a term in the Einstein equations, see Eq.
(\ref{EinsteinEquations}), that corresponds to the energy density of the
vacuum of quantum field theories, $\Lambda \equiv 8\pi G\rho_v$, see
Ref.~\cite{WeinbergCosmo}. These theories predict a value of order
$\rho_v \sim M_{\rm P}^4 \simeq 5\times 10^{93}$ g/cm$^3$, which is
about 123 orders of magnitude larger than the critical density
(\ref{CriticalDensity}). Such a discrepancy is one of the biggest
problems of theoretical physics \cite{CPT}. It has always been assumed
that quantum gravity effects, via some as yet unknown symmetry, would
exactly cancel the cosmological constant, but this remains a downright
speculation. Moreover, one of the difficulties with a non-zero value for
$\Lambda$ is that it appears coincidental that we are now living at a
special epoch when the cosmological constant starts to dominate the
dynamics of the universe, and that it will do so forever after, see
Section 2.1.2 and Eq.~(\ref{OmegaM}). Nevertheless, ever since Einstein
introduced it in 1917, this ethereal constant has been invoked several
times in history to explain a number of apparent crises, always to
disappear under further scrutiny~\cite{Freedman}.

\begin{figure}[htb]
\begin{center}\vspace*{-5mm}
\includegraphics[width=9cm]{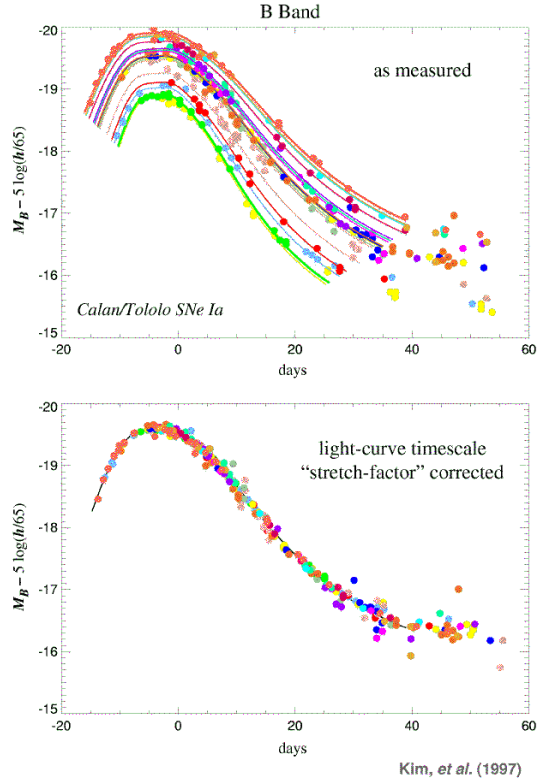}
\vspace*{-5mm}
\caption{The Type Ia supernovae observed nearby show a relationship
between their absolute luminosity and the timescale of their light
curve: the brighter supernovae are slower and the fainter ones are
faster. A simple linear relation between the absolute magnitude and a
``stretch factor'' multiplying the light curve timescale fits the data
quite well. From Ref.~\cite{SCP}.}
\label{fig26}
\end{center}
\end{figure}

In spite of the theoretical prejudice towards $\Lambda=0$, there are new
observational arguments for a non-zero value. The most compelling ones
are recent evidence that we live in a flat universe, from observations
of CMB anisotropies, together with strong indications of a low mass
density universe ($\Omega_{\rm M}<1$), from the large scale distribution
of galaxies, clusters and voids, that indicate that some kind of dark
energy must make up the rest of the energy density up to critical, i.e.
$\Omega_\Lambda = 1 - \Omega_{\rm M}$. In addition, the discrepancy
between the ages of globular clusters and the expansion age of the
universe may be cleanly resolved with $\Lambda\neq0$. Finally, there is
growing evidence for an accelerating universe from observations of
distant supernovae. I will now discuss the different arguments one by
one.

The only known way to reconcile a low mass density with a flat universe
is if an additional ``dark'' energy dominates the universe today. It
would have to resist gravitational collapse, otherwise it would have
been detected already as part of the energy in the halos of galaxies.
However, if most of the energy of the universe resists gravitational
collapse, it is impossible for structure in the universe to grow. This
dilemma can be resolved if the hypothetical dark energy was negligible
in the past and only recently became the dominant component. According
to general relativity, this requires that the dark energy have {\rm
negative} pressure, since the ratio of dark energy to matter density
goes like $a(t)^{-3p/\rho}$. This argument~\cite{OS} would rule out
almost all of the usual suspects, such as cold dark matter, neutrinos,
radiation, and kinetic energy, since they all have zero or positive
pressure. Thus, we expect something like a cosmological constant, with
negative pressure, $p\approx-\rho$, to account for the missing energy. 

This negative pressure would help accelerate the universe and reconcile
the expansion age of the universe with the ages of stars in globular
clusters, see Fig.~\ref{fig11}, where $t_0H_0$ is shown as a function of
$\Omega_{\rm M}$, in a flat universe, $\Omega_\Lambda = 1 - \Omega_{\rm
M}$, and an open one, $\Omega_\Lambda = 0$. For the present age of the
universe of $t_0 = 13\pm 1$ Gyr, and the measured rate of expansion,
$H_0 = 70\pm7$ km/s/Mpc, one finds $t_0H_0=0.93\pm0.12$ (adding errors
in quadrature), which corresponds to $\Omega_{\rm M}= 0.05^{\ +0.24}_{\
-0.10}$ for an open universe, see Fig.~\ref{fig11}, marginally
consistent with observations of large scale structure. On the other
hand, for a flat universe with a cosmological constant,
$t_0H_0=0.93\pm0.12$ corresponds to $\Omega_{\rm M}= 0.34^{\ +0.20}_{\
-0.12}$, which is perfectly compatible with recent observations. These
suggest that we probably live in a flat universe that is accelerating,
dominated today by a vacuum energy density.

\begin{figure}[htb]
\vspace*{-5mm}
\begin{center}
\includegraphics[width=10cm]{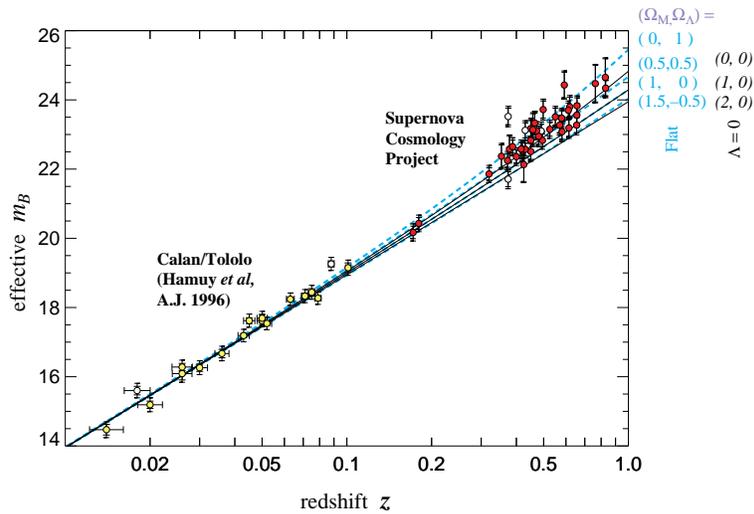}
\end{center}
\vspace*{-5mm}
\caption{Hubble diagram for the high redshift supernovae found by the SN
Cosmology Project.  From Ref.~\cite{SCP}. A similar diagram is found by
the High Redshift Supernova Project~\cite{HRS}. Both groups
conclude that distant supernovae are fainter than expected, and
this could be due to an accelerating universe.}
\label{fig27}
\end{figure}

This conclusions have been supported by growingly robust observational
evidence from distant supernovae. In their quest for the cosmological
parameters, astronomers look for distant astrophysical objects that can
serve as standard candles to determine the distance to the object from
their observed apparent luminosity. A candidate that has recently been
exploited with great success is a certain type of supernova explosions
at large redshifts, called SN of type Ia. These are white dwarf stars at
the end of their life cycle that accrete matter from a companion until
they become unstable and violently explode in a natural thermonuclear
explosion that out-shines their progenitor galaxy. The intensity of the
distant flash varies in time, it takes about three weeks to reach its
maximum brightness and then it declines over a period of months.
Although the maximum luminosity varies from one supernova to another,
depending on their original mass, their environment, etc., there is a
pattern: brighter explosions last longer than fainter ones. By studying
the characteristic light curves, see Fig.~\ref{fig26}, of a reasonably
large statistical sample, cosmologists from two competing groups, the
Supernova Cosmology Project~\cite{SCP} and the High-redshift Supernova
Project~\cite{HRS}, are confident that they can use this type of
supernova as a standard candle. Since the light coming from some of
these rare explosions has travelled for a large fraction of the size of
the universe, one expects to be able to infer from their distribution
the spatial curvature and the rate of expansion of the universe.

\begin{figure}[htb]
\vspace*{-2.5cm}
\hspace*{.3cm}
\includegraphics[height=9cm,width=7.2cm]{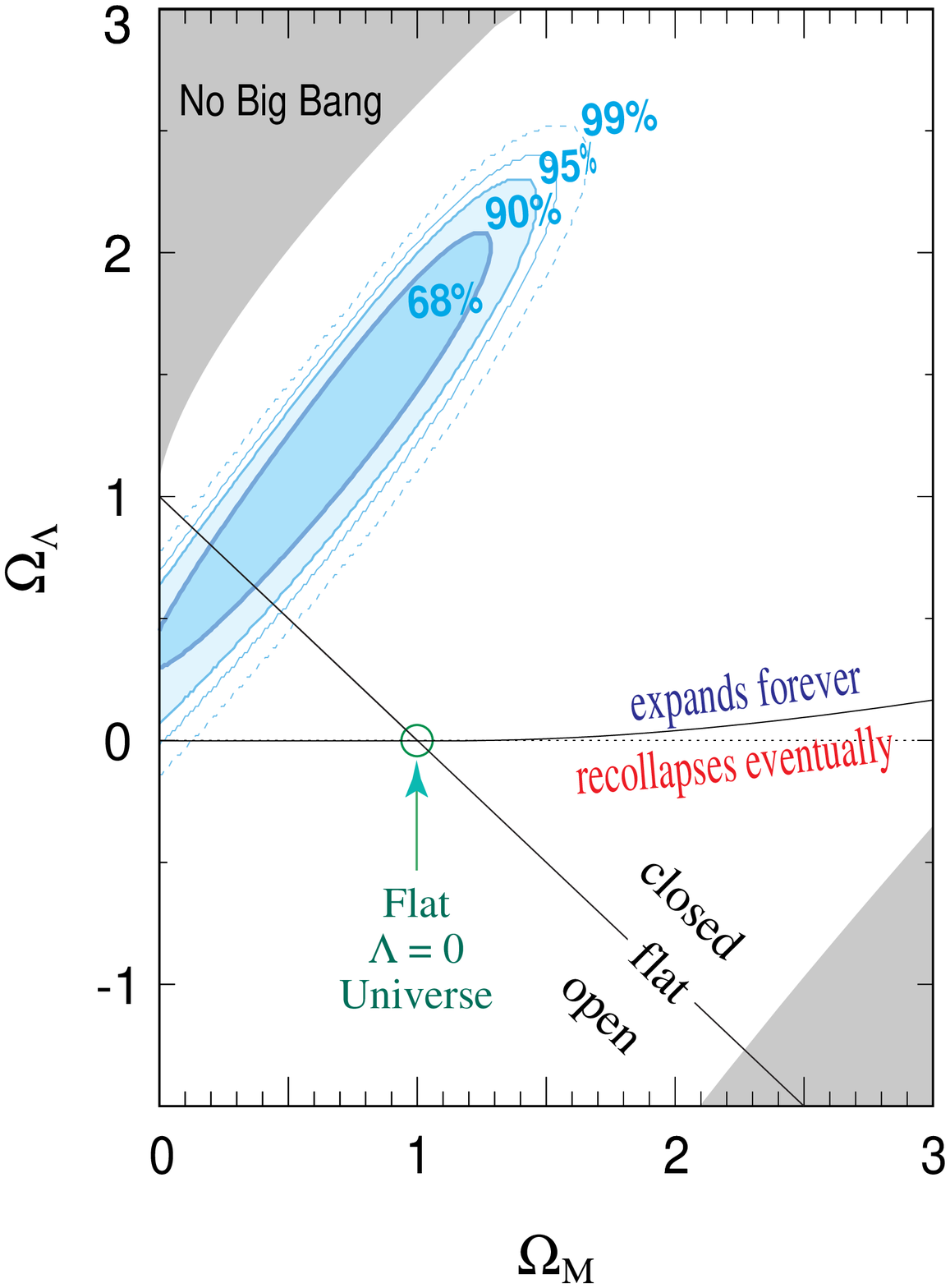}
\vspace*{2cm}
\includegraphics[height=10.1cm,width=7.7cm]{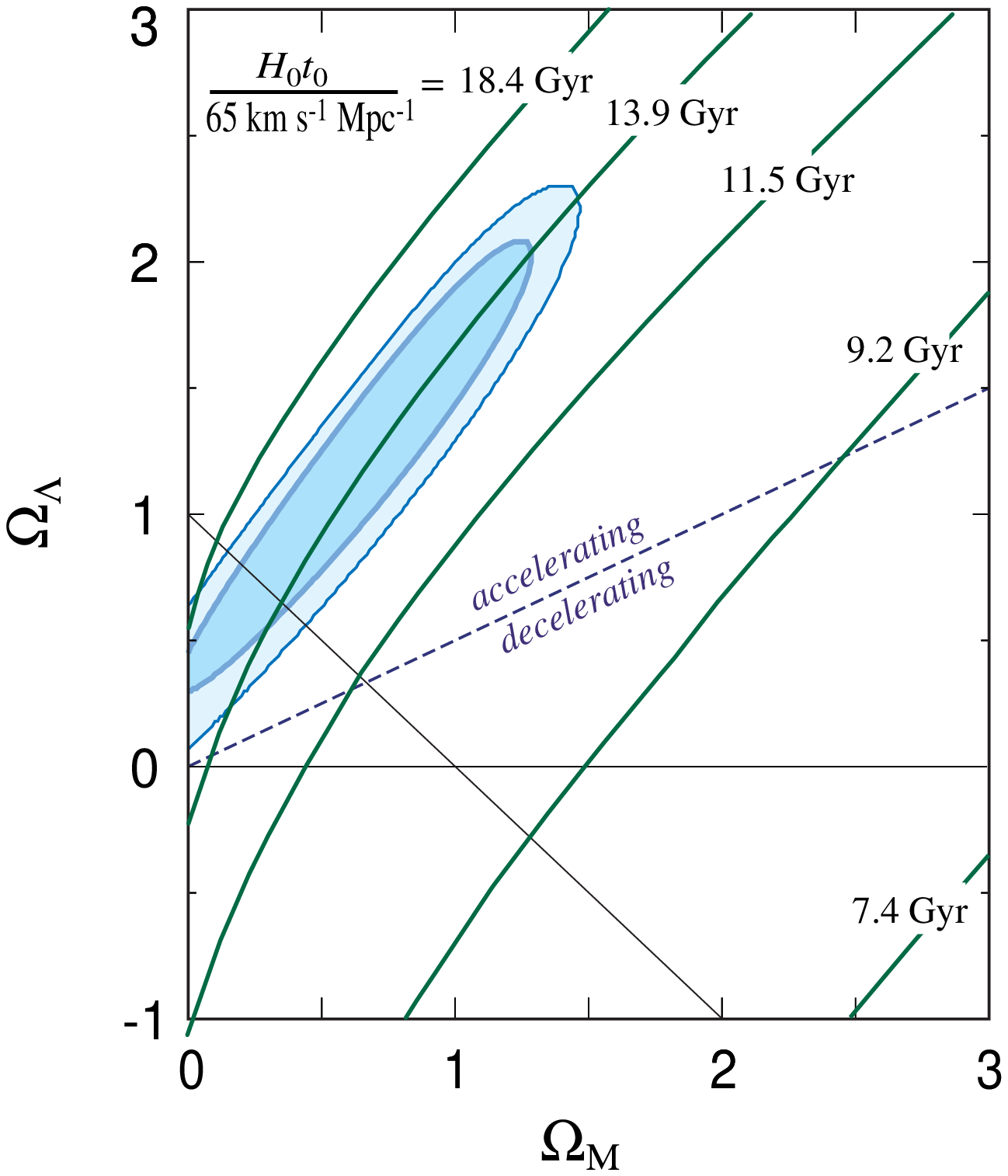}
\vspace*{-2.5cm}
\caption{The left figure shows the best-fit confidence regions (68\% --
99\% c.l.) in the ($\Omega_{\rm M}, \,\Omega_\Lambda$) plane, for the
high redshift supernovae results. The systematic uncertainty is not
shown, and would shift the ellipses vertically. Present observations
disfavour the Eisntein-de Sitter model (circle) by several standard
deviations. The upper-left shaded region represents ``bouncing
universe'' cosmologies with no Big Bang in the past. The lower-right
shaded region corresponds to a universe that is younger than the oldest
heavy elements, for any value of $h\geq0.5$. The right figure shows the
isochrones of constant $H_0t_0$, the age of the universe in units of
the Hubble time, $H_0^{-1}$, with the best-fit 68\% and 90\% confidence
regions in the ($\Omega_{\rm M}, \,\Omega_\Lambda$) plane. From
Ref.~\cite{SCP}.}
\label{fig28}
\end{figure}

One of the surprises revealed by these observations is that high
redshift type Ia supernovae appear fainter than expected for either an
open ($\Omega_{\rm M}<1$) or a flat ($\Omega_{\rm M}=1$) universe, see
Fig.~\ref{fig27}. In fact, the universe appears to be accelerating
instead of decelerating, as was expected from the general attraction of
matter, see Eq.~(\ref{DecelerationParameter}); something seems to be
acting as a repulsive force on very large scales. The most natural
explanation for this is the presence of a cosmological constant, a
diffuse vacuum energy that permeates all space and, as explained above,
gives the universe an acceleration that tends to separate
gravitationally bound systems from each other. The best-fit results from
the Supernova Cosmology Project give a linear combination \
$0.8\Omega_{\rm M} - 0.6\Omega_\Lambda = - 0.2 \pm 0.1$ \ $(1\sigma)$,
and, for a flat universe ($\Omega_{\rm M} + \Omega_\Lambda = 1$), the
best-fit values for the combined analysis of both groups~\cite{SCP,HRS},
are
\begin{eqnarray}\label{OmegaMatter}
\Omega_{\rm M}^{\rm flat} &=& 0.28^{\ +0.09}_{\ -0.08} 
\hspace{2mm} (1\sigma\ {\rm statistical}) 
\ ^{\ +0.05}_{\ -0.04} \ ({\rm identified\ systematics})\,, \\
\Omega_\Lambda^{\rm flat} &=& 0.72^{\ +0.08}_{\ -0.09}
\hspace{2mm} (1\sigma\ {\rm statistical}) 
\ ^{\ +0.04}_{\ -0.05} \ ({\rm identified\ systematics})\,.
\label{OmegaLambda}
\end{eqnarray}

However, one may think that it is still premature to conclude that the
universe is indeed accelerating, because of possibly large systematic
errors inherent to most cosmological measurements, and in particular to
observations of supernovae at large redshifts. There has been attempts
to find crucial systematic effects like evolution, chemical composition
dependence, reddening by dust, etc. in the supernovae observations that
would invalidate the claims, but none of them are now considered as a
serious threat. Perhaps the most critical one today seems to be sampling
effects, since the luminosities of the high-redshift supernovae
($z\sim0.5-1.0$) are all measured relative to the same set of local
supernovae ($z<0.3$). Hence, absolute calibrations, completeness levels,
and any other systematic effects related to both data sets are
critical. For instance, the intense efforts to search for high-redshift
objects have led to the peculiar situation where the nearby sample,
which is used for calibration, is now smaller than the distant
one. Further searches, already underway, for increasing the nearby
supernovae sample will provide an important check.

Moreover, there are bounds on a cosmological constant that come from the
statistics of gravitational lensing, with two different methods.
Gravitational lensing can be due to various accumulations of matter
along the line of sight to the distant light sources. The first method
uses the abundance of multiply imaged sources like quasars, lensed by
intervening galaxies. The probability of finding a lensed image is
directly proportional to the number of galaxies (lenses) along the path
and thus to the distance to the source. This distance, for fixed $H_0$,
increases dramatically for a large value of the cosmological constant:
the age of the universe and the distance to the galaxy become large for
$\Omega_\Lambda\neq0$ because the universe has been expanding for a
longer time; therefore, more lenses are predicted for $\Omega_\Lambda
>0$. Using this method, an upper limit of
\begin{equation}\label{OmegaLensing}
\Omega_\Lambda < 0.75 \hspace{1cm} (95\%\ {\rm c.l.})
\end{equation}
has recently been obtained~\cite{Kochanek}, marginally consistent with
the supernovae results, but there are caveats to this powerful method
due to uncertainties in the number density and lensing cross section of
the lensing galaxies as well as the distant quasars. A second method is
lensing by massive clusters of galaxies, which produces widely separated
lensed images of quasars and distorted images of background galaxies.
The observed statistics, when compared with numerical simulations, rule
out the $\Omega_{\rm M}=1$ models and set an upper bound on the
cosmological constant, $\Omega_\Lambda < 0.7$, see
Ref.~\cite{Bartelmann}. However, this limit is very sensitive to the
resolution of the numerical simulations, which are currently improving.

\subsection{The spatial curvature $\Omega_K$}

As we will discuss in detail in Section~4.4, observations of the
two-point correlation function of temperature anisotropies in the
microwave background provide a crucial test for the spatial curvature
of the universe. From those observations one can tell whether the
photons that left the last scattering surface, at redshift $z=1100$,
have travelled in straight lines, like in a flat universe, or in curved
paths, like in an open one. Very recent observations made by the balloon
experiment BOOMERANG suggest that the universe is indeed spatially flat
($\Omega_K=0$) with about 10\% accuracy~\cite{Boomerang},
\begin{equation}\label{OmegaCurvature}
\Omega_0 = \Omega_{\rm M} + \Omega_\Lambda = 1.0 \pm 0.1
\hspace{1cm} (95\% \ {\rm c.l.})
\end{equation}
These measuremnts are bound to be improved in the near future, by both
balloon experiments and by the Microwave Anisotropy Probe (MAP)
satellite, to be launched by NASA at the end of year 2000~\cite{MAP}. 
Furthermore, with the launch in 2007 of Planck satellite~\cite{Planck}
we will be able to determine $\Omega_0$ with 1\% accuracy.

\subsection{The age of the universe $t_0$}

The universe must be older than the oldest objects it contains. Those
are believed to be the stars in the oldest clusters in the Milky Way,
globular clusters. The most reliable ages come from the application of
theoretical models of stellar evolution to observations of old stars in
globular clusters. For about 30 years, the ages of globular clusters
have remained reasonable stable, at about 15 Gyr~\cite{Vandenberg}.
However, recently these ages have been revised downward~\cite{Krauss}. 

During the 1980s and 1990s, the globular cluster age estimates have
improved as both new observations have been made with CCDs, and since
refinements to stellar evolution models, including opacities,
consideration of mixing, and different chemical abundances have been
incorporated~\cite{Chaboyer}. From the theory side, uncertainties in
globular cluster ages come from uncertainties in convection models,
opacities, and nuclear reaction rates. From the observational side,
uncertainties arise due to corrections for dust and chemical
composition. However, the dominant source of systematic errors in the
globular cluster age is the uncertainty in the cluster distances.
Fortunately, the Hipparcos satellite recently provided geometric
parallax measurements for many nearby old stars with low metallicity,
typical of glubular clusters, thus allowing for a new calibration of the
ages of stars in globular clusters, leading to a downward revision to
$10-13$ Gyr~\cite{Chaboyer}. Moreover, there were very few stars in the
Hipparcos catalog with both small parallax erros and low metal
abundance. Hence, an increase in the sample size could be critical in
reducing the statatistical uncertaintites for the calibration of the
globular cluster ages.  There are already proposed two new parallax
satellites, NASA's Space Interferometry Mission (SIM) and ESA's mission,
called GAIA, that will give 2 or 3 orders of magnitude more accurate
parallaxes than Hipparcos, down to fainter magnitude limits, for several
orders of magnitude more stars. Until larger samples are available,
however, distance errors are likely to be the largest source of
systematic uncertainty to the globular cluster age~\cite{Freedman}.

\begin{figure}[htb]
\begin{center}
\includegraphics[width=8cm]{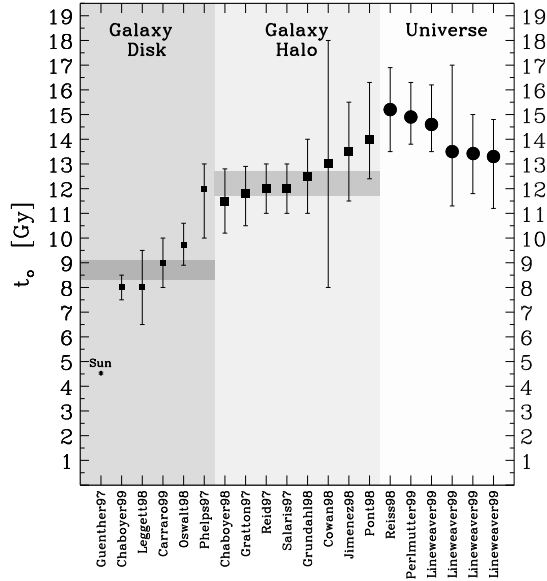}
\end{center}\vspace*{-.5cm}
\caption{The recent estimates of the age of the universe and that of the
oldest objects in our galaxy. The last three points correspond to the
combined analysis of 8 different measurements, for $h=$ 0.64, 0.68 and
7.2, which indicates a relatively weak dependence on $h$. The age of the
Sun is accurately known and is included for reference. Error bars
indicate 1$\sigma$ limits. The averages of the ages of the Galactic Halo
and Disk are shaded in gray. Note that there isn't a single age estimate
more than 2$\sigma$ away from the average. The result $t_0 > t_{\rm
gal}$ is logically inevitable, but the standard EdS model does not
satisfy this unless $h<0.55$. From Ref.~\cite{Charley}.}
\label{fig29}
\end{figure}

The supernovae groups can also determine the age of the universe from
their high redshift observations. Figure~\ref{fig28} shows that the
confidence regions in the $(\Omega_{\rm M}, \Omega_\Lambda)$ plane are
almost parallel to the contours of constant age. For any value of the
Hubble constant less than $H_0 = 70$ km/s/Mpc, the implied age of the
universe is greater than 13 Gyr, allowing enough time for the oldest
stars in globular clusters to evolve~\cite{Chaboyer}. Integrating over
$\Omega_{\rm M}$ and $\Omega_\Lambda$, the best fit value of the age in
Hubble-time units is $H_0t_0=0.93\pm0.06$ or equivalently $t_0 = 14.1
\pm 1.0 \ (0.65\, h^{-1})$ Gyr~\cite{SCP}. The age would be somewhat
larger in a flat universe: $H_0t_0^{\rm flat} =
0.96^{\ +0.09}_{\ -0.07}$ or, equivalently,~\cite{SCP}
\begin{equation}\label{AgeUniverse}
t_0^{\rm flat} = 14.4^{\ +1.4}_{\ -1.1} \hspace{2mm} (0.65\, h^{-1}) 
\ {\rm Gyr}\,.  
\end{equation}

Furthermore, a combination of 8 independent recent measurements: CMB
anisotropies, type Ia SNe, cluster mass-to-light ratios, cluster
abundance evolution, cluster baryon fraction, deuterium-to-hidrogen
ratios in quasar spectra, double-lobed radio sources and the Hubble
constant, can be used to determine the present age of the
universe~\cite{Charley}. The result is shown in Fig.~\ref{fig29},
compared to other recent determinations. The best fit value for the age
of the universe is, according to this analysis, $t_0 = 13.4 \pm 1.6$
Gyr, about a billion years younger than other recent
estimates~\cite{Charley}.

\begin{figure}[htb]
\vspace*{-6cm}
\begin{center}
\includegraphics[width=11cm]{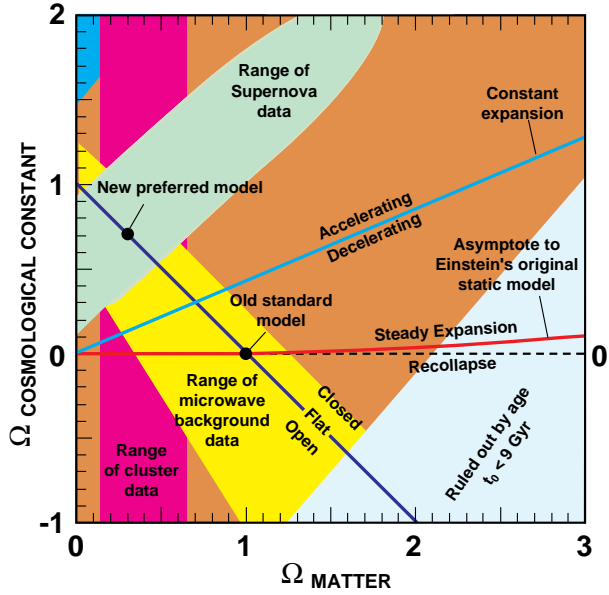}
\end{center}\vspace*{-2cm}
\caption{The concordance region. The sum \ $\Omega_{\rm M} +
\Omega_\Lambda$ gives the total cosmic energy content and determines the
geometry of space-time, whether spatially flat, open or closed. Their
difference, \ $\Omega_{\rm M}/2 - \Omega_\Lambda$, characterizes the
relative strength of expansion and gravity, and determines how the
expansion rate changes with time, whether accelerating or
decelerating. Furthermore, a balance between the two densities
determines the fate of the universe, whether it will expand forever or
recollapse. These three effects have been probed by recent observations,
from large scale structure (cluster data), temperature anisotropies
(microwave background data) and the universe expansion (supernova
data). Surprisingly enough, at present all observations seem to lie
within a naroow region in parameter space. The Einstein-de Sitter model
is no longer the preferred one. The best model today is a flat model
with a third of the energy density in the form of non-relativistic
matter and two thirds in the form of vacuum energy or a cosmological
constant. From Ref.~\cite{Bellido}.}
\label{fig30}
\end{figure}


We can summarize this Section by showing the region in parameter space
where we stand nowadays, thanks to the recent cosmological observations.
We have plotted that region in Fig.~\ref{fig30}. One could also
superimpose the contour lines corresponding to equal $t_0H_0$ lines, as
a cross check.  It is extraordinary that only in the last few months we
have been able to reduce the concordance region to where it stands
today, where all the different observations seem to converge. There are
still many uncertainties, mainly systematic; however, those are quickly
decreasing and becoming predominantly statistical. In the near future,
with precise observations of the anisotropies in the microwave
background temperature and polarization, to be discussed in Section~4.4,
we will be able to reduce those uncertainties to the level of one
percent. This is the reason why cosmologists are so excited and why it
is claimed that we live in the Golden Age of Cosmology.

\section{THE INFLATIONARY PARADIGM}

The hot Big Bang theory is nowadays a very robust edifice, with many
independent observational checks: the expansion of the universe; the
abundance of light elements; the cosmic microwave background; a
predicted age of the universe compatible with the age of the oldest
objects in it, and the formation of structure via gravitational collapse
of initially small inhomogeneities. Today, these observations are
confirmed to within a few percent accuracy, and have helped establish
the hot Big Bang as the preferred model of the universe. All the physics
involved in the above observations is routinely tested in the laboratory
(atomic and nuclear physics experiments) or in the solar system
(general relativity).

However, this theory leaves a range of crucial questions unanswered,
most of which are initial conditions' problems. There is the reasonable
assumption that these cosmological problems will be solved or explained
by {\em new physical principles} at high energies, in the early
universe. This assumption leads to the natural conclusion that accurate
observations of the present state of the universe may shed light onto
processes and physical laws at energies above those reachable by
particle accelerators, present or future. We will see that this is a
very optimistic approach indeed, and that there are many unresolved
issues related to those problems. However, there might be in the near
future reasons to be optimistic.

\subsection{Shortcomings of Big Bang Cosmology}

The Big Bang theory could not explain the origin of matter and structure
in the universe; that is, the origin of the matter--antimatter
asymmetry, without which the universe today would be filled by a uniform
radiation continuosly expanding and cooling, with no traces of matter,
and thus without the possibility to form gravitationally bound systems
like galaxies, stars and planets that could sustain life. Moreover, the
standard Big Bang theory assumes, but cannot explain, the origin of the
extraordinary smoothness and flatness of the universe on the very large
scales seen by the microwave background probes and the largest galaxy
catalogs. It cannot explain the origin of the primordial density
perturbations that gave rise to cosmic structures like galaxies,
clusters and superclusters, via gravitational collapse; the quantity and
nature of the dark matter that we believe holds the universe together;
nor the origin of the Big Bang itself.

A summary~\cite{JGB} of the problems that the Big Bang theory cannot
explain is:

\begin{itemize}

\item The global structure of the universe.

- Why is the universe so close to spatial flatness?

- Why is matter so homogeneously distributed on large scales?

\item The origin of structure in the universe.

- How did the primordial spectrum of density perturbations originate?

\item The origin of matter and radiation.

- Where does all the energy in the universe come from?

- What is the nature of the dark matter in the universe?

- How did the matter-antimatter asymmetry arise?


\item The initial singularity.

- Did the universe have a beginning?

- What is the global structure of the universe beyond our observable 
patch?

\end{itemize}

\noindent
Let me discuss one by one the different issues:

\subsubsection{The Flatness Problem}

The Big Bang theory assumes but cannot explain the extraordinary spatial
flatness of our local patch of the universe. In the general FRW metric
(\ref{FRWmetric}) the parameter $K$ that characterizes spatial curvature
is a free parameter. There is nothing in the theory that determines this
parameter a priori. However, it is directly related, via the Friedmann
equation (\ref{FriedmannEquation}), to the dynamics, and thus the matter
content, of the universe,
\begin{equation}\label{SpatialK}
K = {8\pi G\over3} \rho a^2 - H^2a^2 \ = \ {8\pi G\over3} \rho a^2
\Big({\Omega-1\over\Omega}\Big)\,.
\end{equation}
We can therefore define a new variable,
\begin{equation}\label{NewVariableX}
x \equiv {\Omega-1\over\Omega} = {{\rm const.}\over\rho a^2}\,,
\end{equation}
whose time evolution is given by
\begin{equation}\label{Stability}
x' = {dx\over dN} = (1+3\omega)\,x\,,
\end{equation}
where $N=\ln(a/a_i)$ characterizes the {\em number of $e$-folds} of
universe expansion ($dN=H dt$) and where we have used
Eq.~(\ref{DensityEvolution}) for the time evolution of the total energy,
$\rho a^3$, which only depends on the barotropic ratio $\omega$. It is
clear from Eq.~(\ref{Stability}) that the phase-space diagram $(x,x')$
presents an unstable critical (saddle) point at $x=0$ for $\omega >
-1/3$, i.e. for the radiation ($\omega = 1/3$) and matter ($\omega = 0$)
eras. A small perturbation from $x=0$ will drive the system towards
$x=\pm\infty$. Since we know the universe went through both the radiation
era (because of primordial nucleosynthesis) and the matter era (because
of structure formation), tiny deviations from $\Omega=1$ would have grown
since then, such that today
\begin{equation}\label{x0}
x_0 = {\Omega_0 - 1\over\Omega_0} = x_{\rm in}\,
\Big({T_{\rm in}\over T_{\rm eq}}\Big)^2 (1+z_{\rm eq})\,.
\end{equation}
In order that today's value be in the range $0.1 < \Omega_0 < 1.2$, or
$x_0 \approx {\cal O}(1)$, it is required that at, say, primordial
nucleosynthesis
($T_{_{\rm NS}} \simeq 10^6\, T_{\rm eq}$) \ its value be
\begin{equation}\label{OmegaNucleosynthesis}
\Omega(t_{_{\rm NS}}) = 1 \pm 10^{-15}\,,
\end{equation}
which represents a tremendous finetuning. Perhaps the universe indeed
started with such a peculiar initial condition, but it is
epistemologically more satisfying if we give a fundamental dynamical
reason for the universe to have started so close to spatial flatness.
These arguments were first used by Robert Dicke in the 1960s, much
before inflation. He argued that the most natural initial condition for
the spatial curvature should have been the Planck scale curvature,
$^{(3)}\!R = 6K/l_{\rm P}^2$, where the Planck length is $l_{\rm P} =
(\hbar G/c^3)^{1/2} = 1.62\times 10^{-33}$ cm, that is, 60 orders of
magnitude smaller than the present size of the universe, $a_0 =
1.38\times 10^{28}$ cm. A universe with this immense curvature would
have collapsed within a Planck time, $t_{\rm P} = (\hbar G/c^5)^{1/2} =
5.39\times 10^{-44}$ s, again 60 orders of magnitude smaller than the
present age of the universe, $t_0 = 4.1\times 10^{17}$ s. Therefore, the
flatness problem is also related to the Age Problem, why is it that the
universe is so old and flat when, under ordinary circumstances (based on
the fundamental scale of gravity) it should have lasted only a Planck
time and reached a size of order the Planck length? As we will see,
inflation gives a dynamical reason to such a peculiar initial condition.

\subsubsection{The Homogeneity Problem}

An expanding universe has {\em particle horizons}, that is, spatial
regions beyond which causal communication cannot occur. The horizon
distance can be defined as the maximum distance that light could have
travelled since the origin of the universe~\cite{KT},
\begin{equation}\label{HorizonDistance}
d_{\rm H}(t) \equiv a(t)\int_0^t {dt'\over a(t')} \sim 
H^{-1}(t)\,,
\end{equation}
which is proportional to the Hubble scale.\footnote{For the radiation
era, the horizon distance is equal to the Hubble scale. For the
matter era it is twice the Hubble scale.} For
instance, at the beginning of nucleosynthesis the horizon distance is a
few light-seconds, but grows {\em linearly} with time and by the end of
nucleosynthesis it is a few light-minutes, i.e. a factor 100 larger,
while the scale factor has increased {\em only} a factor of 10. The fact
that the causal horizon increases faster, $d_{\rm H}\sim t$, than the
scale factor, $a\sim t^{1/2}$, implies that at any given time the
universe contains regions within itself that, according to the Big Bang
theory, were {\em never} in causal contact before. For instance, the
number of causally disconnected regions at a given redshift $z$ present 
in our causal volume today, $d_{\rm H}(t_0)\equiv a_0$, is
\begin{equation}\label{DisconnectedRegions}
N_{\rm CD}(z) \sim \left({a(t)\over d_{\rm H}(t)}\right)^3 \simeq
(1+z)^{3/2}\,,
\end{equation}
which, for the time of decoupling, is of order $N_{\rm CD}(z_{\rm dec}) 
\sim 10^5 \gg1$.

\begin{figure}[htb]
\vspace*{-3cm}
\begin{center}
\includegraphics[width=10cm]{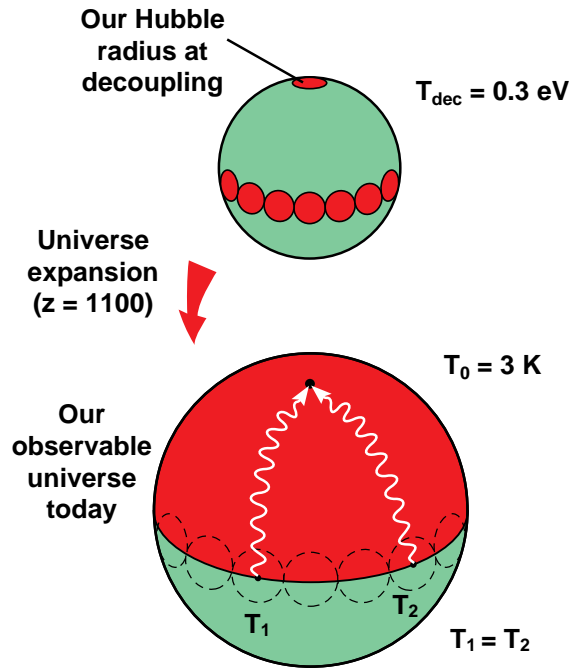}
\vspace*{-2cm}
\caption{Perhaps the most acute problem of the Big Bang theory is
explaining the extraordinary homogeneity and isotropy of the microwave
background, see Fig.~\ref{fig6}. At the time of decoupling, the volume
that gave rise to our present universe contained many causally
disconnected regions (top figure). Today we observe a blackbody spectrum
of photons coming from those regions and they appear to have the same
temperature, $T_1=T_2$, to one part in $10^5$. Why is the universe so
homogeneous?  This constitutes the so-called horizon problem, which is
spectacularly solved by inflation. From Ref.~\cite{Alvaro,Bellido}.}
\label{fig31}
\end{center}
\end{figure}

This phenomenon is particularly acute in the case of the observed
microwave background. Information cannot travel faster than the speed of
light, so the causal region at the time of photon decoupling could not
be larger than $d_{\rm H}(t_{\rm dec}) \sim 3\times 10^5$ light years
across, or about $1^\circ$ projected in the sky today. So why should
regions that are separated by more than $1^\circ$ in the sky today have
exactly the same temperature, to within 10 ppm, when the photons that
come from those two distant regions could not have been in causal
contact when they were emitted? This constitutes the so-called horizon
problem, see Fig.~\ref{fig31}, and was first discussed by Robert Dicke
in the 1970s as a profound inconsistency of the Big Bang theory.

\subsection{Cosmological Inflation}

\begin{figure}[htb]
\vspace*{1cm}
\begin{center}\hspace*{-7cm}
\includegraphics[width=8cm]{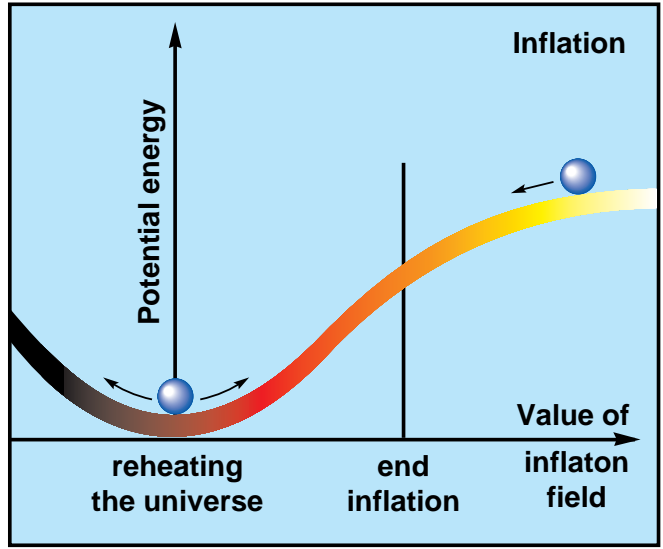}
\vspace*{-6cm}
\caption{The inflaton field can be represented as a ball rolling
down a hill.  During inflation, the energy density is approximately
constant, driving the tremendous expansion of the universe. When the
ball starts to oscillate around the bottom of the hill, inflation ends
and the inflaton energy decays into particles. In certain cases, the
coherent oscillations of the inflaton could generate a resonant
production of particles which soon thermalize, reheating the universe.
From Ref.~\cite{Bellido}.}
\label{fig32}
\end{center}
\end{figure}

In the 1980s, a new paradigm, deeply rooted in fundamental physics, was
put forward by Alan H. Guth~\cite{Guth}, Andrei D. Linde~\cite{Linde}
and others~\cite{Andy,Guthbook,LindeBook}, to address these fundamental
questions. According to the inflationary paradigm, the early universe
went through a period of exponential expansion, driven by the
approximately constant energy density of a scalar field called the
inflaton. In modern physics, elementary particles are represented by
quantum fields, which resemble the familiar electric, magnetic and
gravitational fields. A field is simply a function of space and time
whose quantum oscillations are interpreted as particles. In our case,
the inflaton field has, associated with it, a large potential energy
density, which drives the exponential expansion during inflation, see
Fig.~\ref{fig32}. We know from general relativity that the density of
matter determines the expansion of the universe, but a constant energy
density acts in a very peculiar way: as a repulsive force that makes any
two points in space separate at exponentially large speeds. (This does
not violate the laws of causality because there is no information
carried along in the expansion, it is simply the stretching of
space-time.)

This superluminal expansion is capable of explaining the large scale
homogeneity of our observable universe and, in particular, why the
microwave background looks so isotropic: regions separated today by more
than $1^\circ$ in the sky were, in fact, in causal contact before
inflation, but were stretched to cosmological distances by the
expansion. Any inhomogeneities present before the tremendous expansion
would be washed out. This explains why photons from supposedly causally
disconneted regions have actually the same spectral distribution with
the same temperature, see Fig.~\ref{fig31}. 

\begin{figure}[htb]
\begin{center}\hspace*{-4cm}
\includegraphics[width=8cm]{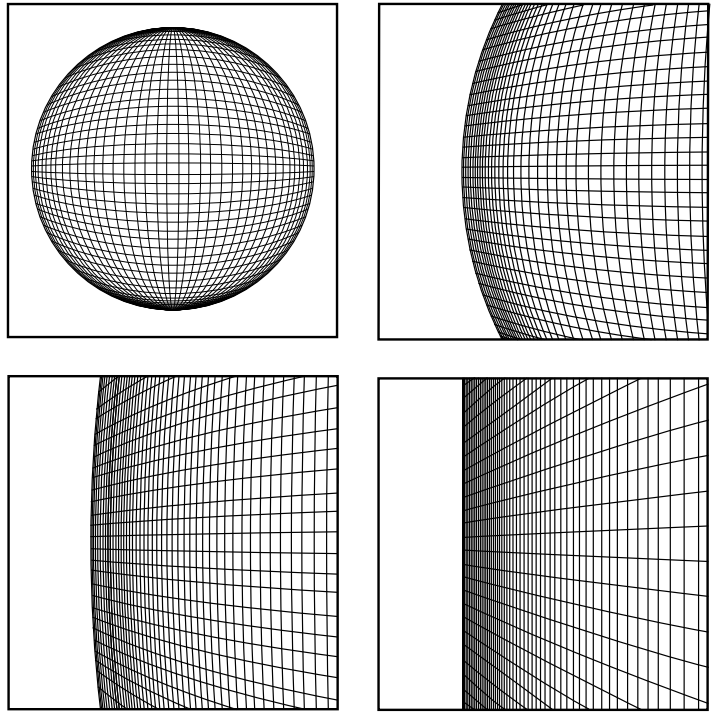}
\vspace*{-5.5cm}
\caption{The exponential expansion during inflation made the
radius of curvature of the universe so large that our observable patch
of the universe today appears essentialy flat, analogous (in three
dimensions) to how the surface of a balloon appears flatter and flatter
as we inflate it to enormous sizes. This is a crucial prediction of
cosmological inflation that will be tested to extraordinary accuracy in
the next few years. From Ref.~\cite{Guthbook,Bellido}.}
\label{fig33}
\end{center}
\end{figure}

Moreover, in the usual Big Bang scenario a flat universe, one in which
the gravitational attraction of matter is exactly balanced by the cosmic
expansion, is unstable under perturbations: a small deviation from
flatness is amplified and soon produces either an empty universe or a
collapsed one. As we discussed above, for the universe to be nearly flat
today, it must have been extremely flat at nucleosynthesis, deviations
not exceeding more than one part in $10^{15}$. This extreme fine tuning
of initial conditions was also solved by the inflationary paradigm, see
Fig.~\ref{fig33}. Thus inflation is an extremely elegant hypothesis that
explains how a region much, much greater that our own observable
universe could have become smooth and flat without recourse to {\em ad
hoc} initial conditions. Furthermore, inflation dilutes away any
``unwanted'' relic species that could have remained from early universe
phase transitions, like monopoles, cosmic strings, etc., which are
predicted in grand unified theories and whose energy density could be so
large that the universe would have become unstable, and collapsed, long
ago. These relics are diluted by the superluminal expansion, which
leaves at most one of these particles per causal horizon, making them
harmless to the subsequent evolution of the universe.

The only thing we know about this peculiar scalar field, the {\em
inflaton}, is that it has a mass and a self-interaction potential
$V(\phi)$ but we ignore everything else, even the scale at which its
dynamics determines the superluminal expansion. In particular, we still
do not know the nature of the inflaton field itself, is it some new {\em
fundamental} scalar field in the electroweak symmetry breaking sector,
or is it just some {\em effective} description of a more fundamental
high energy interaction?  Hopefully, in the near future, experiments in
particle physics might give us a clue to its nature. Inflation had its
original inspiration in the Higgs field, the scalar field supposed to be
responsible for the masses of elementary particles (quarks and leptons)
and the breaking of the electroweak symmetry. Such a field has not been
found yet, and its discovery at the future particle colliders would help
understand one of the truly fundamental problems in physics, the origin
of masses. If the experiments discover something completely new and
unexpected, it would automatically affect the idea of inflation at a
fundamental level.

\subsubsection{Homogeneous scalar field dynamics}

In this subsection I will describe the theoretical basis for the
phenomenon of inflation. Consider a scalar field $\phi$, a singlet under
any given interaction, with an effective potential $V(\phi)$. The
Lagrangian for such a field in a curved background is 
\begin{equation}\label{ScalarLagrangian}
{\cal L}_{\rm inf} = {1\over2}\,g^{\mu\nu}\partial_\mu\phi
\partial_\nu\phi - V(\phi)\,, 
\end{equation}
whose evolution equation in a Friedmann-Robertson-Walker metric
(\ref{FRWmetric}) and for a {\em homogeneous} field $\phi(t)$ is given by
\begin{equation}\label{ScalarEvolution}
\ddot\phi + 3H\dot\phi + V'(\phi)=0\,,
\end{equation}
where $H$ is the rate of expansion, together with the Einstein equations,
\begin{eqnarray}\label{ScalarEinstein}
H^2 &=& {\kappa^2\over3}\Big({1\over2}\,\dot\phi^2 + V(\phi)\Big)\,,\\
\dot H &=& - {\kappa^2\over2}\,\dot\phi^2\,,
\end{eqnarray}
where $\kappa^2\equiv 8\pi G$. The dynamics of inflation can be
described as a perfect fluid (\ref{PerfectFluid}) with a time dependent
pressure and energy density given by
\begin{eqnarray}\label{ScalarFluid}
\rho &=& {1\over2}\,\dot\phi^2 + V(\phi)\,,\\
p &=& {1\over2}\,\dot\phi^2 - V(\phi)\,.
\end{eqnarray}
The field evolution equation (\ref{ScalarEvolution}) can then be written
as the energy conservation equation, 
\begin{equation}\label{ScalarEnergyCons}
\dot\rho + 3H(\rho+p) = 0\,. 
\end{equation}
If the potential energy density of the scalar field dominates the kinetic
energy, $V(\phi) \gg \dot\phi^2$, then we see that
\begin{equation}\label{ConstantRate}
p\simeq -\rho \hspace{5mm} \Rightarrow \hspace{5mm} 
\rho\simeq {\rm const.} \hspace{5mm}
\Rightarrow \hspace{5mm} H(\phi) \simeq {\rm const.} \,,
\end{equation}
which leads to the solution
\begin{equation}\label{ExponentialSolution}
a(t) \sim \exp(Ht) \hspace{5mm} \Rightarrow \hspace{5mm} 
{\ddot a\over a} > 0 \hspace{5mm} {\rm accelerated\ expansion} \,.
\end{equation}
Using the definition of the number of $e$-folds, $N=\ln(a/a_i)$, we see
that the scale factor grows exponentially, $a(N) = a_i \exp(N)$. This
solution of the Einstein equations solves immediately the flatness
problem. Recall that the problem with the radiation and matter eras is
that $\Omega=1$ ($x=0$) is an unstable critical point in phase-space.
However, during inflation, with $p\simeq -\rho \ \Rightarrow \
\omega\simeq -1$, we have that $1+3\omega\geq0$ and therefore $x=0$ is a
stable {\em attractor} of the equations of motion, see
Eq.~(\ref{Stability}). As a consequence, what seemed an {\em ad hoc}
initial condition, becomes a natural {\em prediction} of inflation.
Suppose that during inflation the scale factor increased $N$ $e$-folds,
then
\begin{equation}\label{SolutionFlatness}
x_0 = x_{\rm in}\,e^{-2N}\,
\Big({T_{\rm rh}\over T_{\rm eq}}\Big)^2 (1+z_{\rm eq}) \
\simeq \ e^{-2N}\,10^{56} \leq 1 \hspace{5mm} 
\Rightarrow \hspace{5mm} N \geq 65\,,
\end{equation}
where we have assumed that inflation ended at the scale $V_{\rm end}$,
and the transfer of the inflaton energy density to thermal radiation at
reheating occurred almost instantaneously\footnote{There could be a
small delay in thermalization, due to the intrinsic inefficiency of
reheating, but this does not change significantly the required number of
$e$-folds.} at the temperature \ $T_{\rm rh} \sim V_{\rm end}^{1/4} \sim
10^{15}$ GeV. Note that we can now have initial conditions with a large
uncertainty, $x_{\rm in} \simeq 1$, and still have today $x_0 \simeq 1$,
thanks to the inflationary attractor towards $\Omega=1$.  This can be
understood very easily by realizing that the three curvature evolves
during inflation as
\begin{equation}\label{R3Flatness}
^{(3)}R = {6K\over a^2} = \ ^{(3)}\!R_{\rm in}\,e^{-2N} \hspace{5mm} 
\longrightarrow \hspace{5mm} 0\,, \hspace{5mm} {\rm for} \ \ N\gg 1\,. 
\end{equation}
Therefore, if cosmological inflation lasted over 65 $e$-folds, as most
models predict, then today the universe (or at least our local patch)
should be exactly flat, see Fig.~\ref{fig33}, a prediction that can be
tested with great accuracy in the near future and for which already
seems to be some evidence from observations of the microwave
background~\cite{Boomerang}.

Furthermore, inflation also solves the homogeneity problem in a
spectacular way. First of all, due to the superluminal expansion, 
any inhomogeneity existing prior to inflation will be washed out,
\begin{equation}\label{Homogeneity}
\delta_k \sim \ \left({k\over aH}\right)^2\,
\Phi_k \hspace{2mm} \propto \hspace{2mm} e^{-2N} \hspace{5mm} 
\longrightarrow \hspace{5mm} 0\,, \hspace{5mm} {\rm for} \ \ N\gg 1\,. 
\end{equation}
Moreover, since the scale factor grows exponentially, while the horizon
distance remains essentially constant, $d_H(t) \simeq H^{-1} = $ const.,
any scale within the horizon during inflation will be stretched by the
superluminal expansion to enormous distances, in such a way that at
photon decoupling all the causally disconnected regions that encompass
our present horizon actually come from a single region during inflation,
about 65 $e$-folds before the end. This is the reason why two points
separated more than $1^\circ$ in the sky have the same backbody
temperature, as observed by the COBE satellite: they were actually in
causal contact during inflation. There is at present no other proposal
known that could solve the homogeneity problem without invoquing an
acausal mechanism like inflation.

Finally, any relic particle species (relativistic or not) existing prior
to inflation will be diluted by the expansion,
\begin{eqnarray}\label{RelicSpecies}
\rho_{\rm M} &\propto& a^{-3} \ \sim \ e^{-3N} \hspace{5mm} 
\longrightarrow \hspace{5mm} 0\,, \hspace{5mm} {\rm for} \ \ N\gg 1\,,\\
\rho_{\rm R} &\propto& a^{-4} \ \sim \ e^{-4N} \hspace{5mm} 
\longrightarrow \hspace{5mm} 0\,, \hspace{5mm} {\rm for} \ \ N\gg 1\,.
\end{eqnarray}
Note that the vacuum energy density $\rho_v$ remains constant under the
expansion, and therefore, very soon it is the only energy density
remaining to drive the expansion of the universe.

\subsubsection{The slow-roll approximation}

In order to simplify the evolution equations during inflation, we will
consider the slow-roll approximation (SRA). Suppose that, during
inflation, the scalar field evolves very slowly down its effective
potential, then we can define the slow-roll parameters~\cite{LL93},
\begin{eqnarray}\label{SlowRollParameters}
\epsilon&\equiv&-\,{\dot H\over H^2} \ = \ 
{\kappa^2\over2}\,{\dot\phi^2\over H^2} \ \ll \ 1\,,\\
\delta &\equiv&-\,{\ddot\phi\over H\dot\phi} \ \ll \ 1\,.
\end{eqnarray}
It is easy to see that the condition 
\begin{equation}\label{Inflation}
\epsilon < 1 \hspace{5mm} \Longleftrightarrow \hspace{5mm} 
{\ddot a\over a} > 0
\end{equation}
characterizes inflation: it is all you need for superluminal expansion,
i.e. for the horizon distance to grow more slowly than the scale factor,
in order to solve the homogeneity problem, as well as for the spatial
curvature to decay faster than usual, in order to solve the flatness
problem.

The number of $e$-folds during inflation can be written with the help of
Eq.~(\ref{SlowRollParameters}) as
\begin{equation}\label{Nefolds}
N \ = \ \ln{a_{\rm end}\over a_i} \ = \ \int_{t_i}^{t_e} H dt \ = \
\int_{\phi_i}^{\phi_e} {\kappa d\phi\over\sqrt{2\epsilon(\phi)}} \,,
\end{equation}
which is an exact expression in terms of $\epsilon(\phi)$.

In the limit given by Eqs.~(\ref{SlowRollParameters}), the evolution
equations (\ref{ScalarEvolution}) and (\ref{ScalarEinstein}) become
\begin{eqnarray}\label{SlowRollEvolution}
H^2\Big(1- {\epsilon\over3}\Big) &\simeq& H^2 \ = \
{\kappa^2\over3}\,V(\phi)\,,\\
3H\dot\phi\,\Big(1- {\delta\over3}\Big) &\simeq& 3H\dot\phi \ = \
-\,V'(\phi)\,.
\end{eqnarray}
Note that this corresponds to a reduction of the dimensionality of 
phase-space from two to one dimensions, $H(\phi,\dot\phi) \rightarrow
H(\phi)$.  In fact, it is possible to prove a theorem, for single-field
inflation, which states that the slow-roll approximation is an attractor
of the equations of motion, and thus we can always evaluate the
inflationary trajectory in phase-space within the SRA, therefore reducing
the number of initial conditions to just one, the initial value of the
scalar field. If $H(\phi)$ only depends on $\phi$, then $H'(\phi) =
-\kappa^2\dot\phi/2$ and we can rewrite the slow-roll parameters 
(\ref{SlowRollParameters}) as
\begin{eqnarray}\label{neweps}
\epsilon&=&{2\over\kappa^2}\left({H'(\phi)\over H(\phi)}\right)^2 \ 
\simeq \ {1\over2\kappa^2}\left({V'(\phi)\over V(\phi)}\right)^2 \ 
\ll \ 1\,,\\
\delta &=& {2\over\kappa^2}\,{H''(\phi)\over H(\phi)} \ 
\simeq \ {1\over\kappa^2}\,{V''(\phi)\over V(\phi)} -
{1\over2\kappa^2}\left({V'(\phi)\over V(\phi)}\right)^2 \ 
\equiv \ \eta - \epsilon \ \ll \ 1\,. \label{neweta}
\end{eqnarray}
The last expression defines the new slow-roll parameter $\eta$, not to
be confused with conformal time (see next Section). The number of
$e$-folds can also be rewritten in this approximation as
\begin{equation}\label{newefolds}
N \ = \ \kappa^2\, \int_{\phi_i}^{\phi_e} {V(\phi)\,d\phi\over 
V'(\phi)} \,,
\end{equation}
a very useful expression for evaluating $N$ for a given effective
scalar potential $V(\phi)$.

\subsection{The origin of density perturbations}

If cosmological inflation made the universe so extremely flat and
homogeneous, where did the galaxies and clusters of galaxies come from?
One of the most astonishing predictions of inflation, one that was not
even expected, is that quantum fluctuations of the inflaton field are
stretched by the exponential expansion and generate large-scale
perturbations in the metric. Inflaton fluctuations are small wave
packets of energy that, according to general relativity, modify the
space-time fabric, creating a whole spectrum of curvature
perturbations. The use of the word spectrum here is closely related to
the case of light waves propagating in a medium: a spectrum
characterizes the amplitude of each given wavelength. In the case of
inflation, the inflaton fluctuations induce waves in the space-time
metric that can be decomposed into different wavelengths, all with
approximately the same amplitude, that is, corresponding to a
scale-invariant spectrum. These patterns of perturbations in the metric
are like fingerprints that unequivocally characterize a period of
inflation. When matter fell in the troughs of these waves, it created
density perturbations that collapsed gravitationally to form galaxies,
clusters and superclusters of galaxies, with a spectrum that is also
scale invariant. Such a type of spectrum was proposed in the early 1970s
(before inflation) by Harrison and Zel'dovich~\cite{HZ}, to explain
the distribution of galaxies and clusters of galaxies on very large
scales in our observable universe. Perhaps the most interesting aspect
of structure formation is the possibility that the detailed knowledge of
what seeded galaxies and clusters of galaxies will allow us to test the
idea of inflation.

\subsubsection{Gauge invariant perturbation theory}

Until now we have considered only the unperturbed FRW metric described
by a scale factor $a(t)$ and a homogeneous scalar field $\phi(t)$,
\begin{eqnarray}
ds^2 &=& a^2(\eta) [d\eta^2 - \gamma_{ij}\,dx^i dx^j]\,,
\label{HomogeneousBackground}\\ \phi &=& \phi(\eta)\,,
\end{eqnarray}
where $\eta=\int dt/a(t)$ is the conformal time, under which the
background equations of motion can be written as
\begin{eqnarray}
{\cal H}^2 &=& {\kappa^2\over3}\left({1\over2}{\phi'}^2 + 
a^2 V(\phi)\right)\,,\\
{\cal H}'-{\cal H}^2 &=& {\kappa^2\over2}{\phi'}^2 \,,\\[1mm]
\phi''+2{\cal H}\phi'+ a^2V'(\phi)&=&0\,,\label{HomogeneousEquations}
\end{eqnarray}
where ${\cal H}=aH$ and $\phi'=a\dot\phi$.

During inflation, the quantum fluctuations of the scalar field will
induce metric perturbations which will backreact on the scalar field.
Let us consider, in linear perturbation theory, the most general line
element with both scalar and tensor metric 
perturbations~\cite{Bardeen},\footnote{Note
that inflation cannot generate, to linear order, a vector perturbation.}
together with the scalar field perturbations
\begin{eqnarray}\label{MetricPerturbations}
ds^2 &=& a^2(\eta)\left[(1+2A)d\eta^2 - 2B_{|i} dx^i d\eta - \Big\{
(1+2{\cal R})\gamma_{ij}+2E_{|ij}+2h_{ij}\Big\} dx^i dx^j\right]\,,\\
\phi&=&\phi(\eta)+\delta\phi(\eta,x^i)\,.\label{PerturbedBackground}
\end{eqnarray}
The indices $\{i,j\}$ label the three-dimensional spatial coordinates
with metric $\gamma_{ij}$, and the $|i$ denotes covariant derivative
with respect to that metric. The gauge invariant tensor perturbation
$h_{ij}$ corresponds to a transverse traceless gravitational wave,
$\nabla^ih_{ij} = h^i_i=0$. The four scalar perturbations $(A, B, {\cal
R}, E)$ are {\em gauge dependent} functions of $(\eta,x^i)$.
Under a general coordinate (gauge) transformation~\cite{Bardeen,MFB}
\begin{eqnarray}
&&\tilde\eta = \eta + \xi^0(\eta,x^i)\,,\\
&&\tilde x^i = x^i + \gamma^{ij}\xi_{|j}(\eta,x^i)\,,
\label{GaugeTransformation}
\end{eqnarray}
with arbitrary functions $(\xi^0,\xi)$, the scalar and tensor 
perturbations transform, to linear order, as
\begin{eqnarray}
&&\tilde A = A - {\xi^0}' - {\cal H}\xi^0\,,\hspace{1cm}
\tilde B = B + \xi^0 - \xi'\,,\\
&&\tilde {\cal R} = {\cal R} - {\cal H}\xi^0\,,\hspace{1.9cm}
\tilde E = E - \xi\,,\\
&&\hspace{3cm} \tilde h_{ij} = h_{ij}\,, \label{TransformedPerturbations}
\end{eqnarray}
where a prime denotes derivative with respect to conformal time. It is
possible to construct, however, two gauge-invariant gravitational
potentials~\cite{Bardeen,MFB},
\begin{eqnarray}
&& \Phi = A + (B-E')' + {\cal H}(B-E')\,,\\
&& \Psi = {\cal R} + {\cal H}(B-E')\,, 
\label{GravitationalPotentials}
\end{eqnarray}
which are related through the perturbed Einstein equations,
\begin{eqnarray}
 \Phi &=& \Psi\,, \\
 {k^2-3K\over a^2}\Psi &=& {\kappa^2\over2} \delta\rho\,,
\label{PertEinsteinEqs}
\end{eqnarray}
where $\delta\rho$ is the gauge-invariant density perturbation,
and the latter expression is nothing but the Poisson equation for 
the gravitational potential, written in relativistic form.

During inflation, the energy density is given in terms of a scalar
field, and thus the gauge-invariant equations for the perturbations on
comoving hypersurfaces (constant energy density hypersurfaces) are
\begin{eqnarray}
\Phi''+3{\cal H}\Phi'+({\cal H}'+{\cal H}^2)\Phi &=& 
{\kappa^2\over2} [\phi'\delta\phi'- a^2V'(\phi)\delta\phi]\,,\\
-\nabla^2\Phi+3{\cal H}\Phi'+({\cal H}'+{\cal H}^2)\Phi &=& 
- {\kappa^2\over2} [\phi'\delta\phi'+ a^2V'(\phi)\delta\phi]\,,\\
\Phi'+{\cal H}\Phi &=& {\kappa^2\over2} \phi'\delta\phi\,,\\[1mm]
\delta\phi''+2{\cal H}\delta\phi'-\nabla^2\delta\phi &=& 4\phi'\Phi'
-2a^2V'(\phi)\Phi-a^2V''(\phi)\delta\phi\,.
\label{PerturbedEquations}
\end{eqnarray}

This system of equations seem too difficult to solve at first sight. 
However, there is a gauge invariant combination of variables that
allows one to find exact solutions. Let us define~\cite{MFB}
\begin{eqnarray}
&& u \equiv a\delta\phi + z\Phi\,,\\
&& z \equiv a{\phi'\over{\cal H}}\,.
\label{RedefinitionVariables}
\end{eqnarray}
Under this redefinition, the above equations simplify enormously to
just three independent equations,
\begin{eqnarray}
&&u'' - \nabla^2u - {z''\over z}u = 0\,,\label{EquationU}\\
&&\nabla^2\Phi = {\kappa^2\over2} {{\cal H}\over a^2} (zu'-z'u)\,,
\label{NablaPhi}\\
&& \Big({a^2\Phi\over{\cal H}}\Big)' = {\kappa^2\over2} zu\,.
\label{EquationPhi}
\end{eqnarray}
From Equation (\ref{EquationU}) we can find a solution $u(z)$,
which substituted into (\ref{EquationPhi}) can be integrated to give
$\Phi(z)$, and together with $u(z)$ allow us to obtain
$\delta\phi(z)$. 

\subsubsection{Quantum Field Theory in curved space-time}

Until now we have treated the perturbations as classical, but we should
in fact consider the perturbations $\Phi$ and $\delta\phi$ as quantum
fields. Note that the perturbed action for the scalar mode $u$
can be written as
\begin{equation}
\delta S = {1\over2} \int d^3x\,d\eta\,\Big[(u')^2 - (\nabla u)^2
+ {z''\over z}u^2\Big]\,.
\end{equation}
In order to quantize the field $u$ in the curved background defined
by the metric (\ref{HomogeneousBackground}), we can write the operator
\begin{equation}
\hat u(\eta,{\bf x}) = \int {d^3{\bf k}\over(2\pi)^{3/2}}
\Big[u_k(\eta)\,\hat a_{\bf k}\,e^{i{\bf k}\cdot{\bf x}} + 
u_k^*(\eta)\,\hat a_{\bf k}^\dagger\,e^{-i{\bf k}\cdot{\bf x}} \Big]\,,
\label{OperatorU}
\end{equation}
where the creation and annihilation operators satisfy the commutation
relation of bosonic fields, and the scalar field's Fock space is defined
through the vacuum condition,
\begin{eqnarray}\label{CommutationRelation}
[\hat a_{\bf k}, \hat a_{{\bf k}'}^\dagger] &=& \delta^3({\bf k - k}')
\,,\\ \hat a_{\bf k}|0\rangle &=& 0\,.
\end{eqnarray}
Note that we are not assuming that the inflaton is a fundamental scalar
field, but that is can be written as a quantum field with its
commutation relations (as much as a pion can be described as a quantum
field).

The equations of motion for each mode $u_k(\eta)$ are decoupled in
linear perturbation theory,
\begin{equation}
u''_k +\Big(k^2-{z''\over z}\Big)u_k = 0\,.
\label{EquationModek}
\end{equation}
The ratio $z''/z$ acts like a time-dependent potential for this
Schr\"odinger like equation. In order to find exact solutions to the
mode equation, we will use the slow-roll parameters
(\ref{SlowRollParameters}), see Ref.~\cite{LL93}
\begin{eqnarray}
&&\epsilon = 1 - {{\cal H}'\over{\cal H}^2} = {\kappa^2\over2}
{z^2\over a^2}\,,\label{ConfEpsilon}\\
&&\delta = 1 - {\phi''\over{\cal H}\phi'} = 1 + \epsilon - 
{z'\over{\cal H}z}\,.
\label{ConfDelta}
\end{eqnarray}
In terms of these parameters, the conformal time and the effective
potential for the $u_k$ mode can be written as
\begin{eqnarray}
&&\eta = {-1\over{\cal H}} + \int {\epsilon da\over a{\cal H}}\,,\\
&&{z''\over z} = {\cal H}^2[(1+\epsilon-\delta)(2-\delta) +
{\cal H}^{-1}(\epsilon'-\delta')]\,.\label{PotentialZ}
\end{eqnarray}
Note that the slow-roll parameters, (\ref{ConfEpsilon}) and
(\ref{ConfDelta}), can be taken as {\em constant},\footnote{For
instance, there are models of inflation, like power-law inflation,
$a(t)\sim t^p$, where $\epsilon=\delta=1/p<1$, that give constant 
slow-roll parameters.} to order $\epsilon^2$,
\begin{eqnarray}
&&\epsilon' = 2{\cal H}\,\epsilon(\epsilon-\delta) = 
{\cal O}(\epsilon^2)\,,\\
&&\delta' = {\cal H}\,\delta\Big(\epsilon+\delta+{\stackrel{\dots}{\phi}
\over H\ddot\phi}\Big) = {\cal O}(\epsilon^2)\,.
\label{ConstantSlowRoll}
\end{eqnarray}
In that case, for constant parameters, we can write
\begin{eqnarray}
&&\eta = {-1\over{\cal H}}{1\over1-\epsilon}\,,\label{EtaEpsilon}\\
{z''\over z} = {1\over\eta^2}\Big(\nu^2-{1\over4}\Big)\,, && 
\hspace{5mm} {\rm where} \hspace{7mm} \nu =
{1+\epsilon-\delta\over1-\epsilon} + {1\over2}\,.\label{NuParam}
\end{eqnarray}

We are now going to search for approximate solutions of the mode
equation (\ref{EquationModek}), where the effective potential
(\ref{PotentialZ}) is of order $z''/z \simeq 2{\cal H}^2$ in the
slow-roll approximation. In quasi-de Sitter there is a characteristic
scale given by the (event) horizon size or Hubble scale during
inflation, $H^{-1}$. There will be modes $u_k$ with physical wavelengths
much smaller than this scale, $k/a\gg H$, that are well within the
de Sitter horizon and therefore do not feel the curvature of space-time.
On the other hand, there will be modes with physical wavelengths much
greater than the Hubble scale, $k/a\ll H$. In these two asymptotic
regimes, the solutions can be written as
\begin{eqnarray}
&&u_k = {1\over\sqrt{2k}}\,e^{-ik\eta} \hspace{1cm} k\gg aH\,, 
\label{MinkowskyModes}\\
&&u_k = C_1\,z \hspace{2cm} k\ll aH\,.\label{deSitterModes}
\end{eqnarray}
In the limit $k\gg aH$ the modes behave like ordinary quantum modes in
Minkowsky space-time, appropriately normalized, while in the opposite
limit, $u/z$ becomes constant on superhorizon scales. For approximately
constant slow-roll parameters one can find exact solutions to
(\ref{EquationModek}), with the effective potential given by
(\ref{NuParam}), that interpolate between the two asymptotic solutions,
\begin{equation}
u_k(\eta) = {\sqrt\pi\over2}\,e^{i(\nu+{1\over2}){\pi\over2}}\,
(-\eta)^{1/2}\,H_\nu^{_{(1)}}(-k\eta)\,,
\label{ExactSolution}
\end{equation}
where $H_\nu^{_{(1)}}(z)$ is the Hankel function of the first
kind~\cite{AS}, and $\nu$ is given by (\ref{NuParam}) in terms of the
slow-roll parameters. In the limit $k\eta\to0$, the solution becomes
\begin{eqnarray}
&&|u_k| = {2^{\nu-{3\over2}}\over\sqrt{2k}}\,
{\Gamma(\nu)\over\Gamma({3\over2})}\,(-k\eta)^{{1\over2}-\nu} \equiv
{C(\nu)\over\sqrt{2k}}\,\Big({k\over aH}\Big)^{\nu-{1\over2}}\,, 
\label{LimitSolution}\\[2mm]
&&C(\nu) = 2^{\nu-{3\over2}}\,
{\Gamma(\nu)\over\Gamma({3\over2})}\,(1-\epsilon)^{\nu-{1\over2}}
\ \simeq 1 \hspace{1cm} {\rm for} \hspace{.5cm} \epsilon, \delta\ll1\,.
\end{eqnarray}

We can now compute $\Phi$ and $\delta\phi$ from the super-Hubble-scale
mode solution (\ref{deSitterModes}), for $k\ll aH$. Substituting into
Eq.~(\ref{EquationPhi}), we find
\begin{eqnarray}
&&\Phi = C_1\,\Big(1 - {{\cal H}\over a^2}\,\int a^2 d\eta\Big) + 
C_2\,{{\cal H}\over a^2}\,, 
\label{PhiSolution}\\
&&\delta\phi = {C_1\over a^2}\,\int a^2 d\eta - {C_2\over a^2}\,.
\label{DeltaphiSolution}
\end{eqnarray}
The term proportional to $C_1$ corresponds to the growing solution,
while that proportional to $C_2$ corresponds to the decaying solution,
which can soon be ignored. These quantities are gauge invariant but
evolve with time outside the horizon, during inflation, and before
entering again the horizon during the radiation or matter eras. We would
like to write an expression for a gauge invariant quantity that is also
{\em constant} for superhorizon modes. Fortunately, in the case of
adiabatic perturbations, there is such a quantity:
\begin{equation}
\zeta \equiv \Phi + {1\over\epsilon{\cal H}}\,(\Phi'+
{\cal H}\Phi) = {u\over z}\,,\label{Zeta}
\end{equation}
which is constant, see Eq.~(\ref{deSitterModes}), for $k\ll aH$. In
fact, this quantity $\zeta$ is identical, for superhorizon modes, to the
gauge invariant curvature metric perturbation ${\cal R}_c$ on comoving
(constant energy density) hypersurfaces, see Ref.~\cite{Bardeen,GBW},
\begin{equation}
\zeta = {\cal R}_c + {1\over\epsilon{\cal H}^2}\,
\nabla^2\Phi\,.\label{ZetaRc}
\end{equation}
Using Eq.~(\ref{NablaPhi}) we can write the evolution equation for
$\zeta={u\over z}$ as \ $\zeta' = {1\over\epsilon{\cal
H}}\,\nabla^2\Phi$, which confirms that $\zeta$ is constant for
(adiabatic\footnote{This conservation fails for entropy or isocurvature
perturbations, see Ref.~\cite{GBW}.}) superhorizon modes, $k\ll
aH$. Therefore, we can evaluate the Newtonian potential $\Phi_k$ when
the perturbation reenters the horizon during radiation/matter eras in
terms of the curvature perturbation ${\cal R}_k$ when it left the Hubble
scale during inflation,
\begin{equation}
\Phi_k = \Big(1 - {{\cal H}\over a^2}\,\int a^2 d\eta\Big)\,
{\cal R}_k = {3+3\omega\over5+3\omega}\,{\cal R}_k = 
\hspace{1mm}\left\{\begin{array}{cl}
{2\over3}\,{\cal R}_k&\hspace{5mm}{\rm radiation 
\hspace{2mm} era}\,,\\[3mm]
{3\over5}\,{\cal R}_k&\hspace{5mm}{\rm matter 
\hspace{2mm} era}\,.\end{array}\right.
\label{PhiRc}
\end{equation}

Let us now compute the tensor or gravitational wave metric perturbations
generated during inflation. The perturbed action for the tensor mode can
be written as
\begin{equation}
\delta S = {1\over2} \int d^3x\,d\eta\,{a^2\over2\kappa^2}\Big[
(h_{ij}')^2 - (\nabla h_{ij})^2\Big]\,,
\end{equation}
with the tensor field $h_{ij}$ considered as a quantum field,
\begin{equation}
\hat h_{ij}(\eta,{\bf x}) = \int {d^3{\bf k}\over(2\pi)^{3/2}}
\sum_{\lambda=1,2}\Big[h_k(\eta)\,e_{ij}({\bf k},\lambda)\,
\hat a_{\bf k,\lambda}\,e^{i{\bf k}\cdot{\bf x}} + h.c.\Big]\,,
\label{TensorField}
\end{equation}
where $e_{ij}({\bf k},\lambda)$ are the two polarization tensors, 
satisfying symmetric, transverse and traceless conditions
\begin{eqnarray}
&&e_{ij}=e_{ji}\,, \hspace{5mm} k^ie_{ij}=0\,, \hspace{5mm} e_{ii} =0\,, 
\label{PolarCons}\\[2mm]
&&e_{ij}(-{\bf k},\lambda) = e_{ij}^*({\bf k},\lambda)\,, \hspace{5mm} 
\sum_{\lambda}e_{ij}^*({\bf k},\lambda)e^{ij}({\bf k},\lambda)=4\,,
\label{polarSum}
\end{eqnarray}
while the creation and annihilation operators satisfy the usual
commutation relation of bosonic fields, Eq.~(\ref{CommutationRelation}).
We can now redefine our gauge invariant tensor amplitude as
\begin{equation}
v_k(\eta)={a\over\sqrt2\kappa}\,h_k(\eta)\,,
\label{FieldV}
\end{equation}
which satisfies the following evolution equation, decoupled for each 
mode $v_k(\eta)$ in linear perturbation theory,
\begin{equation}
v''_k +\Big(k^2-{a''\over a}\Big)v_k = 0\,.
\label{ModeVk}
\end{equation}
The ratio $a''/a$ acts like a time-dependent potential for this
Schr\"odinger like equation, analogous to the term $z''/z$ for the
scalar metric perturbation. For constant slow-roll parameters, the
potential becomes
\begin{eqnarray}\label{Potentialappa}
&&{a''\over a} = 2{\cal H}^2\Big(1-{\epsilon\over2}\Big) =
{1\over\eta^2}\Big(\mu^2-{1\over4}\Big)\,,\\[1mm]
&&\mu = {1\over1-\epsilon} + {1\over2}\,.\label{MuParam}
\end{eqnarray}
We can solve equation (\ref{ModeVk}) in the two asymptotic regimes,
\begin{eqnarray}
&&v_k = {1\over\sqrt{2k}}\,e^{-ik\eta} \hspace{1cm} k\gg aH\,, 
\label{MinkowskyVkModes}\\[1mm]
&&v_k = C\,a \hspace{2.2cm} k\ll aH\,.\label{deSitterVkModes}
\end{eqnarray}
In the limit $k\gg aH$ the modes behave like ordinary quantum modes in
Minkowsky space-time, appropriately normalized, while in the opposite
limit, the metric perturbation $h_k$ becomes {\em constant} on
superhorizon scales. For constant slow-roll parameters one can find
exact solutions to (\ref{ModeVk}), with effective potential given by
(\ref{Potentialappa}), that interpolate between the two asymptotic
solutions. These are identical to Eq.~(\ref{ExactSolution}) except for
the substitution $\nu\to\mu$.  In the limit $k\eta\to0$, the solution
becomes
\begin{equation}
|v_k| = {C(\mu)\over\sqrt{2k}}\,\Big({k\over aH}\Big)^{\mu-{1\over2}}\,.
\label{LimitVkSolution}
\end{equation}
Since the mode $h_k$ becomes constant on superhorizon scales, we can
evaluate the tensor metric perturbation when it reentered during the
radiation or matter era directly in terms of its value during inflation.

\subsubsection{Power spectrum of scalar and tensor metric perturbations}

Not only do we expect to measure the amplitude of the metric
perturbations generated during inflation and responsible for the
anisotropies in the CMB and density fluctuations in LSS, but we should
also be able to measure its power spectrum, or two-point correlation
function in Fourier space. Let us consider first the scalar metric
perturbations ${\cal R}_k$, which enter the horizon at $a=k/H$. Its
correlator is given by~\cite{LL93}
\begin{eqnarray}
&&\langle0|{\cal R}_k^*{\cal R}_{k'}|0\rangle = {|u_k|^2\over z^2}\,
\delta^3({\bf k - k'}) \equiv {{\cal P}_{\cal R}(k)\over4\pi k^3}\,
(2\pi)^3\,\delta^3({\bf k - k'})\,, 
\label{CorrelatorRc}\\[2mm]
&&{\cal P}_{\cal R}(k)={k^3\over2\pi^2}\,{|u_k|^2\over z^2}=
{\kappa^2\over2\epsilon}\,\Big({H\over2\pi}\Big)^2\,
\Big({k\over aH}\Big)^{3-2\nu}\equiv A_S^2\,
\Big({k\over aH}\Big)^{n-1}\,,
\label{PowerSpectrumRc}
\end{eqnarray}
where we have used \ ${\cal R}_k=\zeta_k={u_k\over z}$ and
Eq.~(\ref{LimitSolution}). This last equation determines the power
spectrum in terms of its amplitude at horizon-crossing, $A_S$, and a
tilt,
\begin{equation}
n-1\equiv{d\ln{\cal P}_{\cal R}(k)\over d\ln k} = 3-2\nu = 
2 \Big({\delta-2\epsilon\over1-\epsilon}\Big)\simeq 2\eta-6\epsilon\,,
\label{ScalarTilt}
\end{equation}
see Eqs.~(\ref{neweps}), (\ref{neweta}).  Note from this equation that
it is possible, in principle, to obtain from inflation a scalar tilt
which is either positive ($n>1$) or negative ($n<1$). Furthermore,
depending on the particular inflationary model~\cite{LR99}, we
can have significant departures from scale invariance.

Let us consider now the tensor (gravitational wave) metric perturbation,
which enter the horizon at \ $a=k/H$,
\begin{eqnarray}
&&\sum_\lambda\langle0|h^*_{k,\lambda}h_{k',\lambda}|0\rangle = 4\,
{2\kappa^2\over a^2}\,|v_k|^2
\delta^3({\bf k - k'}) \equiv {{\cal P}_g(k)\over4\pi k^3}\,
(2\pi)^3\,\delta^3({\bf k - k'})\,, 
\label{CorrelatorHk}\\[2mm]
&&{\cal P}_g(k)=8\kappa^2\,\Big({H\over2\pi}\Big)^2\,
\Big({k\over aH}\Big)^{3-2\mu}\equiv A_T^2\,
\Big({k\over aH}\Big)^{n_T}\,,
\label{TensorSpectrum}
\end{eqnarray}
where we have used Eqs.~(\ref{FieldV}) and (\ref{LimitVkSolution}).
Therefore, the power spectrum can be approximated by a power-law
expression, with amplitude $A_T$ and tilt
\begin{equation}
n_T\equiv{d\ln{\cal P}_g(k)\over d\ln k} = 3-2\mu = -
\Big({2\epsilon\over1-\epsilon}\Big)\simeq -2\epsilon < 0\,,
\label{TensorTilt}
\end{equation}
which is always negative. In the slow-roll approximation, $\epsilon\ll1$,
the tensor power spectrum is scale invariant.

\subsection{The anisotropies of the microwave background}

The metric fluctuations generated during inflation are not only
responsible for the density perturbations that gave rise to galaxies via
gravitational collapse, but one should also expect to see such ripples
in the metric as temperature anisotropies in the cosmic microwave
background, that is, minute deviations in the temperature of the
blackbody spectrum when we look at different directions in the sky. Such
anisotropies had been looked for ever since Penzias and Wilson's
discovery of the CMB, but had eluded all detection, until COBE satellite
discovered them in 1992, see Fig.~\ref{fig6}. The reason why they took
so long to be discovered was that they appear as perturbations in
temperature of only one part in $10^5$. Soon after COBE, other groups
quickly confirmed the detection of temperature anisotropies at around
30\,$\mu$K, at higher multipole numbers or smaller angular scales. There
are at this moment dozens of ground and balloon-borne experiments
analysing the anisotropies in the microwave background with angular
resolutions from $10^\circ$ to a few arc minutes in the sky, see
Fig.~\ref{fig34}.

\subsubsection{Acoustic oscillations in the plasma}

\begin{figure}[htb]
\vspace*{-1.3cm}
\begin{center}
\includegraphics[width=10cm]{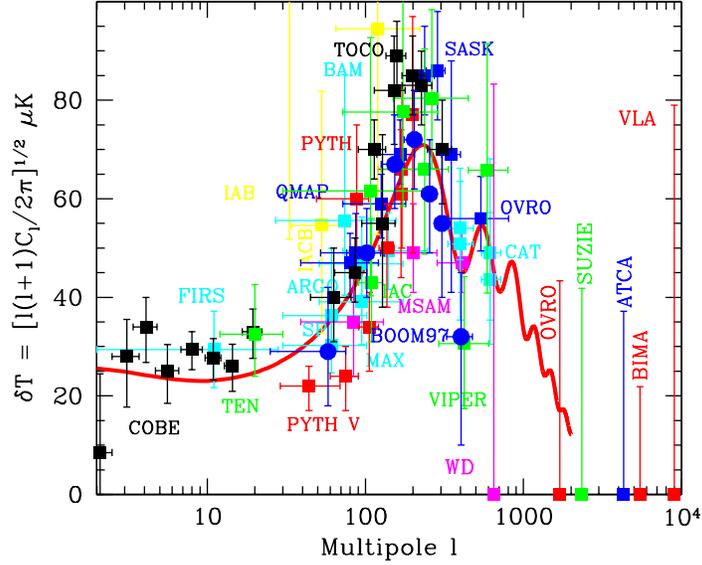}
\vspace*{-1.3cm}
\caption{There are at present dozens of ground and balloon-borne
experiments looking at the microwave background temperature anisotropies
with angular resolutions from $10^\circ$ to a few arc minutes in the
sky, corresponding to multipole numbers $l=2 - 3000$. Present
observations suggest the existence of a peak in the angular
distribution, as predicted by inflation. The theoretical curve (thick
line) illustrates a particular model which fits the data. From
Ref.~\cite{Tegmark}.}
\label{fig34}
\end{center}
\end{figure}

The physics of the CMB anisotropies is relatively simple~\cite{CMB}. The
universe just before recombination is a very tightly coupled fluid, due
to the large electromagnetic Thomson cross section
(\ref{ThomsonCrossSection}).  Photons scatter off charged particles
(protons and electrons), and carry energy, so they feel the
gravitational potential associated with the perturbations imprinted in
the metric during inflation. An overdensity of baryons (protons and
neutrons) does not collapse under the effect of gravity until it enters
the causal Hubble radius. The perturbation continues to grow until
radiation pressure opposes gravity and sets up acoustic oscillations in
the plasma, very similar to sound waves. Since overdensities of the same
size will enter the Hubble radius at the same time, they will oscillate
in phase. Moreover, since photons scatter off these baryons, the
acoustic oscillations occur also in the photon field and induces a
pattern of peaks in the temperature anisotropies in the sky, at
different angular scales, see Fig.~\ref{fig34}.

There are three different effects that determine the temperature
anisotropies we observe in the CMB. First, gravity: photons fall in and
escape off gravitational potential wells, characterized by $\Phi$ in the
comoving gauge, and as a consequence their frequency is gravitationally
blue- or red-shifted, $\delta\nu/\nu = \Phi$. If the gravitational
potential is not constant, the photons will escape from a larger or
smaller potential well than they fell in, so their frequency is also
blue- or red-shifted, a phenomenon known as the Rees-Sciama effect.
Second, pressure: photons scatter off baryons which fall into
gravitational potential wells and the two competing forces create
acoustic waves of compression and rarefaction. Finally, velocity:
baryons accelerate as they fall into potential wells. They have minimum
velocity at maximum compression and rarefaction. That is, their velocity
wave is exactly $90^\circ$ off-phase with the acoustic waves. These
waves induce a Doppler effect on the frequency of the photons.

The temperature anisotropy induced by these three effects is therefore
given by~\cite{CMB}
\begin{equation}
{\delta T\over T}({\bf r}) = \Phi({\bf r},t_{\rm dec}) + 
2\int_{t_{\rm dec}}^{t_0} \dot\Phi({\bf r},t) dt \ + \
{1\over3}{\delta\rho\over\rho} \ - \ {{\bf r}\cdot{\bf v}\over c}\,.
\label{TemperatureAnisotropy}
\end{equation}
Metric perturbations of different wavelengths enter the horizon at
different times. The largest wavelengths, of size comparable to our
present horizon, are entering now. There are perturbations with
wavelengths comparable to the size of the horizon at the time of last
scattering, of projected size about $1^\circ$ in the sky today, which
entered precisely at decoupling. And there are perturbations with
wavelengths much smaller than the size of the horizon at last
scattering, that entered much earlier than decoupling, all the way to
the time of radiation-matter equality, which have gone through several
acoustic oscillations before last scattering. All these perturbations
of different wavelengths leave their imprint in the CMB anisotropies.

The baryons at the time of decoupling do not feel the gravitational
attraction of perturbations with wavelength greater than the size of the
horizon at last scattering, because of causality.  Perturbations with
exactly that wavelength are undergoing their first contraction, or
acoustic compression, at decoupling. Those perturbations induce a large
peak in the temperature anisotropies power spectrum, see
Fig.~\ref{fig34}. Perturbations with wavelengths smaller than these will
have gone, after they entered the Hubble scale, through a series of
acoustic compressions and rarefactions, which can be seen as secondary
peaks in the power spectrum. Since the surface of last scattering is not
a sharp discontinuity, but a region of $\Delta z \sim 100$, see
Fig.~\ref{fig4}, there will be scales for which photons, travelling from
one energy concentration to another, will erase the perturbation on that
scale, similarly to what neutrinos or HDM do for structure on small
scales.  That is the reason why we don't see all the acoustic
oscillations with the same amplitude, but in fact they decay
exponentialy towards smaller angular scales, an effect known as Silk
damping, due to photon diffusion~\cite{Silk,CMB}.

\subsubsection{The Sachs-Wolfe effect}

The anisotropies corresponding to large angular scales are only
generated via gravitational red-shift and density perturbations through
the Einstein equations, $\delta\rho/\rho = -2\Phi$ for adiabatic
perturbations; we can ignore the Doppler contribution, since the
perturbation is non-causal. In that case, the temperature anisotropy 
in the sky today is given by~\cite{SW}
\begin{equation}
{\delta T\over T}(\theta,\phi) = {1\over3}\Phi(\eta_{\rm LS})\,
Q(\eta_0, \theta, \phi) + 2\int_{\eta_{\rm LS}}^{\eta_0} 
dr\,\Phi'(\eta_0-r)\,Q(r, \theta, \phi)\,,
\label{SachsWolfe}
\end{equation}
where $\eta_0$ is the {\em coordinate distance} to the last scattering
surface, i.e. the present conformal time, while $\eta_{\rm LS}\simeq0$
determines that comoving hypersurface. The above expression is known as
the Sachs-Wolfe effect~\cite{SW}, and contains two parts, the intrinsic
and the Integrated Sachs-Wolfe (ISW) effect, due to integration along
the line of sight of time variations in the gravitational potential.

In linear perturbation theory, the scalar metric perturbations can be
separated into $\Phi(\eta, {\bf x}) \equiv \Phi(\eta)\,Q({\bf x})$,
where $Q({\bf x})$ are the scalar harmonics, eigenfunctions of the
Laplacian in three dimensions, $\nabla^2 Q_{klm}(r, \theta, \phi) =
-k^2\,Q_{klm}(r, \theta, \phi)$. These functions have the general 
form~\cite{Harrison}
\begin{equation}
Q_{klm}(r, \theta, \phi) = \Pi_{kl}(r)\,Y_{lm}(\theta, \phi) \,,
\label{ScalarHarmonics}
\end{equation}
where $Y_{lm}(\theta, \phi)$ are the usual spherical
harmonics~\cite{AS}.

In order to compute the temperature anisotropy associated with the
Sachs-Wolfe effect, we have to know the evolution of the metric
perturbation during the matter era,
\begin{equation}
\Phi''+3{\cal H}\,\Phi' + a^2 \Lambda\,\Phi - 2K\,\Phi = 0\,.
\label{PhiEvEq}
\end{equation}
In the case of a flat universe without cosmological constant, the
Newtonian potential remains constant during the matter era and only the
intrinsic SW effect contributes to $\delta T/T$. In case of a
non-vanishing $\Lambda$, since its contribution is negligible in the
past, see Eq.~(\ref{OmegaL}), most of the photon's trajectory towards us
is unperturbed, and the only difference with respect to the $\Lambda=0$
case is an overall factor~\cite{CPT}. We will consider here the
approximation $\Phi={\rm const}$. during the matter era and ignore that
factor, see Ref.~\cite{BLW}.

In a flat universe, the radial part of the eigenfunctions
(\ref{ScalarHarmonics}) can be written as~\cite{Harrison}
\begin{equation}
\Pi_{kl}(r) = \sqrt{2\over\pi}\,k\,j_l(kr)\,,
\label{RadialEigenfunctions}
\end{equation}
where $j_l(z)$ are the spherical Bessel functions~\cite{AS}. The growing
mode solution of the metric perturbation that left the Hubble scale
during inflation contributes to the temperature anisotropies on large
scales (\ref{SachsWolfe}) as
\begin{equation}\label{alm}
{\delta T\over T}(\theta,\phi) \ = \ {1\over3}\Phi(\eta_{\rm LS})\,Q \ 
= \ {1\over5}\,{\cal R}\,Q(\eta_0, \theta, \phi) \ \equiv \
\sum_{l=2}^\infty \sum_{m=-l}^l\,a_{lm}\,Y_{lm}(\theta, \phi)\,,
\end{equation}
where we have used the fact that at reentry (at the surface of last
scattering) the gauge invariant Newtonian potential $\Phi$ is related to
the curvature perturbation ${\cal R}$ at Hubble-crossing during
inflation, see Eq.~(\ref{PhiRc}); and we have expanded $\delta T/T$ in
spherical harmonics. 

We can now compute the two-point correlation function or angular power
spectrum, $C(\theta)$, of the CMB anisotropies on large scales, defined
as an expansion in multipole number,
\begin{equation}
C(\theta)=
\left\langle{\delta T\over T}^*\!({\bf n}){\delta T\over T}({\bf n}')
\right\rangle_{{\bf n}\cdot{\bf n}'=\cos\theta} = {1\over4\pi}
\sum_{l=2}^\infty (2l+1)\,C_l\,P_l(\cos\theta)\,,
\label{Ctheta}
\end{equation}
where $P_l(z)$ are the Legendre polynomials~\cite{AS}, and we have
averaged over different universe realizations. Since the coefficients
$a_{lm}$ are isotropic (to first order), we can compute the $C_l =
\langle|a_{lm}|^2\rangle$ as
\begin{equation}
C_l^{(S)} = {4\pi\over25}\,\int_0^\infty\,{dk\over k}\,
{\cal P}_{\cal R}(k)\,j_l^2(k\eta_0)\,,
\label{ClScalar}
\end{equation}
where we have used Eqs.~(\ref{alm}) and (\ref{CorrelatorRc}). In the
case of scalar metric perturbation produced during inflation, the scalar
power spectrum at reentry is given by ${\cal P}_{\cal R}(k)=
A_S^2(k\eta_0)^{n-1}$, in the power-law approximation, see
Eq.~(\ref{PowerSpectrumRc}). In that case, one can integrate
(\ref{ClScalar}) to give
\begin{eqnarray}
&&C_l^{(S)}={2\pi\over25}\,A_S^2\,
{\Gamma[{3\over2}]\,\Gamma[1-{n-1\over2}]\,\Gamma[l+{n-1\over2}]\over
\Gamma[{3\over2}-{n-1\over2}]\,\Gamma[l+2-{n-1\over2}]}\,,\\[2mm]
&&{l(l+1)\,C_l^{(S)}\over2\pi}={A_S^2\over25} \ = \ {\rm constant}\,,
\hspace{1cm}{\rm for} \hspace{3mm} n=1\,.\label{CLS}
\end{eqnarray}
This last expression corresponds to what is known as the Sachs-Wolfe
plateau, and is the reason why the coefficients $C_l$ are always plotted
multiplied by $l(l+1)$, see Fig.~\ref{fig34}. 

Tensor metric perturbations also contribute with an approximately
constant angular power spectrum, $l(l+1)C_l$. The Sachs-Wolfe effect for
a gauge invariant tensor perturbation is given by~\cite{SW}
\begin{equation}
{\delta T\over T}(\theta,\phi) = \int_{\eta_{\rm LS}}^{\eta_0} 
dr\,h'(\eta_0-r)\,Q_{rr}(r, \theta, \phi)\,,
\label{SachsWolfeTensor}
\end{equation}
where $Q_{rr}$ is the $rr$-component of the tensor harmonic along the
line of sight~\cite{Harrison}. The tensor perturbation $h$ during the
matter era satisfies the following evolution equation
\begin{equation}
h''_k+3{\cal H}\,h'_k+(k^2+2K)\,h_k=0\,,
\end{equation}
which depends on the wavenumber $k$, contrary to what happens with the
scalar modes, see Eq.~(\ref{PhiEvEq}). For a flat ($K=0$) universe, the
solution to this equation is \ $h_k(\eta)=h\,G_k(\eta)$, where $h$ is
the constant tensor metric perturbation at horizon crossing and
$G_k(\eta)=3\,j_1(k\eta)/k\eta$, normalized so that $G_k(0)=1$ at the
surface of last scattering. The radial part of the tensor harmonic
$Q_{rr}$ in a flat universe can be written as~\cite{Harrison}
\begin{equation}
Q^{rr}_{kl}(r) = \left[{(l-1)l(l+1)(l+2)\over\pi k^2}\right]^{1/2}\,
{j_l(kr)\over r^2}\,.
\end{equation}
The tensor angular power spectrum can finally be expressed as
\begin{eqnarray}\label{CLTensor}
&&C_l^{(T)}={9\pi\over4}\,(l-1)l(l+1)(l+2)\,\int_0^\infty\,{dk\over k}\,
{\cal P}_g(k)\,I_{kl}^2\,,\\[1mm]
&&I_{kl} = \int_0^{x_0}\,dx\,{j_2(x_0-x)j_l(x)\over(x_0-x)x^2}\,,
\end{eqnarray}
where $x\equiv k\eta$, and ${\cal P}_g(k)$ is the primordial tensor spectrum
(\ref{TensorSpectrum}). For a scale invariant spectrum, $n_T=0$, we can
integrate (\ref{CLTensor}) to give~\cite{Staro}
\begin{equation}\label{CLT}
l(l+1)\,C_l^{(T)} = {\pi\over36}\Big(1 + {48\pi^2\over385}\Big)\,A_T^2\,
B_l\,,
\end{equation}
with \ $B_l = (1.1184, 0.8789, \dots, 1.00)$ \ for $l=2, 3, \dots,30$.
Therefore, $l(l+1)\,C_l^{(T)}$ also becomes constant for large $l$. Beyond
$l\sim30$, the Sachs-Wolfe expression is not a good approximation and
the tensor angular power spectrum decays very quickly at large $l$,
see Fig.\ref{fig40}.

\subsubsection{The consistency relation}

In spite of the success of inflation in predicting a homogeneous and
isotropic background on which to imprint a scale-invariant spectrum of
inhomogeneities, it is difficult to test the idea of inflation. A CMB
cosmologist before the 1980s would have argued that {\em ad hoc} initial
conditions could have been at the origin of the homogeneity and flatness
of the universe on large scales, while a LSS cosmologist would have
agreed with Harrison and Zel'dovich that the most natural spectrum
needed to explain the formation of structure was a scale-invariant
spectrum. The surprise was that inflation incorporated an understanding
of {\em both} the globally homogeneous and spatially flat background, and
the approximately scale-invariant spectrum of perturbations in the same
formalism. But that could have been a coincidence, and is not
epistemologically testable.

What is {\em unique} to inflation is the fact that inflation determines
not just one but {\em two} primordial spectra, corresponding to the
scalar (density) and tensor (gravitational waves) metric perturbations,
from a single continuous function, the inflaton potential $V(\phi)$. In
the slow-roll approximation, one determines, from $V(\phi)$, two
continuous functions, ${\cal P}_{\cal R}(k)$ and ${\cal P}_g(k)$, that
in the power-law approximation reduces to two amplitudes, $A_S$ and
$A_T$, and two tilts, $n$ and $n_T$. It is clear that there must be a
relation between the four parameters. Indeed, one can see from
Eqs.~(\ref{CLT}) and (\ref{CLS}) that the ratio of the tensor to scalar
contribution to the angular power spectrum is proportional to the tensor
tilt~\cite{LL93}, 
\begin{equation}\label{Consistency}
R\equiv {C_l^{(T)}\over C_l^{(S)}} = {25\over9}
\Big(1 + {48\pi^2\over385}\Big)\,2\epsilon \simeq - 2\pi\,n_T\,.
\end{equation}
This is a unique prediction of inflation, which could not have been
postulated a priori by any cosmologist. If we finally observe a tensor
spectrum of anisotropies in the CMB, or a stochastic gravitational wave
background in laser interferometers like LIGO or VIRGO~\cite{LIGO}, with
sufficient accuracy to determine their spectral tilt, one might have
some chance to test the idea of inflation, via the consistency relation
(\ref{Consistency}). For the moment, observations of the microwave
background anisotropies suggest that the Sachs-Wolfe plateau exists, see
Fig.~\ref{fig34}, but it is still premature to determine the tensor
contribution. Perhaps in the near future, from the analysis of
polarization as well as temperature anisotropies, with the CMB
satellites MAP and Planck, we might have a chance of determining the
validity of the consistency relation.

Assuming that the scalar contribution dominates over the tensor on large
scales, i.e. $R \ll 1$, one can actually give a measure of the amplitude
of the scalar metric perturbation from the observations of the
Sachs-Wolfe plateau in the angular power spectrum~\cite{BLW},
\begin{eqnarray}\label{MeasuredPower}
\left[{l(l+1)\,C_l^{(S)}\over2\pi}\right]^{1/2}\hspace{-2mm}&=&
{A_S\over5} \ = \ (1.03\pm0.07)\times10^{-5}\,,\\[2mm]
n&=&1.02\pm0.12\,.
\label{MeasuredTilt}
\end{eqnarray}
These measurements can be used to normalize the primordial spectrum and
determine the parameters of the model of inflation~\cite{LR99}.
In the near future these parameters will be determined with much better
accuracy, as described in Section~4.4.5.

\begin{figure}[htb]
\vspace*{-.5cm}
\hspace*{3cm}
\includegraphics[width=10cm]{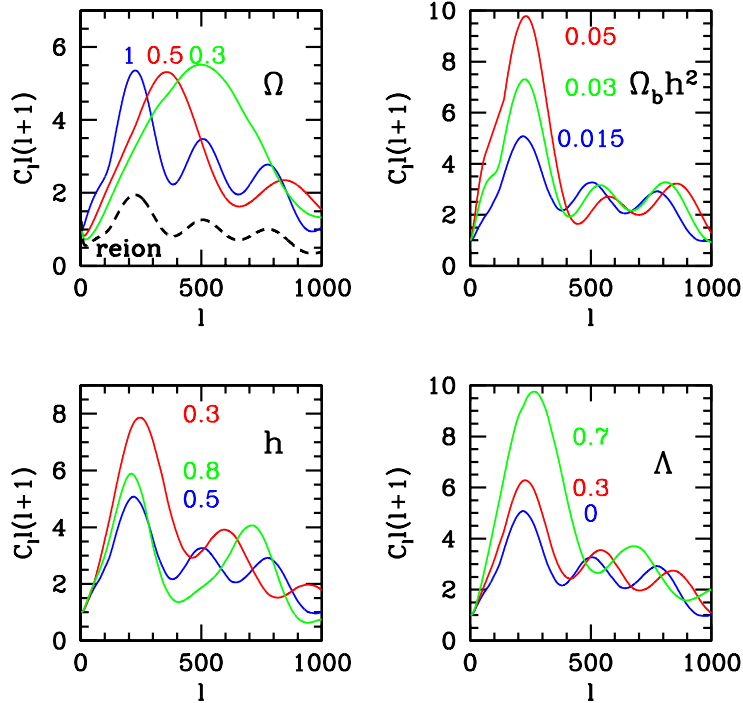}
\vspace*{-.5cm}
\caption{Theoretical predictions for CMB temperature angular power
spectra as a function of multipole number $l$ for models with primordial
adiabatic perturbations. Each graph shows the effect of a variation in
one of these parameters. From Ref.~\cite{KK}.}
\label{fig35}
\end{figure}

\subsubsection{The acoustic peaks}

The Sachs-Wolfe plateau is a distinctive feature of Fig.~\ref{fig34}.
These observations confirm the existence of a primordial spectrum of
scalar (density) perturbations on all scales, otherwise the power
spectrum would have started from zero at $l=2$. However, we see that the
spectrum starts to rise around $l=20$ towards the first acoustic peak,
where the SW approximation breaks down and the above formulae are no
longer valid.  

\begin{figure}[htb]
\hspace*{5mm}
\includegraphics[width=7.1cm]{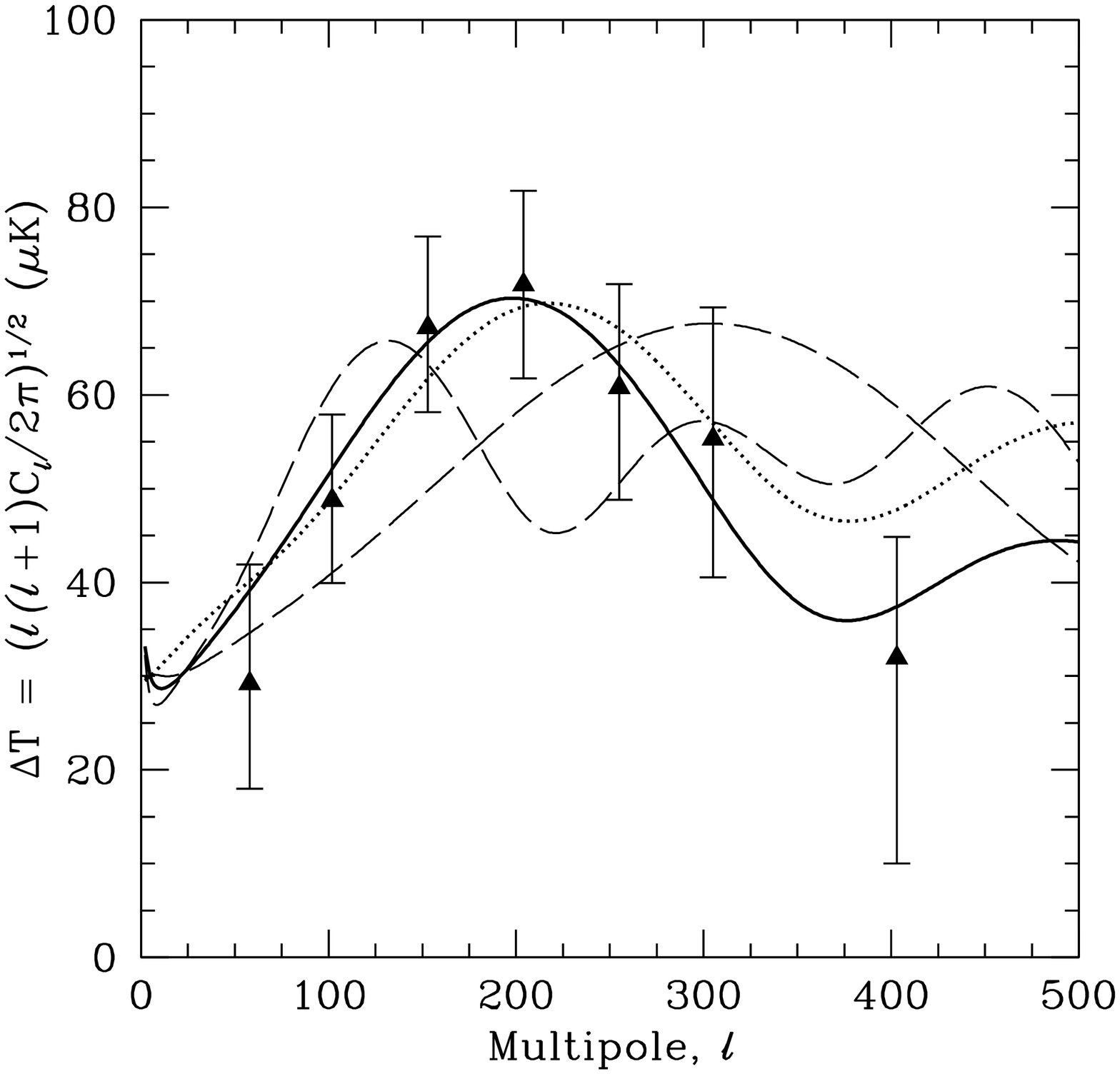}
\includegraphics[width=7.3cm]{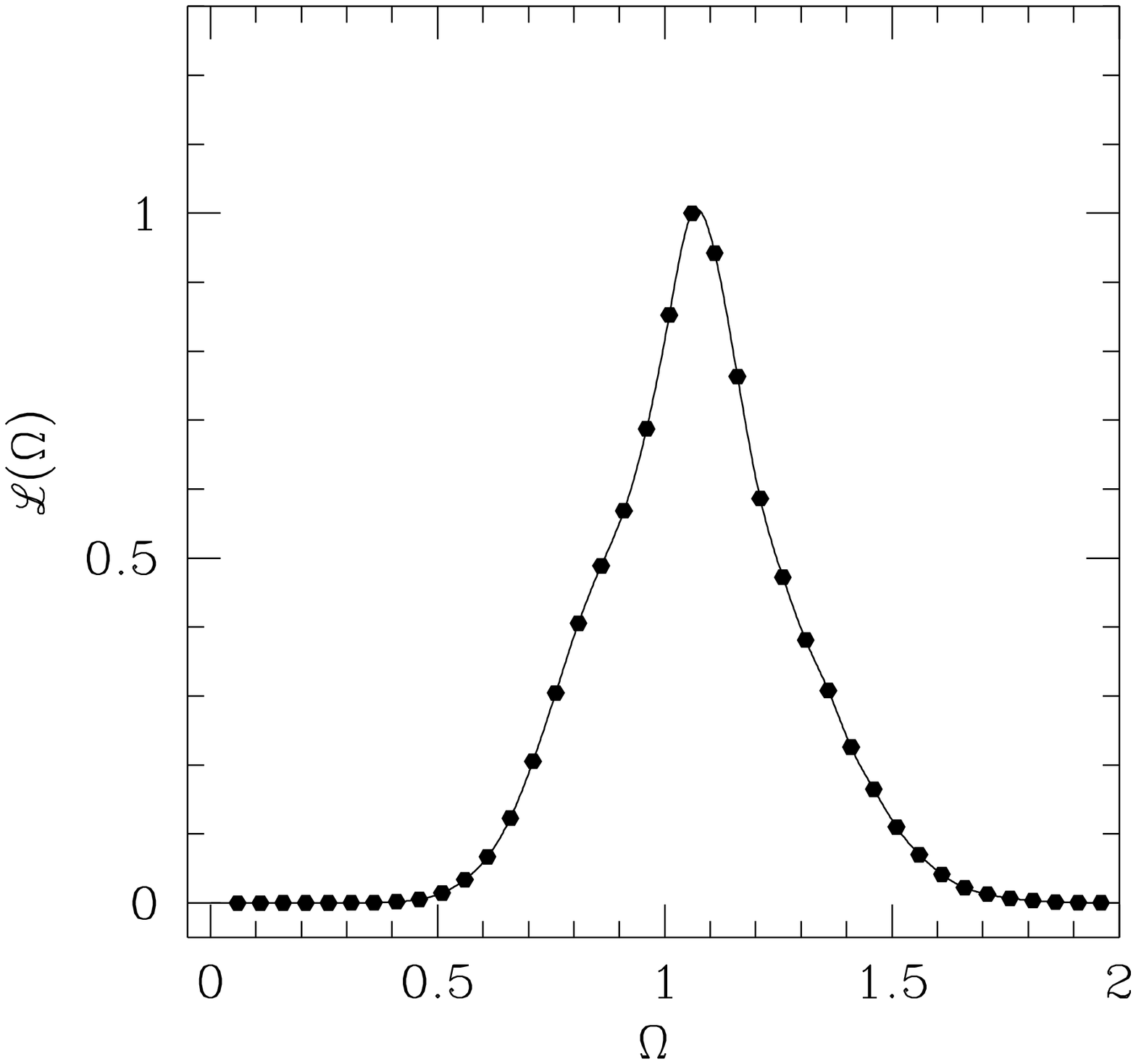}
\vspace*{-.2cm}
\caption{The left figure shows the power spectrum of the BOOMERANG
experiment with 6 arcminute pixelization. The solid curve is a
marginally closed model with $(\Omega_{\rm B}, \Omega_{\rm M},
\Omega_\Lambda, n, h) = (0.05, 0.26, 0.75, 0.95, 0.7)$. The dotted curve
is Standard CDM with $(0.05, 0.95, 0.0, 1.0, 0.65)$. The dashed curves
are open and closed models with fixed $\Omega_0 = \Omega_{\rm M} +
\Omega_\Lambda = 0.66$ and 1.55, respectively. The right figure shows
the likelihood function of $\Omega_0$ normalized to unity at the peak,
after marginalizing over the $\Omega_{\rm M}-\Omega_\Lambda$
direction. From Ref.~\cite{Boomerang}.}
\label{fig36}
\end{figure}

As mentioned above, the first peak in the photon distribution
corresponds to overdensities that have undergone half an oscillation,
that is, a compression, and appear at a scale associated with the size
of the horizon at last scattering, about $1^\circ$ projected in the sky
today. Since photons scatter off baryons, they will also feel the
acoustic wave and create a peak in the correlation function. The height
of the peak is proportional to the amount of baryons: the larger the
baryon content of the universe, the higher the peak. The position of the
peak in the power spectrum depends on the geometrical size of the
particle horizon at last scattering. Since photons travel along
geodesics, the projected size of the causal horizon at decoupling
depends on whether the universe is flat, open or closed. In a flat
universe the geodesics are straight lines and, by looking at the angular
scale of the first acoustic peak, we would be measuring the actual size
of the horizon at last scattering. In an open universe, the geodesics
are inward-curved trajectories, and therefore the projected size on the
sky appears smaller. In this case, the first acoustic peak should occur
at higher multipoles or smaller angular scales. On the other hand, for a
closed universe, the first peak occurs at smaller multipoles or larger
angular scales. The dependence of the position of the first acoustic
peak on the spatial curvature can be approximately given by~\cite{CMB}
\begin{equation}
l_{\rm peak} \simeq 220\,\Omega_0^{-1/2}\,,
\label{lPeak}
\end{equation}
where $\Omega_0=\Omega_{\rm M} + \Omega_\Lambda = 1-\Omega_K$.  Present
observations, specially the ones of the Mobile Anisotropy Telescope
(MAT) in Cerro Tololo, Chile, which produced two data sets, TOCO97 and
TOCO98~\cite{TOCO}, and the recent balloon-borne experiment
BOOMERANG~\cite{Boomerang}, suggest that the peak is between $l=180$ and
250 at 95\% c.l., with an amplitude $\delta T = 80\pm10\ \mu$K, and
therefore the universe is most probably flat, see Fig.~\ref{fig36}, and
Ref.~\cite{KP}. In particular, these measuremts determine that
\begin{equation}
0.85 \leq \Omega_0 \leq 1.25 \hspace{5mm}(68\% \ {\rm c.l.})
\label{Omega0}
\end{equation}
That is, the universe is flat, within 10\% uncertainty, which is much
better than we could ever do before. In the near future we will
measure $\Omega_0$ to within 1\%, with the new microwave anisotropy
satellites.

At the moment there is not enough information at small angular scales,
or large multipole numbers, to determine the existence or not of the
secondary acoustic peaks. These peaks should occur at harmonics of the
first one, but are typically much lower because of Silk damping. Since
the amplitude and position of the primary and secondary peaks are
directly determined by the sound speed (and, hence, the equation of
state) and by the geometry and expansion of the universe, they can be
used as a powerful test of the density of baryons and dark matter, and
other cosmological parameters, see Fig.~\ref{fig35}.

By looking at these patterns in the anisotropies of the microwave
background, cosmologists can determine not only the cosmological
parameters, see Fig.~\ref{fig35}, but also the primordial spectrum of
density perturbations produced during inflation. It turns out that the
observed temperature anisotropies are compatible with a scale-invariant
spectrum, see Eq.~(\ref{MeasuredTilt}), as predicted by inflation. This
is remarkable, and gives very strong support to the idea that inflation
may indeed be responsible for both the CMB anisotropies and the
large-scale structure of the universe. Different models of inflation
have different specific predictions for the fine details associated with
the spectrum generated during inflation. It is these minute differences
that will allow cosmologists to differentiate between alternative models
of inflation and discard those that do not agree with observations.
However, most importantly, perhaps, the pattern of anisotropies
predicted by inflation is completely different from those predicted by
alternative models of structure formation, like cosmic defects: strings,
vortices, textures, etc. These are complicated networks of energy
density concentrations left over from an early universe phase
transition, analogous to the defects formed in the laboratory in certain
kinds of liquid crystals when they go through a phase transition. The
cosmological defects have spectral properties very different from those
generated by inflation. That is why it is so important to launch more
sensitive instruments, and with better angular resolution, to determine
the properties of the CMB anisotropies.

\begin{figure}[htb]
\begin{center}\hspace{-1mm}
\includegraphics[width=7.9cm,height=5cm]{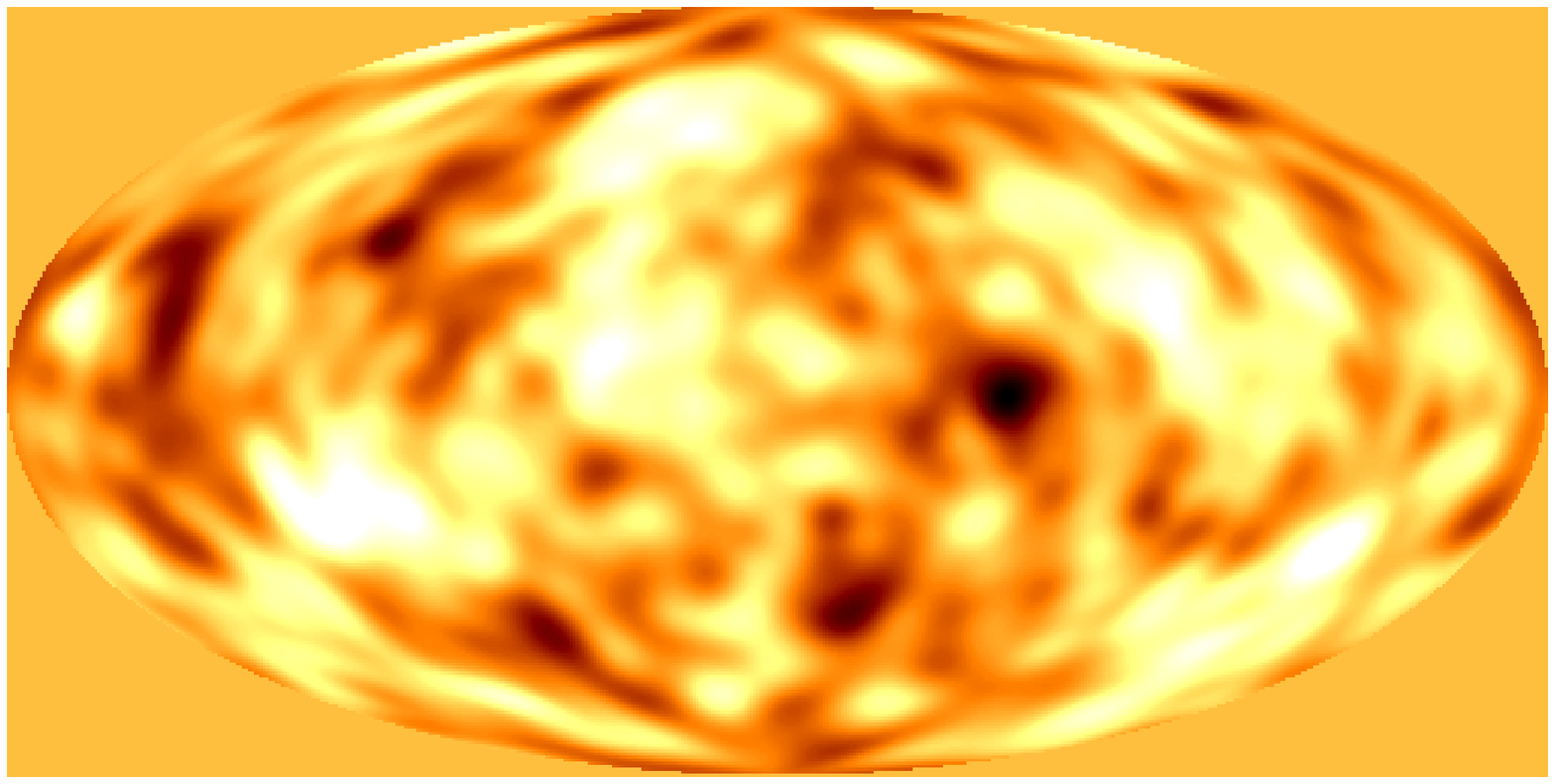}
\includegraphics[width=7.9cm,height=5cm]{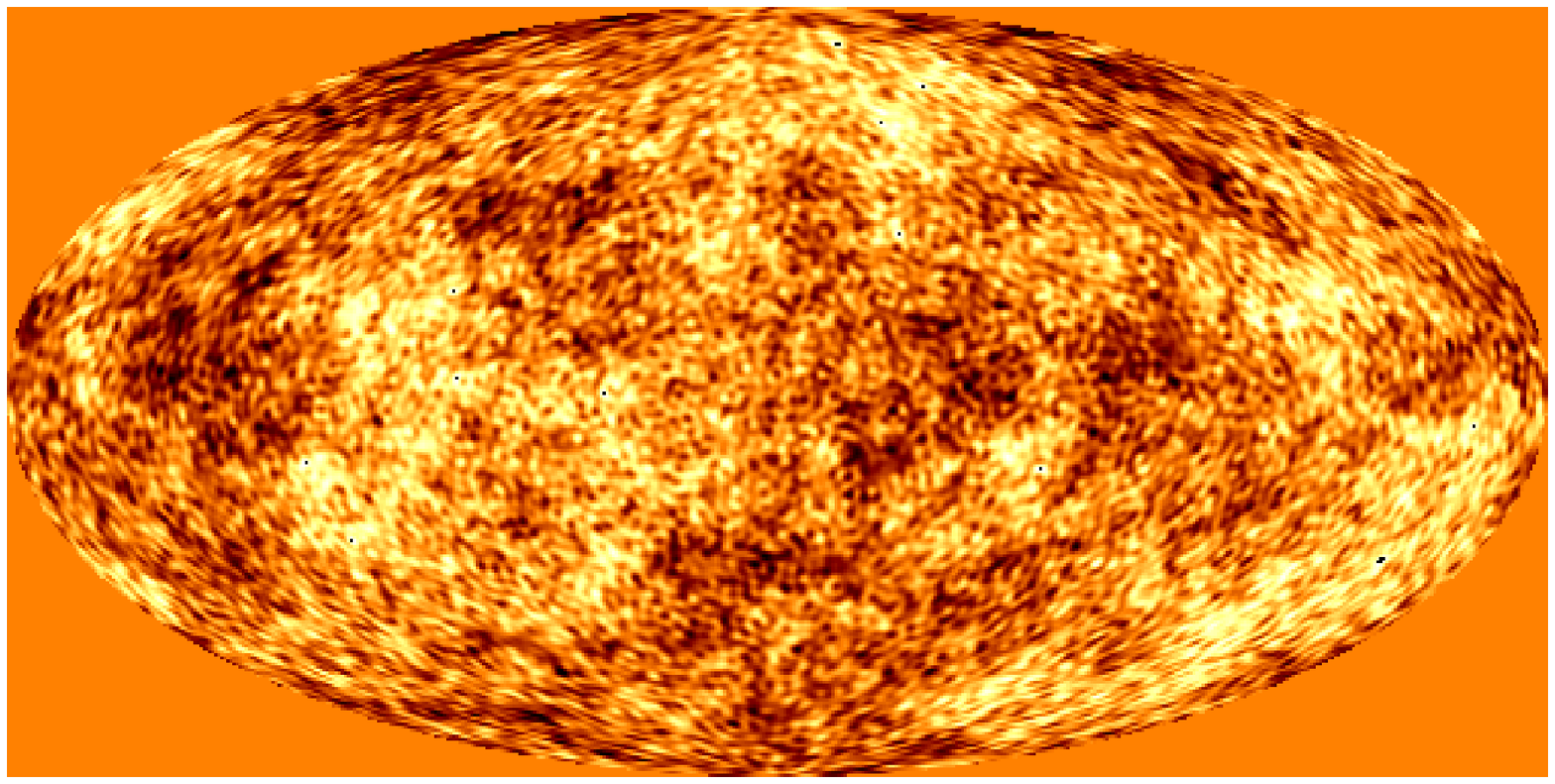}
\end{center}\vspace*{-3mm}
\caption{The left figure shows a simulation of the temperature
anisotropies predicted by a generic model of inflation, as would be seen
by a satellite like COBE with angular resolution of $7^\circ$. The right
figure shows the same, but with a satellite like Planck, with a
resolution 100 times better. From Ref.~\cite{Planck}.}
\label{fig37}
\end{figure}

\subsubsection{The new microwave anisotropy satellites, MAP and Planck}

The large amount of information encoded in the anisotropies of the
microwave background is the reason why both NASA and the European Space
Agency have decided to launch two independent satellites to measure the
CMB temperature and polarization anisotropies to unprecendented
accuracy. The Microwave Anisotropy Probe~\cite{MAP} will be launched by
NASA at the end of 2000, and Planck~\cite{Planck} is expected in 2007.

As we have emphasized before, the fact that these anisotropies have such
a small amplitude allow for an accurate calculation of the predicted
anisotropies in linear perturbation theory. A particular cosmological
model is characterized by a dozen or so parameters: the rate of
expansion, the spatial curvature, the baryon content, the cold dark
matter and neutrino contribution, the cosmological constant (vacuum
energy), the reionization parameter (optical depth to the last
scattering surface), and various primordial spectrum
parameters like the amplitude and tilt of the adiabatic and isocurvature
spectra, the amount of gravitational waves, non-Gaussian effects,
etc. All these parameters can now be fed into a fast code called
CMBFAST~\cite{CMBFAST} that computes the predicted temperature and
polarization anisotropies to 1\% accuracy, and thus can be used to
compare with observations.

\begin{figure}[htb]
\vspace*{-1.5cm}
\hspace{-2mm}
\includegraphics[width=8cm]{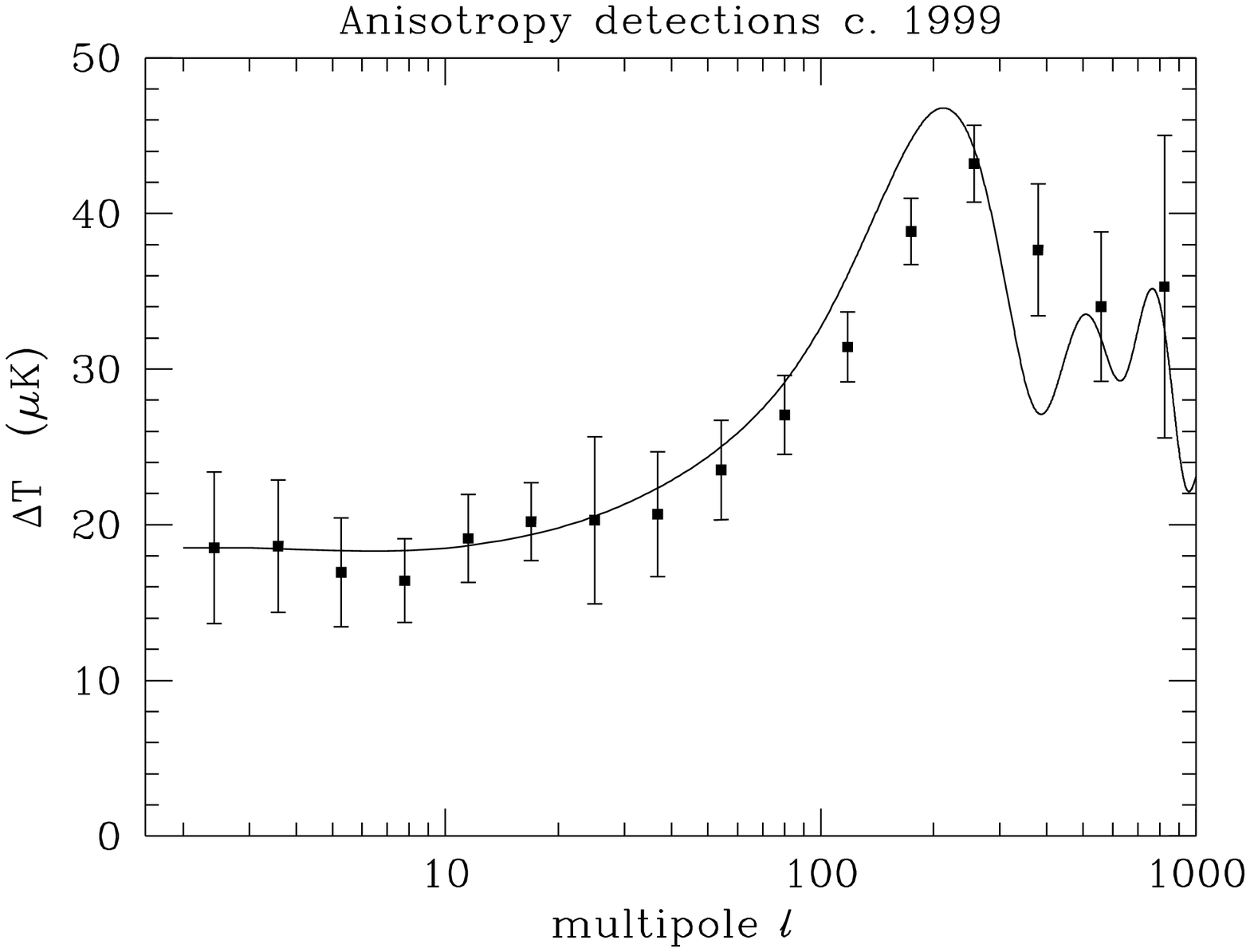}
\includegraphics[width=8cm]{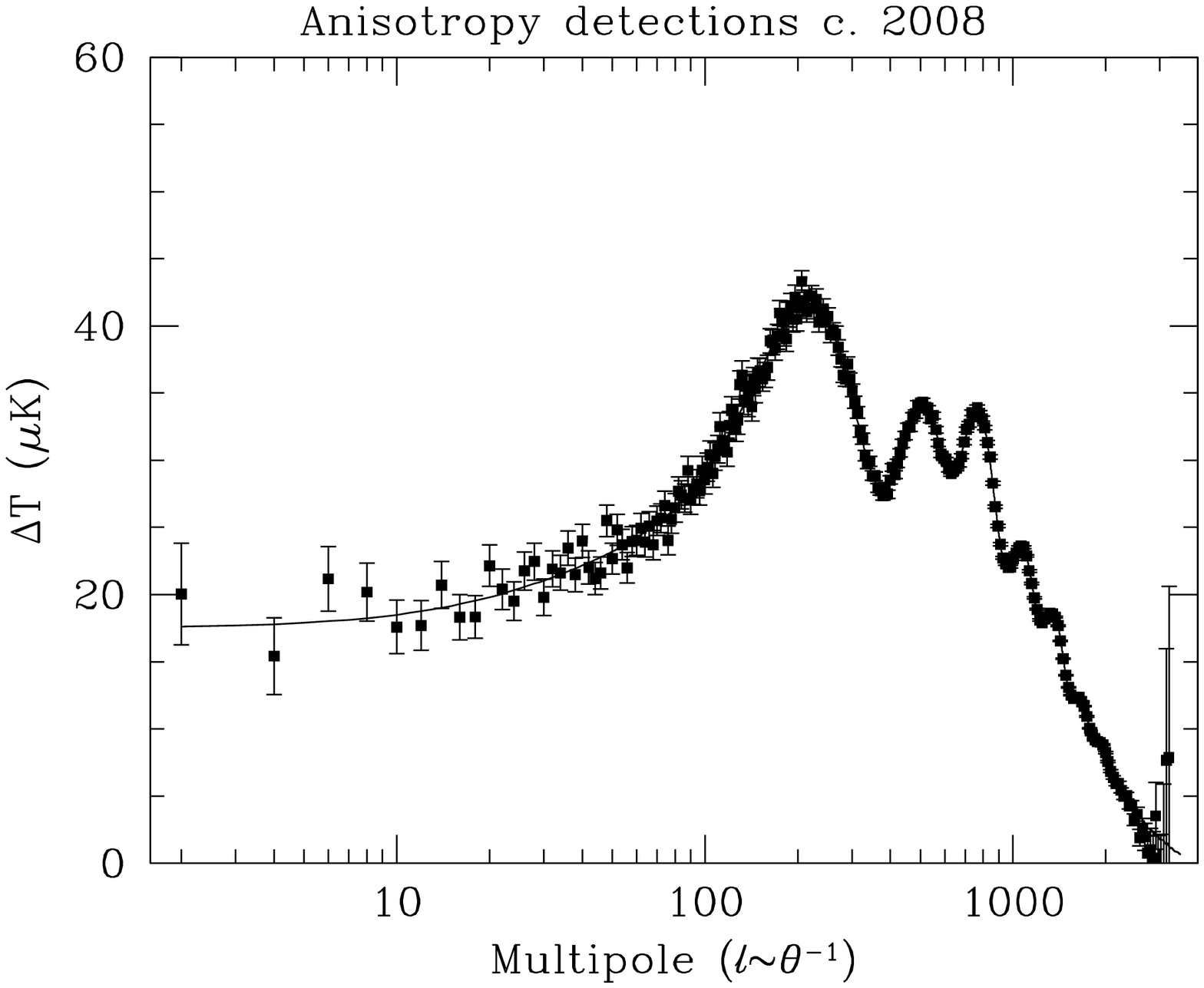}
\vspace*{-8mm}
\caption{The predicted angular power spectrum of temperature
anisotropies, compared with the present data, see Fig.~\ref{fig34},
binned into 16 logarithmic intervals in multipole number between $l=2$
and $l=1000$. The right figure gives an estimate of the accuracy with
which the power spectrum will be measured by Planck. It is only limited
by cosmic variance on all the angular scales relevant to primary
anisotropies. From Ref.~\cite{Scott}.}
\label{fig38}
\end{figure}

These two satellites will improve both the sensitivity, down to $\mu$K,
and the resolution, down to arc minutes, with respect to the previous
COBE satellite, thanks to large numbers of microwave horns of various
sizes, positioned at specific angles, and also thanks to recent advances
in detector technology, with high electron mobility transistor
amplifiers (HEMTs) for frequencies below 100 GHz and bolometers for
higher frequencies. The primary advantage of HEMTs is their ease of use
and speed, with a typical sensitivity of 0.5 mKs$^{1/2}$, while the
advantage of bolometers is their tremendous sensitivity, better than 0.1
mKs$^{1/2}$, see Ref.~\cite{Page}. For instance, to appreciate the
difference, compare the resolution in the temperature anisotropies that
COBE and Planck would observe for the same simulated sky in
Fig.~\ref{fig37}. This will allow cosmologists to extract information
from around 3000 multipoles! Since most of the cosmological parameters
have specific signatures in the height and position of the first few
acoustic peaks, the higher the resolution, the more peaks one is
expected to see, and thus the better the accuracy with which one will be
able to measure those parameters, see Table~1. As an example of the kind
of data that these two satellites will be able to provide, see
Fig.~\ref{fig38}, which compares the present observational status with
that which will become available around 2008.

\begin{figure}[htb]
\vspace*{-4cm}
\hspace{1cm}
\includegraphics[width=13cm]{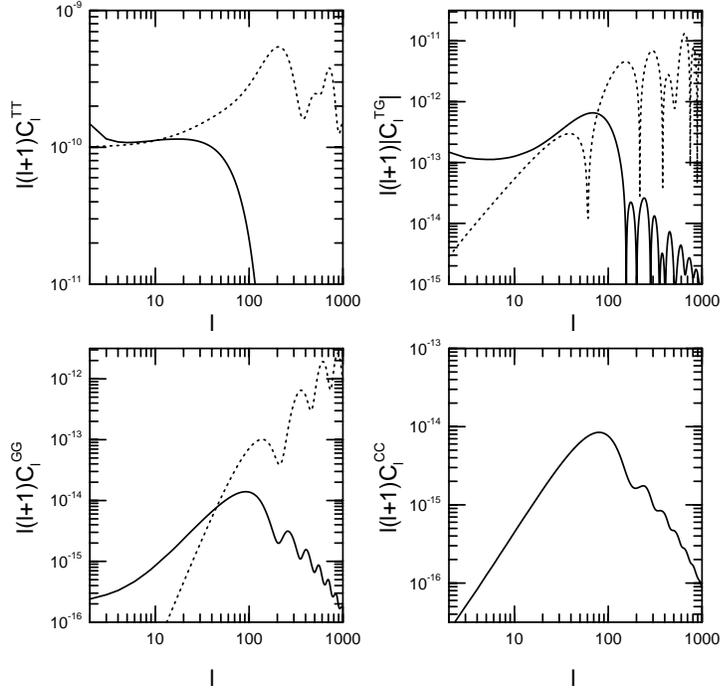}
\vspace*{-2mm}
\caption{Theoretical predictions for the four non-zero CMB
temperature-polarization spectra as a function of multipole moment. The
dotted curves are the predictions for a COBE-normalized scalar
perturbation from an inflationary model with no reionization and no
gravitational waves for $h=0.65$, $\Omega_{\rm B}h^2=0.024$, and
$\Lambda=0$. The solid curves are the corresponding predictions if the
COBE anisotropy were entirely due to a stochastic gravitational wave
background with a flat scale-invariant spectrum (with the same
cosmological parameters). The panel for $C_l^{\rm CC}$ contains no
dotted curve because scalar perturbations produce no curl component of
the polarization vector. From Ref.~\cite{KK}.}
\label{fig39}
\end{figure}

Although the satellite probes were designed for the accurate measurement
of the CMB temperature anisotropies, there are other experiments, like
balloon-borne and ground interferometers, which will probably accomplish
the same results with similar resolution (in the case of MAP), before
the satellites start producing their own results~\cite{Page}. Probably
the most important objective of the future satellites will be the
measurement of the CMB polarization anisotropies, yet to be discovered.
These anisotropies are predicted by models of structure formation and
are expected to arise at the level of microKelvin sensitivities, where
the new satellites are aiming at. The complementary information
contained in the polarization anisotropies will provide much more
stringent constraints on the cosmological parameters than from the
temperature anisotropies alone. In particular, the curl-curl component
of the polarization power spectra is nowadays the only means we have to
determine the tensor (gravitational wave) contribution to the metric
perturbations responsible for temperature anisotropies, see
Fig.~\ref{fig39}. If such a component is found, one could constraint
very precisely the model of inflation from its spectral properties,
specially the tilt~\cite{KK}.

\begin{table}[htb]
\begin{center}
\begin{tabular}{|l|c|c|c|c|}
\hline
{\bf physical quantity} & {\bf symbol} & 
{\bf present range} & {\bf MAP} & {\bf Planck} \\
\hline 
\hline 
luminous matter & $\Omega_{\rm lum}h^2$ & $0.001 - 0.005$ & $-$ & $-$ \\
\hline
baryonic matter & $\Omega_{\rm B}h^2$ & $0.01 - 0.03$ & 5\% & 0.6\% \\
\hline
cold dark matter & $\Omega_{\rm M}h^2$ & $0.2 - 1.0$ & 10\% & 0.6\% \\
\hline
hot dark matter & $\Omega_\nu h^2$ & $0 - 0.3$ & 5\% & 2\% \\
\hline
cosmological constant & $\Omega_\Lambda h^2$ & $0 - 0.8$ & 8\% & 0.5\% \\
\hline
spatial curvature & $\Omega_0 h^2$ & $0.2 - 1.5$ & 4\% & 0.7\% \\
\hline
rate of expansion & $h$ & $0.4 - 0.8$ & 11\% & 2\% \\
\hline
age of the universe & $t_0$ & $11 - 17$ Gyr & 10\% & 2\% \\
\hline
spectral amplitude & $Q_{\rm rms}$ & $20 - 30\ \mu$K & 0.5\% & 0.1\% \\
\hline
spectral tilt & $n_{_{\rm S}}$ & $0.5 - 1.5$ & 3\% & 0.5\% \\
\hline
tensor-scalar ratio & $r_{\rm ts}$ & $0 - 1.0$ & 25\% & 10\% \\
\hline
reionization & $\tau$ & $0.01 - 1.0$ & 20\% & 15\% \\
\hline
\end{tabular}
\end{center}
\caption{ 
{\bf The parameters of the standard cosmological model}.
The standard model of cosmology has around 12 different parameters,
needed to describe the background space-time, the matter content and the
spectrum of density perturbations. We include here the present range of
the most relevant parameters, and the percentage error with which the
microwave background probes MAP and Planck (without polarization) will
be able to determine them in the near future.
The rate of expansion is in units of $H_0=100\,h$ km/s/Mpc.}
\label{table1}
\end{table}

\subsection{From metric perturbations to large scale structure}

If inflation is responsible for the metric perturbations that gave rise
to the temperature anisotropies observed in the microwave background,
then the primordial spectrum of density inhomogeneities induced by the
same metric perturbations should also be responsible for the present
large scale structure~\cite{LiddleLyth}. This simple connection allows
for more stringent tests on the inflationary paradigm for the generation
of metric perturbations, since it relates the large scales (of order the
present horizon) with the smallest scales (on galaxy scales). This
provides a very large lever arm for the determination of primordial
spectra parameters like the tilt, the nature of the perturbations,
whether adiabatic or isocurvature, the geometry of the universe, as well
as its matter and energy content, whether CDM, HDM or mixed CHDM.

\subsubsection{The galaxy power spectrum}

As metric perturbations enter the causal horizon during the radiation or
matter era, they create density fluctuations via gravitational
attraction of the potential wells. The density contrast $\delta$ can be
deduced from the Einstein equations in linear perturbation theory, see
Eq.~(\ref{PertEinsteinEqs}),
\begin{equation}\label{deltaRc}
\delta_k \equiv {\delta\rho_k\over\rho} = 
\left({k\over aH}\right)^2\,{2\over3}\,\Phi_k =
\left({k\over aH}\right)^2\,{2+2\omega\over5+3\omega}\,{\cal R}_k\,,
\end{equation}
where we have assumed $K=0$, and used Eq.~(\ref{PhiRc}). From this
expression one can compute the power spectrum, at horizon crossing, of
matter density perturbations induced by inflation, see
Eq.~(\ref{CorrelatorRc}),
\begin{equation}
P(k) = \langle|\delta_k|^2\rangle = A\,\left({k\over aH}\right)^n\,,
\end{equation}
with $n$ given by the scalar tilt (\ref{ScalarTilt}), $n = 1 + 2\eta -
6\epsilon$. This spectrum reduces to a Harrison-Zel'dovich spectrum 
(\ref{HarrisonZeldovich}) in the slow-roll approximation: \ $\eta, \,
\epsilon \ll 1$.

Since perturbations evolve after entering the horizon, the power
spectrum will not remain constant. For scales entering the horizon well
after matter domination ($k^{-1} \gg k^{-1}_{\rm eq} \simeq 81$ Mpc),
the metric perturbation has not changed significantly, so that ${\cal
R}_k({\rm final}) = {\cal R}_k({\rm initial})$. Then Eq.~(\ref{deltaRc})
determines the final density contrast in terms of the initial one. On
smaller scales, there is a linear transfer function $T(k)$, which may be
defined as~\cite{LL93}
\begin{equation}\label{TransferFunction}
{\cal R}_k({\rm final}) = T(k)\,{\cal R}_k({\rm initial})\,.
\end{equation}
To calculate the transfer function one has to specify the initial
condition with the relative abundance of photons, neutrinos, baryons and
cold dark matter long before horizon crossing. The most natural
condition is that the abundances of all particle species are uniform on
comoving hypersurfaces (with constant total energy density). This is
called the {\em adiabatic} condition, because entropy is conserved
independently for each particle species $X$, i.e. $\delta\rho_X =
\dot\rho_X\delta t$, given a perturbation in time from a comoving
hypersurface, so
\begin{equation}\label{AdiabaticCondition}
{\delta\rho_X\over\rho_X+p_X} = {\delta\rho_Y\over\rho_Y+p_Y}\,,
\end{equation}
where we have used the energy conservation equation for each species,
$\dot\rho_X=-3H(\rho_X+p_X)$, valid to first order in perturbations. It
follows that each species of radiation has a common density contrast
$\delta_r$, and each species of matter has also a common density
contrast $\delta_m$, with the relation \ $\delta_m={3\over4}\delta_r$.

Within the horizon, the density perturbation amplitude evolves according
to the following equation, see Ref.~\cite{LL93},
\begin{equation}\label{EvolutionDensityContrast}
H^{-2}\ddot\delta_k + [2-3(2\omega-c_s^2)]\,H^{-1}\dot\delta_k-{3\over2}\,
(1-6c_s^2+8\omega-3\omega^2)\,\delta_k = -\left({k\over aH}\right)^2
{\delta p_k\over\rho}\,,
\end{equation}
where \ $\omega=p/\rho$ \ is the barotropic ratio, and \ $c_s^2 = 
\dot p/\dot\rho$ \ is the speed of sound of the fluid. 

Given the adiabatic condition, the transfer function is determined by
the physical processes occuring between horizon entry and matter
domination. If the radiation behaves like a perfect fluid, its density
perturbation oscillates during this era, with decreasing amplitude. The
matter density contrast living in this background does not grow
appreciably before matter domination because it has negligible
self-gravity. The transfer function is therefore given roughly by,
see Eq.~(\ref{PowerSpectrum}),
\begin{equation}\label{Tk}
T(k) = \left\{\begin{array}{ll}
1\,,&\hspace{1cm}k\ll k_{\rm eq}\\[2mm]
(k/k_{\rm eq})^2\,,&
\hspace{1cm}k\gg k_{\rm eq}\end{array}\right.
\end{equation}

The perfect fluid description of the radiation is far from being correct
after horizon entry, because roughly half of the radiation consists of
neutrinos whose perturbation rapidly disappears through free
streeming. The photons are also not a perfect fluid because they diffuse
significantly, for scales below the Silk scale, $k_S^{-1} \sim 1$ Mpc.
One might then consider the opposite assumption, that the radiation has
zero perturbation after horizon entry. Then the matter density
perturbation evolves according to Eq.~(\ref{EvolutionDensityContrast}),
with $\delta$ and $\rho$ now referring to the matter alone,
\begin{equation}\label{EvolutionDensity}
\ddot\delta_k + 2H\dot\delta_k + (c_s^2\,k^2_{\rm ph}-4\pi G\rho)\,
\delta_k = 0\,,
\end{equation}
which corresponds to the equation of a damped harmonic oscillator.  The
zero-frequency oscillator defines the Jeans wavenumber, $k_J =
\sqrt{4\pi G\rho/c_s^2}$. For $k\ll k_J$, $\delta_k$ grows exponentially
on the dynamical timescale, $\tau_{\rm dyn} = {\rm Im}\,\omega^{-1} =
(4\pi G\rho)^{-1/2} = \tau_{\rm grav}$, which is the time scale for
gravitational collapse. One can also define the Jeans length,
\begin{equation}\label{JeansLength}
\lambda_J = {2\pi\over k_J} = c_s\,\sqrt{\pi\over G\rho}\,,
\end{equation}
which separates gravitationally stable from unstable modes. If we define
the pressure response timescale as the size of the perturbation over the
sound speed, $\tau_{\rm pres} \sim \lambda/c_s$, then, if $\tau_{\rm pres} 
> \tau_{\rm grav}$, gravitational collapse of a perturbation can occur
before pressure forces can response to restore hydrostatic equilibrium
(this occurs for $\lambda > \lambda_J$). On the other hand, if
$\tau_{\rm pres} < \tau_{\rm grav}$, radiation pressure prevents
gravitational collapse and there are damped acoustic oscillations
(for $\lambda < \lambda_J$).

We will consider now the behaviour of modes within the horizon during
the transition from the radiation ($c_s^2=1/3$) to the matter era
($c_s^2=0$). The growing and the decaying solutions of
Eq.~(\ref{EvolutionDensity}) are
\begin{eqnarray}\label{GrowingSolution}
&&\delta=A\,\Big(1+{3\over2}\,y\Big)\,,\\[2mm]
&&\delta=B\,\left[\Big(1+{3\over2}\,y\Big)\,\ln{\sqrt{1+y}+1\over
\sqrt{1+y}-1} - 3\,\sqrt{1+y}\right]\,,
\end{eqnarray}
where $A$ and $B$ are constants, and \ $y=a/a_{\rm eq}$. The growing mode
solution (\ref{GrowingSolution}) increases only by a factor of 2 between
horizon entry and the epoch when matter starts to dominate, i.e. $y=1$.
The transfer function is therefore again roughly given by Eq.~(\ref{Tk}).

\begin{figure}[htb]
\begin{center}
\hspace{-5mm}
\includegraphics[width=9cm]{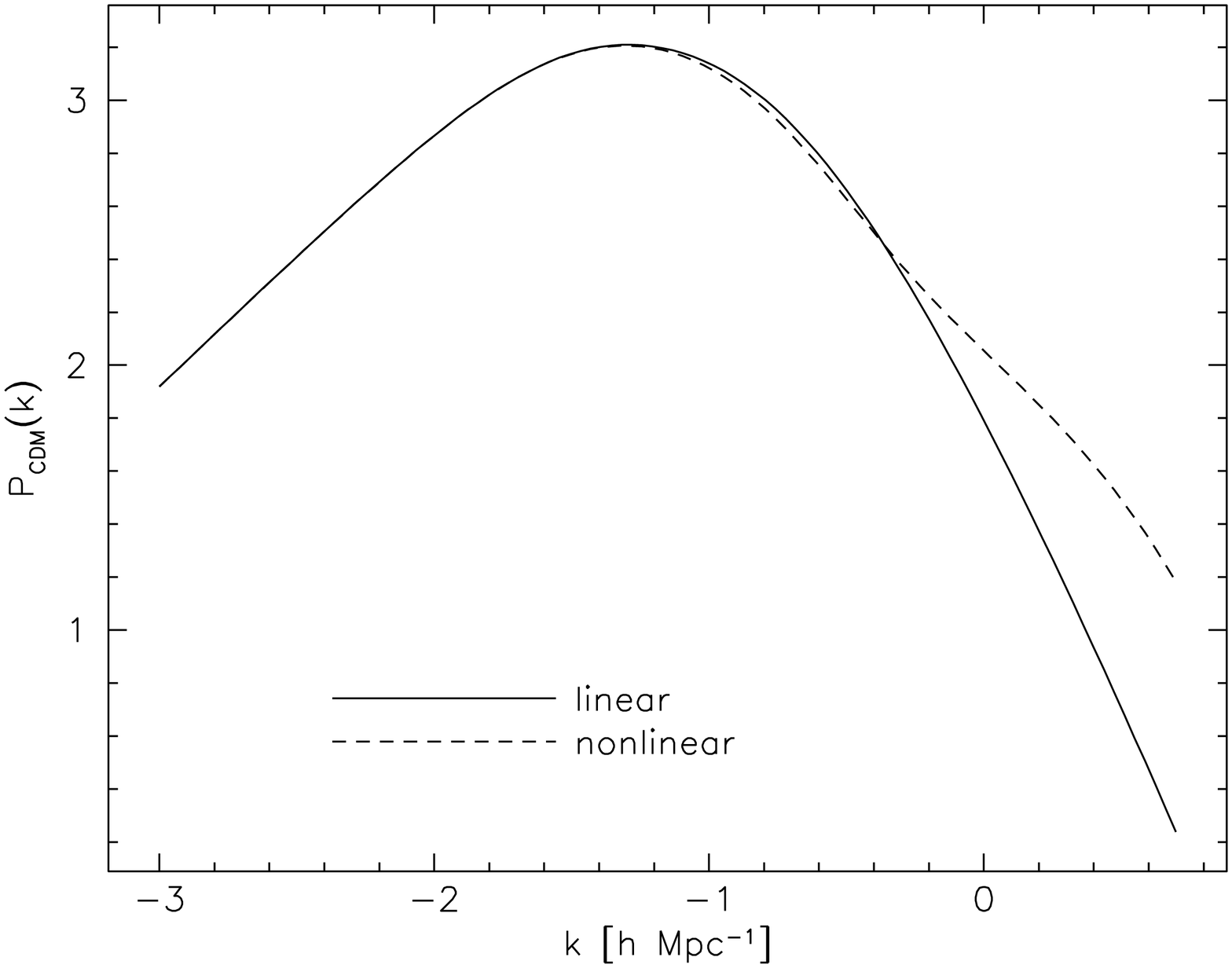}
\end{center}
\vspace*{-8mm}
\caption{The CDM power spectrum $P(k)$ as a function of wavenumber $k$,
in logarithmic scale, normalized to the local abundance of galaxy
clusters, for an Einstein-de Sitter universe with $h=0.5$. The solid
(dashed) curve shows the linear (non-linear) power spectrum. While the
linear power spectrum falls off like $k^{-3}$, the non-linear
power-spectrum illustrates the increased power on small scales due to
non-linear effects, at the expense of the large-scale structures.  From
Ref.~\cite{Bartelmann}.}
\label{fig40}
\end{figure}

Since the radiation consists roughly half of neutrinos, which free
streem, and half of photons, which either form a perfect fluid or just
diffuse, neither the perfect fluid nor the free-streeming approximation
looks very sensible. A more precise calculation is needed, including:
neutrino free streeming around the epoch of horizon entry; the diffusion
of photons around the same time, for scales below Silk scale; the
diffusion of baryons along with the photons, and the establishment after
matter domination of a common matter density contrast, as the baryons
fall into the potential wells of cold dark matter. All these effects
apply separately, to first order in the perturbations, to each Fourier
component, so that a linear transfer function is produced. There are
several parametrizations in the literature, but the one which is more
widely used is that of Ref.~\cite{Bond},
\begin{eqnarray}\label{TFk}
&&T(k) = \left[1+\Big(ak+(bk)^{3/2}+(ck)^2\Big)^\nu
\right]^{-1/\nu}\,, \hspace{1cm} \nu=1.13\,,\\[2mm]
&&a=6.4\,(\Omega_{\rm M}h)^{-1}\,h^{-1} \ {\rm Mpc}\,, \\
&&b=3.0\,(\Omega_{\rm M}h)^{-1}\,h^{-1} \ {\rm Mpc}\,, \\
&&c=1.7\,(\Omega_{\rm M}h)^{-1}\,h^{-1} \ {\rm Mpc}\,.
\end{eqnarray}
We see that the behaviour estimated in Eq.~(\ref{Tk}) is roughly
correct, although the break at $k=k_{\rm eq}$ is not at all sharp, see
Fig.~\ref{fig40}. The transfer function, which encodes the soltion to
linear equations, ceases to be valid when the density contrast becomes
of order 1. After that, the highly nonlinear phenomenon of gravitational
collapse takes place, see Fig.~\ref{fig40}.

\subsubsection{The new redshift catalogs, 2dF and Sloan Digital Sky Survey}

Our view of the large-scale distribution of luminous objects in the
universe has changed dramatically during the last 25 years~\cite{Guzzo}:
from the simple pre-1975 picture of a distribution of field and cluster
galaxies, to the discovery of the first single superstructures and
voids, to the most recent results showing an almost regular web-like
network of interconnected clusters, filaments and walls, separating huge
nearly empty volumes. The increased efficiency of redshift surveys, made
possible by the development of spectrographs and -- specially in the
last decade -- by an enormous increase in multiplexing gain (i.e. the
ability to collect spectra of several galaxies at once, thanks to
fibre-optic spectrographs), has allowed us not only to do {\em
cartography} of the nearby universe, but also to statistically
characterize some of its properties, see Ref.~\cite{royalsoc}. At the
same time, advances in theoretical modeling of the development of
structure, with large high-resolution gravitational simulations coupled
to a deeper yet limited understanding of how to form galaxies within the
dark matter halos, have provided a more realistic connection of the
models to the observable quantities~\cite{Moore}. Despite the large
uncertainties that still exist, this has transformed the study of
cosmology and large-scale structure into a truly quantitative science,
where theory and observations can progress side by side.

\begin{figure}[htb]
\vspace*{-5mm}
\begin{center}
\includegraphics[width=7.9cm]{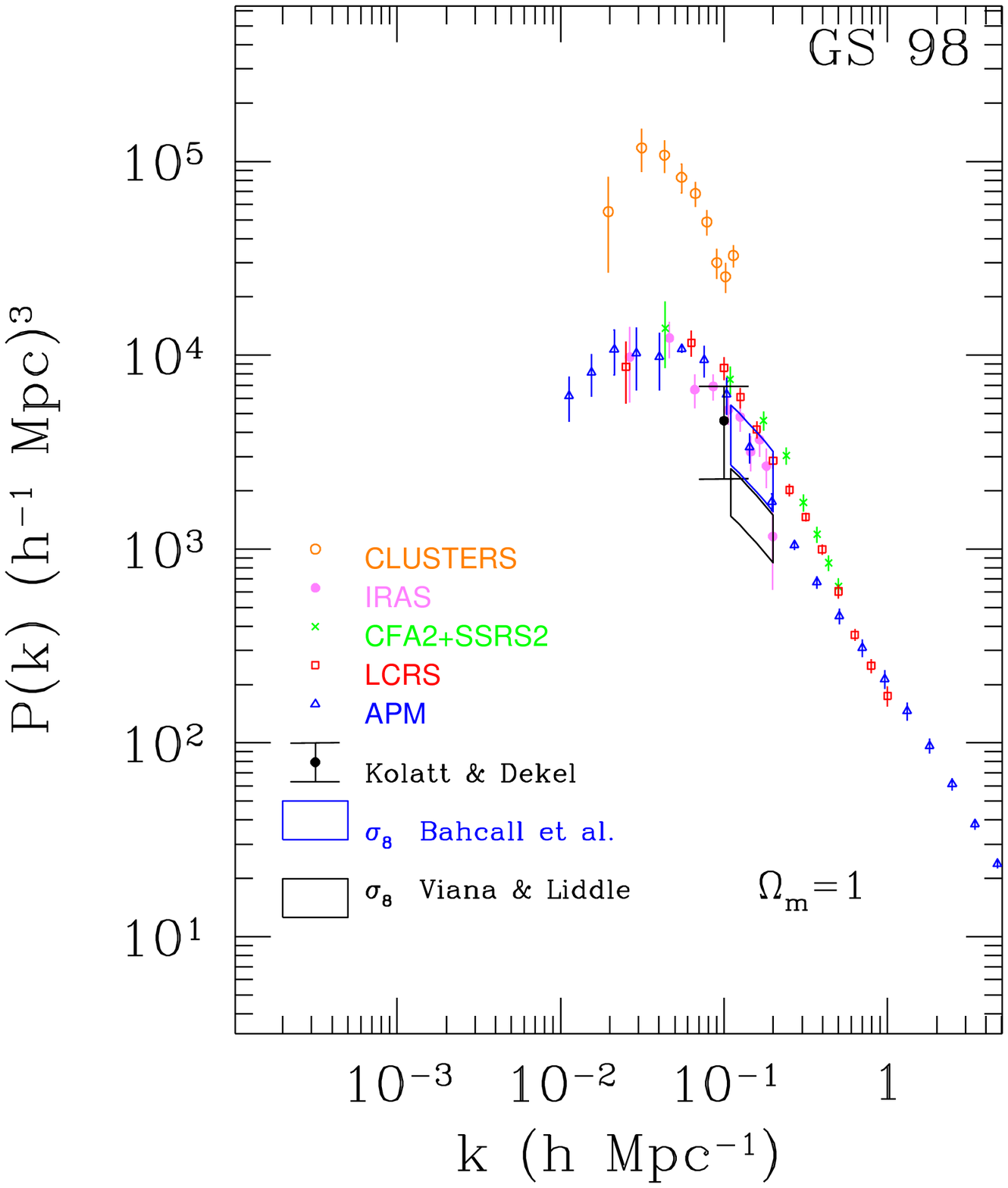}
\includegraphics[width=7.9cm]{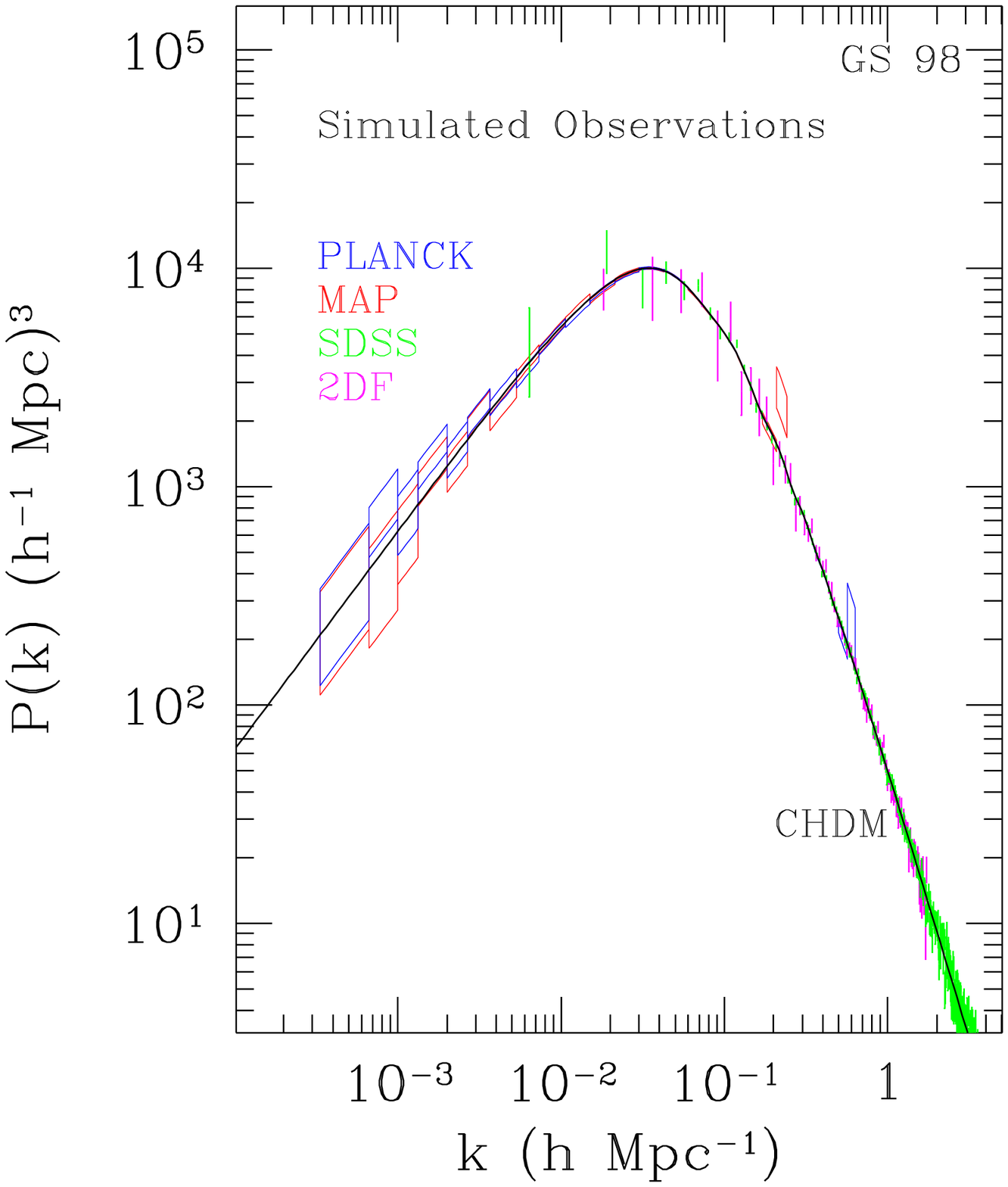}
\end{center}
\vspace*{-1cm}
\caption{Compilation of large-scale structure observations, showing the
power spectrum $P(k)$ as a function of wavenumber $k$. No corrections
for bias, redshift distortions, or non-linear evolution have been made.
Some of the redshift surveys have been rebinned to make the points
nearly independent. The black box comes from measurements of $\sigma_8$
from present-day number abundances of rich clusters, and the black point
with error bars is from peculiar velocities. The height shows the 68\%
confidence interval. $\Omega_{\rm M}=1$ is assumed. The right panel
shows a simulation of high-precision future CMB and LSS observations.
MAP (red boxes) and Planck (blue boxes) are simulated assuming that CHDM
is the correct model. Green error bars show the accuracy of the Sloan
Digital Sky Survey and magenta error bars are for the 2 Degree Field
Survey. No corrections are made for redshift distortions or non-linear
evolution. The simulated data are indistinguishable from the underlying
CHDM model for a wide range of $k$. From Ref.~\cite{Gawiser}.}
\label{fig41}
\end{figure}

For a review of the variety and details about the different existing
redshift catalogs, see Ref.~\cite{Guzzo}, and Fig.~\ref{fig41}. Here I
will concentrate on two of the new catalogs, which are taking data at
the moment and which will revolutionize the field, the 2-degree-Field
(2dF) Catalog and the Sloan Digital Sky Survey (SDSS). The advantages of
multi-object fibre spectroscopy have been pushed to the extreme with the
construction of the 2dF spectrograph for the prime focus of the
Anglo-Australian Telescope~\cite{2dF}. This instrument is able to
accommodate 400 automatically positioned fibres over a 2 degree in
diameter field. This implies a density of fibres on the sky of
approximately 130 deg$^{-2}$, and an optimal match to the galaxy counts
for a magnitude $b_J\simeq 19.5$, similar to that of previous surveys
like the ESP, with the difference that with such an area yield, the same
number of redshifts as in the ESP survey can be collected in about 10
exposures, or slightly more than one night of telescope time with
typical 1 hour exposures.  This is the basis of the 2dF galaxy redshift
survey. Its goal is to measure redshifts for more than 250,000 galaxies
with $b_J<19.5$.  In addition, a faint redshift survey of 10,000
galaxies brighter than $R=21$ will be done over selected fields within
the two main strips of the South and North Galactic Caps.  The survey is
steadily collecting redshifts, and there were about 93,000 galaxies
measured by January 2000. See also Ref.~\cite{2dF}, where the survey is
continuously updated.

\begin{figure}[htb]
\begin{center}
\includegraphics[width=7.8cm]{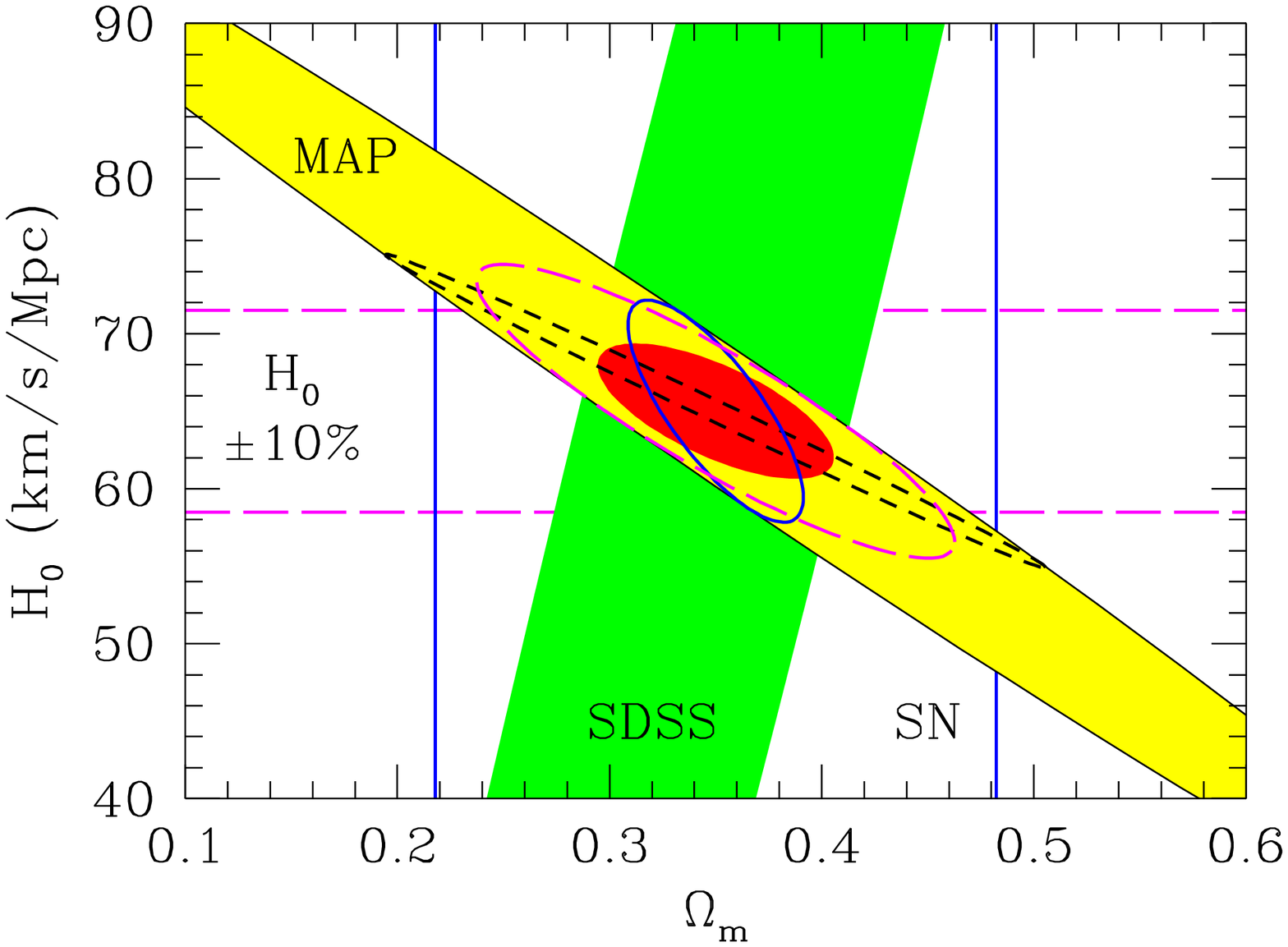}
\includegraphics[width=7.8cm]{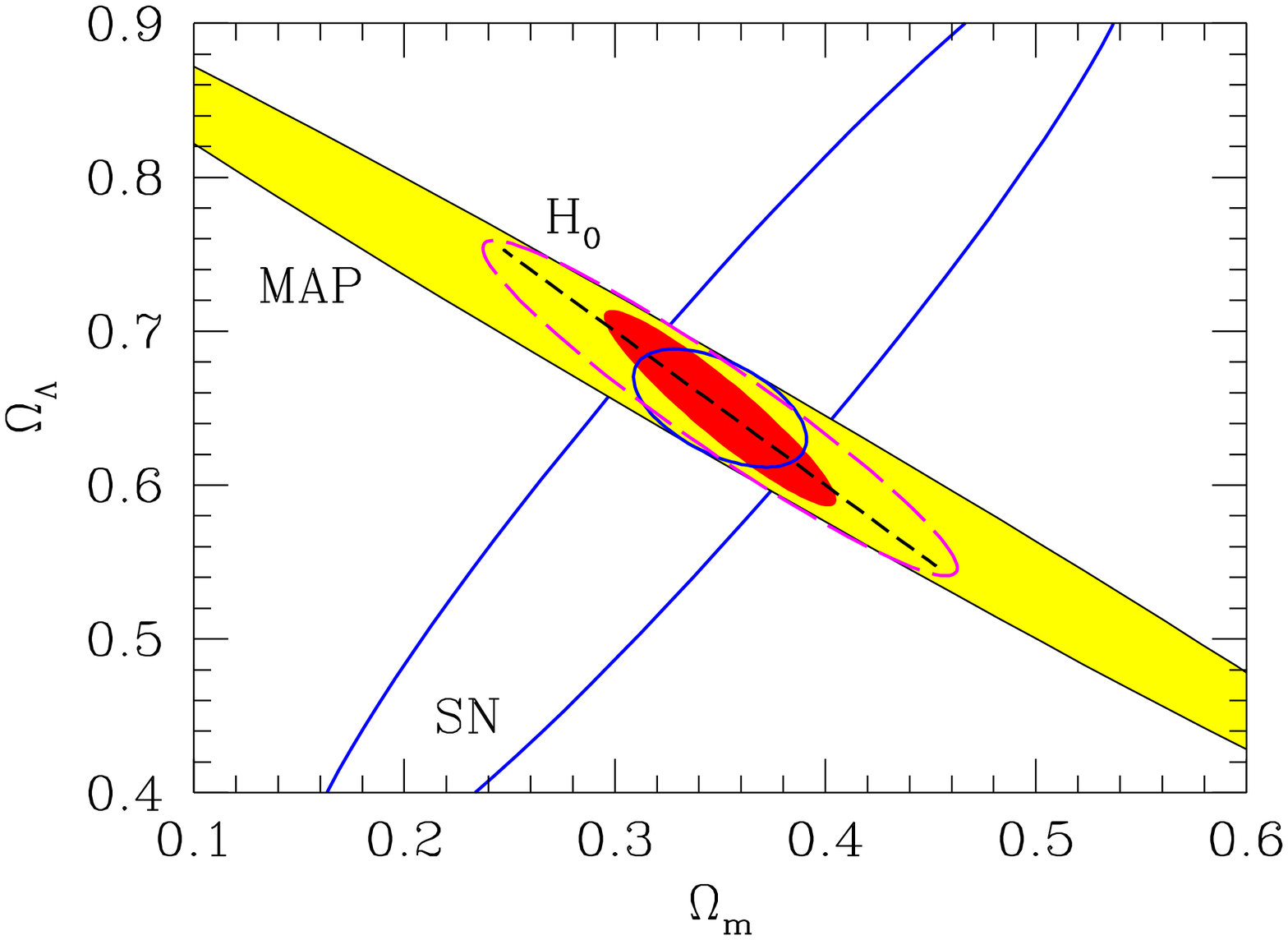}
\end{center}\vspace*{-5mm}
\caption{Constraint regions in the $\Omega_{\rm M}-H_0$ plane from 
various combinations of data sets. MAP data with polarization yields
the ellipse from upper left to lower right; assuming the universe flat
gives a smaller region (short-dashed line). SDSS ($k_{\rm max} = 0.1h$
Mpc$^{-1}$) gives the vertical shaded region; combined with MAP gives 
the small filled ellipse. A projecton of future supernovae Ia results
gives the solid vertical lines as bounds; combined with MAP gives the 
solid ellipse. A direct 10\% measurement of $H_0$ gives the long-dashed
lines and ellipse. All regions are 68\% confidence. The fiducial model 
is the $\Omega_{\rm M}=0.35$ flat $\Lambda$CDM model. The right figure
shows the same as before, but for constraints in the $\Omega_{\rm M}-
\Omega_\Lambda$ plane. From Ref.~\cite{Complementarity}.}
\label{fig42}
\end{figure}

The most ambitious and comprehensive galaxy survey currently in progress
is without any doubt the Sloan Digital Sky Survey~\cite{SDSS}. The aim
of the project is first of all to observe photometrically the whole
Northern Galactic Cap, 30$^\circ$ away from the galactic plane (about
$10^4$ deg$^2$) in five bands, at limiting magnitudes from 20.8 to
23.3. The expectation is to detect around 50 million galaxies and around
$10^8$ star-like sources. This has already led to the discovery of
several high-redshift ($z>4$) quasars, including the highest-redshift
quasar known, at $z=5.0$, see Ref.~\cite{SDSS}. Using two fibre
spectrographs carrying 320 fibres each, the spectroscopic part of the
survey will then collect spectra from about $10^6$ galaxies with $r'<18$
and $10^5$ AGNs with $r'<19$. It will also select a sample of about
$10^5$ red luminous galaxies with $r'<19.5$, which will be observed
spectroscopically, providing a nearly volume-limited sample of
early-type galaxies with a median redshift of $z\simeq0.5$, that will be
extremely valuable to study the evolution of clustering. The data
expected to arise from these new catalogs is so outstanding that already
cosmologists are making simulations and predicting what will be the
scientific outcome of these surveys, together with the future CMB
anisotropy probes, for the determination of the cosmological parameters
of the standard model of cosmology, see Figs.~\ref{fig41} and
\ref{fig42}.

As often happens in particle physics, not always are observations from a
single experiment sufficient to isolate and determine the precise value
of the parameters of the standard model. We mentioned in the previous
Section that some of the cosmological parameters created similar effects
in the temperature anisotropies of the microwave background. We say that
these parameters are {\em degenerate} with respect to the observations.
However, often one finds combinations of various
experiments/observations which break the degeneracy, for example by
depending on a different combination of parameters. This is precisely
the case with the cosmological parameters, as measured by a combination
of large-scale structure observations, microwave background
anisotropies, Supernovae Ia observations and Hubble Space Telescope
measurements, a feature named somewhat idiosyncratically as ``cosmic
complementarity'', see Ref.~\cite{Complementarity}. It is expected that
in the near future we will be able to determine the parameters of the
standard cosmological model with great precision from a combination of
several different experiments, as shown in Fig.~\ref{fig42}.

\section{CONCLUSION}

We have entered a new era in cosmology, were a host of high-precision
measurements are already posing challenges to our understanding of the
universe: the density of ordinary matter and the total amount of energy
in the universe; the microwave background anisotropies on a fine-scale
resolution; primordial deuterium abundance from quasar absorption lines;
the acceleration parameter of the universe from high-redshift supernovae
observations; the rate of expansion from gravitational lensing; large
scale structure measurements of the distribution of galaxies and their
evolution; and many more, which already put constraints on the parameter
space of cosmological models, see Fig.~\ref{fig30}. However, these are
only the forerunners of the precision era in cosmology that will
dominate the new millennium, and will make cosmology a phenomenological
science.

It is important to bear in mind that all physical theories are
approximations of reality that can fail if pushed too far. Physical
science advances by incorporating earlier theories that are
experimentally supported into larger, more encompassing frameworks. The
standard Big Bang theory is supported by a wealth of evidence, nobody
really doubts its validity anymore. However, in the last decade it has
been incorporated into the larger picture of cosmological inflation,
which has become the new standard cosmological model. All cosmological
issues are now formulated in the context of the inflationary paradigm.
It is the best explanation we have at the moment for the increasing
set of cosmological observations.

In the next few years we will have an even larger set of high-quality
observations that will test inflation and the cold dark matter paradigm
of structure formation, and determine most of the 12 or more parameters
of the standard cosmological model to a few percent accuracy (see
table~1). It may seem that with such a large number of parameters one
can fit almost anything. However, that is not the case when there is
enough quantity and quality of data. An illustrative example is the
standard model of particle physics, with around 21 parameters and a host
of precise measurements from particle accelerators all over the world.
This model is, nowadays, rigurously tested, and its parameters measured
to a precision of better than 1\% in some cases. It is clear that
high-precision measurements will make the standard model of cosmology as
robust as that of particle physics. In fact, it has been the
technological advances of particle physics detectors that are mainly
responsible for the burst of new data coming from cosmological
observations. This is definitely a very healthy field, but there is
still a lot to do. With the advent of better and larger precision
experiments, cosmology is becoming a mature science, where speculation
has given way to phenomenology.

There are still many unanswered fundamental questions in this emerging
picture of cosmology. For instance, we still do not know the nature of
the inflaton field, is it some new fundamental scalar field in the
electroweak symmetry breaking sector, or is it just some effective
description of a more fundamental high energy interaction? Hopefully, in
the near future, experiments in particle physics might give us a clue to
its nature. Inflation had its original inspiration in the Higgs field,
the scalar field supposed to be responsible for the masses of elementary
particles (quarks and leptons) and the breaking of the electroweak
symmetry. Such a field has not been found yet, and its discovery at the
future particle colliders would help understand one of the truly
fundamental problems in physics, the origin of masses. If the
experiments discover something completely new and unexpected, it would
automatically affect inflation at a fundamental level.

One of the most difficult challenges that the new cosmology will have to
face is understanding the origin of the cosmological constant, if indeed
it is confirmed by independent sets of observations. Ever since Einstein
introduced it as a way to counteract gravitational attraction, it has
haunted cosmologists and particle physicists for decades. We still do
not have a mechanism to explain its extraordinarily small value, 120
orders of magnitude below what is predicted by quantum physics. For
several decades there has been the reasonable speculation that this
fundamental problem may be related to the quantization of gravity.
General relativity is a classical theory of space-time, and it has proved
particularly difficult to construct a consistent quantum theory of
gravity, since it involves fundamental issues like causality and the
nature of space-time itself. 

The value of the cosmological constant predicted by quantum physics is
related to our lack of understanding of gravity at the microscopic
level. However, its effect is dominant at the very largest scales of
clusters or superclusters of galaxies, on truly macroscopic scales. This
hints at what is known in quantum theory as an anomaly, a quantum
phenomenon relating both ultraviolet (microscopic) and infrared
(macroscopic) divergences. We can speculate that perhaps general
relativity is not the correct description of gravity on the very largest
scales. In fact, it is only in the last few billion years that the
observable universe has become large enough that these global effects
could be noticeable. In its infancy, the universe was much smaller than
it is now, and, presumably, general relativity gave a correct description
of its evolution, as confirmed by the successes of the standard Big Bang
theory. As it expanded, larger and larger regions were encompassed, and,
therefore, deviations from general relativity would slowly become
important. It may well be that the recent determination of a
cosmological constant from observations of supernovae at high redshifts
is hinting at a fundamental misunderstanding of gravity on the very
large scales.

If this were indeed the case, we should expect that the new generation
of precise cosmological observations will not only affect our
cosmological model of the universe but also a more fundamental
description of nature.

\vskip5mm

\section*{ACKNOWLEDGEMENTS}
I thank the organizers of the CERN-JINR European School of High Energy
Physics for a very warm and friendly atmosphere. I also would like to
thank my friends and collaborators Andrei Linde, Andrew Liddle, David
Wands, David Lyth, Jaume Garriga, Xavier Montes, Enrique Gazta\~naga,
Elena Pierpaoli, Stefano Borgani, and many others, for sharing with me
their insight about this fascinating science of cosmology. This work was
supported by the Royal Society.


\end{document}